\documentclass[12pt]{article}
\usepackage{latexsym}
\usepackage{amssymb}
\usepackage{amsmath}

\newcommand{\be}{\begin{equation}}
\newcommand{\ee}{\end{equation}}
\def\tr{{\rm tr}}
\newcommand{\Tr}{{\rm Tr}}

\def\cN{{\cal N}}
\def\bea{\begin{eqnarray}}
\def\eea{\end{eqnarray}}
\def\nn{\nonumber}

\begin{document}
\begin{titlepage}

\phantom{x}

\vspace{1cm}

\begin{center}
{\LARGE\bf Superfield theories on $S^3$\\[3mm] and their localization}

\vspace{1.5cm}

{\large\bf I.B.~Samsonov \footnote{On leave from
Tomsk Polytechnic University, 634050 Tomsk, Russia}  and
D.~Sorokin}
\\[10pt]
{\it INFN, Sezione di Padova, 35131 Padova, Italy}\\[1mm]
{\tt samsonov, sorokin@pd.infn.it}
\end{center}
\vspace{0.5cm}

\begin{abstract}
We consider the superfield formulation of supersymmetric gauge and
matter field theories on a three--dimensional sphere with rigid
$\cN=2$ supersymmetry, as well as with $\cN> 2$. The
construction is based on a supercoset $SU(2|1)/U(1)$ containing
$S^3$ as the bosonic subspace. We derive an explicit form of
$SU(2|1)/U(1)$ supervielbein and covariant derivatives, and use
them to construct classical superfield actions for gauge and
matter supermultiplets in this superbackground. We then apply
superfield methods for computing one--loop partition functions of
these theories and demonstrate how the localization technique
works directly in the superspace.
\end{abstract}

\end{titlepage}


\numberwithin{equation}{section}

\section{Introduction}
Supersymmetric field theories on curved backgrounds
with rigid supersymmetries are in an intermediate position between
locally supersymmetric field theories coupled to supergravity and those in flat
space. Although these theories describe field dynamics in curved
space--time, they share many properties of corresponding field
theories in flat space, in particular, when the theory is (super)conformal and the background is conformally flat. In such cases results of quantum computations performed in curved (compact) backgrounds can be extrapolated to the flat--space field theory.

For field theories on curved backgrounds with
rigid supersymmetry there is a special tool which allows one to compute quantum objects, such as the partition function,
correlators or Wilson loops {\it exactly}, beyond the perturbation
theory. This is the
so--called localization method (see \emph{e.g.} \cite{BT} for a review and references) whose efficiency was exploited by Pestun \cite{Pestun} for studying
non-perturbative aspects of four--dimensional superconformal field
theories on $S^4$. Subsequently, this technique was
extended to field theories in diverse dimensions and to other interesting curved
supersymmetric backgrounds. It has proved to be one of the most powerful approaches to
study quantum dynamics of supersymmetric field theories non--perturbatively.

These developments brought into the foreground the problem of a
systematic construction of classical actions for field models on
curved backgrounds with rigid supersymmetry, which until recently
was mainly of an academic interest. Within the component field
formulation, the systematic approach for solving this problem was
developed in \cite{FS,Closset1,Closset2,Closset3}. The prescription is to
couple a supersymmetric field model to off--shell supergravity (which requires the presence of auxiliary fields) and then to
`freeze' a supergravity background such that it preserves some
number of supersymmetries. In the limit of large Plank mass the
gravity fluctuations decouple and one is left
with the field theory model on the fixed curved background
which, by construction, respects the supersymmetries of the background.

Within the superfield formulation of supergravity and supersymmetric field theories (see, \emph{e.g.}, \cite{BKbook,GGRS})
the prescription of  \cite{FS} is carried out straightforwardly, since the superfield
formulations include all the necessary auxiliary fields which
automatically receive correct values when one fixes the
superfield background. So, in superspace one can, in principle, construct any field
theory on curved background with rigid supersymmetries when the
corresponding superfield actions in flat superspace are available and a curved
superbackground possessing superisometries is chosen.

The problem is to solve superfield supergravity constraints for a
given superbackground and to find an explicit form of the
superfield objects, such as supervielbeins and super--connections,
which encode its geometry. This problem is drastically simplified
when the background superspace has the structure of a supercoset
manifold $G/H$ (as \emph{e.g.} a supersphere, or an $AdS$
superspace) with $G$ being the isometry supergroup and $H$ being
its stability subgroup. In these cases the superbackground
geometry is described by Cartan superforms on $G/H$, which satisfy
corresponding Maurer--Cartan equations. The derivation of an
explicit form of the $G/H$ Cartan superforms as series expansions
in powers of Grassmann--odd coordinates  is carried out by
conventional group--theoretical methods. Once this is done, it is
straightforward to consider field models in such a curved
superspace. We will follow exactly this strategy and develop basic
methods for studying some classical and quantum aspects of such
theories.

We will mainly consider
three-dimensional gauge and matter field theories with $\cN=2$
supersymmetry, \emph{i.e.}
with four supercharges, on the round  $S^3$ sphere, but will also
discuss $\cN=2$ superfield formulations of $\cN=4,6$ and 8 supersymmetric theories.
The appropriate superspace with four Grassmann--odd directions,
whose bosonic subspace is $S^3$, is the supercoset $SU(2|1)/U(1)$. For this supercoset
we construct explicitly all the basic geometric objects such as
supervielbeins, superconnection, supertorsion and supersymmetric
covariant derivatives.
We consider
superfield actions on $SU(2|1)/U(1)$ which are, in fact, Euclidean counterparts of superfield models in an $AdS_3$
superspace considered in \cite{Kuzenko1,Kuzenko2,Kuzenko3,Kuzenko4}. Next, we develop
methods of quantum one--loop  computations for such superfield
theories and show how to apply the localization technique to
the Chern--Simons theory in $\cN=2$ superspace which was considered originally in
\cite{KWY,KWY1} employing conventional component fields.

The superspace and superfield techniques allow us to
make several simple observations about field theories on $S^3$
with rigid supersymmetries.
\if{}
First, it is possible to choose
superspace coordinates such that one half of supersymmetry
generators ${\mathbb Q}_\alpha$ are given simply by the operators
of supertranslations along Grassmann directions such that
\be
{\mathbb Q}^2 = {\mathbb Q}^\alpha{\mathbb Q}_\alpha = \partial^\alpha \partial_\alpha\,,\qquad
\partial_\alpha = \frac\partial{\partial\theta^\alpha}\,.
\ee
As a consequence, any full superspace action is Q-exact since
$\partial^2$ is a part of the full superspace measure. This is
useful for the needs of the localization method since it is
essentially based on deforming the original theory by Q-exact
terms.
\fi
For instance, we find that the supervolume of $SU(2|1)/U(1)$ vanishes,
\be
\int d^3x d^2\theta d^2\bar\theta\,E = 0\,,
\ee
where $E= {\rm
Ber}E_M{}^A$ is the Berezinian of the $SU(2|1)/U(1)$ supervielbein. In particular, for the $\cN=2$ super--Yang--Mills theory this fact
trivializes the problem of finding critical points, \emph{i.e.} the values of (super)fields for which the SYM action vanishes,
\be
0=S_{\rm SYM} \propto \tr\int d^3x d^2\theta d^2\bar\theta\,E\, G^2\quad
\Rightarrow \quad
G = const.
\ee
Here $G$ is the superfield strength of the $\cN=2$ gauge superfield
$V(x,\theta,\bar\theta)$. In components this superfield starts with the scalar $\sigma(x)$
which is part of the $\cN=2$, $d=3$ gauge supermultiplet. As we will show, in a certain supersymmetric gauge the vanishing of the $SU(2|1)/U(1)$ supervolume also trivializes the contribution of the gauge supermultiplet into the SYM partition function which acquires non--trivial structure due to Faddeev--Popov and Nielsen--Kallosh ghosts.

We will also show that the geometry of the supercoset $SU(2|1)/U(1)$ is superconformally flat.
This property is useful for extending quantum superfield methods from flat superspace to $SU(2|1)/U(1)$.

When constructing an $\cN=4$ supersymmetric extension of the $\cN=2$ SYM
theory on $S^3$ by adding to the latter a chiral matter superfield,
we come across the fact that when the chiral superfield carries a
non--zero $\cN=2$ R--charge, the invariance of the SYM action under
$\cN=4$ supersymmetry, in general, requires the presence of a Chern--Simons term
(see Section \ref{Sect4.1} for details).
In the component formulation this fact was first noticed in \cite{HHL}
using $SU(2)_R$ symmetry arguments. In this paper we will present
an explicit form of the $\cN=4$ supersymmetry transformations on
$S^3$, which to the best of our knowledge have not been given
in the literature before.

Finally, we point out that the superfield approach is quite useful
at the quantum level. The localization method effectively reduces functional integrals
to the problem of computing one--loop determinants of operators of quadratic
fluctuations of bosonic and fermionic fields around
critical points (see, \emph{e.g.}, \cite{Marino11} for a review).
As a rule, these one--loop determinants are given by simple
elementary functions since many bosonic and fermionic modes
cancel against each other due to supersymmetry. As we will show, in the
superfield gauge theories on $S^3$ the one--loop determinants
correspond to supersymmetric operators acting on superfields propagating on
the coset $SU(2|1)/U(1)$. For such operators the pairing of
bosonic and fermionic modes is automatic, since the gauge fixing is
supersymmetric. This is a useful feature of the superspace approach.

The main part of this paper is organized as follow. In Section 2 we consider the geometry of the supercoset
$SU(2|1)/U(1)$. In particular, we construct in a suitable chiral basis the
supervielbeins, supercurvature, supercovariant derivatives and the Killing supervector.
The geometry of $SU(2|1)/U(1)$ is shown to be superconformally flat. In Section 3 we
introduce classical $\cN=2$ superfield actions for gauge and matter fields
on $SU(2|1)/U(1)$. Section 4 is devoted to constructing
$\cN=2$ superfield actions for models with
extended supersymmetry, such as $\cN=4$ SYM and Gaiotto--Witten theories, $\cN=8$ SYM and $\cN=6$ ABJM
theory. In Section 5 we develop
superfield methods of one--loop quantum computations in $\cN=2$ SYM and
chiral matter models on $SU(2|1)/U(1)$  and use them, in particular, for computing one--loop
partition functions. In Section 6 we consider how the localization techniques works for
the $\cN=2$ Chern--Simons theory in the superfield form. Section 7
is devoted to discussions of the results and perspectives. In appendices we collect details of
direct computations of determinants of supersymmetric operators
and revisit component field calculations of the SYM partition
function.

\section{$SU(2|1)/U(1)$ supergeometry}
\subsection{$su(2|1)$ superalgebra}

We are interested in field theories on $S^3$ which are invariant under the $SU(2|1)$ supergroup.
We would like to describe these theories in a superspace whose isometries include
$SU(2|1)$ \footnote{For the construction of quantum mechanical models on
different cosets of $SU(2|1)$ see \emph{e.g.} \cite{Ivanov:2013ova,Ivanov:2013cea} and
references therein.}.
So, we need a superspace with three bosonic
variables $x^m$, $m=1,2,3$ and four Grassmann--odd variables $\theta^\mu$,
$\bar\theta^\mu$, $\mu =1,2$ such that its bosonic body is the sphere $S^3$. The $SU(2)\times SU(2)$
isometry of $S^3$ naturally embeds into the supergroup
$SU(2|1)\times SU(2)$, so one can realize the superspace in question as the supercoset
\be
\frac{SU(2|1)\times SU(2)}{U(1) \times SU(2)}\,.
\label{big-coset}
\ee
Formally, the $SU(2)$ factors cancel against each other. Hence, we
can obtain the same superspace by considering a simpler coset
\be
\frac{SU(2|1)}{U(1)}\,.
\label{small-coset}
\ee
The only price for this is that not all the $SU(2)\times SU(2)$ isometries of $S^3$ are
explicit in this case. However, the second $SU(2)$ symmetry is realized as the group of external automorphisms of the $su(2|1)$ algebra and, hence, can be easily included in the construction.

The $su(2|1)$ (anti)commutation relations are
\bea
&&
{}[M_a,M_b]=\frac{2 i}{r}\varepsilon_{abc}M_c\,,\nn\\
&&{}[M_a, Q_\alpha]=-\frac1{r}(\gamma_a)_\alpha^\beta Q_\beta\,,
\quad
[M_a, \bar Q_\alpha]=-\frac1{r}(\gamma_a)_\alpha^\beta \bar Q_\beta\,,
\nn\\&&
\{ Q_\alpha ,\bar Q_\beta \}=\gamma^a_{\alpha\beta}M_a
+\frac1{r}\varepsilon_{\alpha\beta} R\,,\quad
[R,Q_\alpha]=-Q_\alpha\,,\quad
[R,\bar Q_\alpha]=\bar Q_\alpha\,
\label{superalgebra}
\eea
(all other (anti)commutators vanish.) Here
$M_a$, $(a=1,2,3)$ are three generators of the $SU(2)$
subgroup, while $R$ is the $U(1)$ R--symmetry generator and $Q_\alpha$ and $\bar Q_\alpha$, ($\alpha=1,2$) are the Grassmann--odd supersymmetry generators.
The parameter $r$ is the radius of the sphere and $(\gamma^a)^\beta_{\alpha}$ are the Pauli matrices. For the details on our notation and conventions see Appendix A.

The group of the external $SU(2)$ automorphisms of the $su(2)$ algebra is generated by an independent set of three generators $L_a$
\be\label{2su2}
[L_a,L_b]={2i}\varepsilon_{abc}L_c\,,
\ee
whose commutation relations with the $SU(2|1)$ generators are
\be\label{auto}
[L_a,M_b]=2i\varepsilon_{abc}M_c\,,\quad [L_a, Q_\alpha]=-(\gamma_a)_\alpha^\beta Q_\beta\,,\quad [L_a, \bar Q_\alpha]=-(\gamma_a)_\alpha^\beta \bar Q_\beta\,\quad [L_a,R]=0\,.
\ee
The generators $M_a$ and $L_a$ form the $SO(4)\sim SU(2)\times SU(2)$ isometry
of $S^3$ with the two $SU(2)$'s being generated by $M_a$ and $(L_a-\sqrt{r} M_a)$, respectively.
Note that the latter commute with the whole $SU(2|1)$.

In the limit
$r\to\infty$ the algebra (\ref{superalgebra}), \eqref{2su2} and \eqref{auto} reduces to the
standard three--dimensional Euclidean ``Poincar\' e'' superalgebra
in which $M_a$ play the role of commuting momenta
operators and $L_a$ generate the $SO(3) \sim SU(2)$ rotations in flat $3d$ space.

The superalgebra (\ref{superalgebra}) is invariant under the following
Hermitian conjugation of the generators
\be
(M_a)^\dag = M_a\,,\qquad R^\dag = R\,, \qquad (Q_\alpha)^\dag =
\bar Q^\alpha\,.
\label{conj}
\ee
Note that the spinor index changes its position under the
conjugation since the spinor group is $SU(2)$.

In the rest of this section we will derive, using the superalgebra
(\ref{superalgebra}), an explicit form of supersymmetric
vielbeins, connections, torsion, curvature and
covariant derivatives on the supercoset $SU(2|1)/U(1)$ with the aim of using them afterwards for the construction of superfield actions.

\subsection{Supervielbein}\label{sv}
Let $z^M=(x^m,\theta^\mu,\bar\theta^\mu)$ be local coordinates
parametrizing the supercoset $SU(2|1)/U(1)$. In principle, the
coordinates $\theta^\mu$ and $\bar\theta^\mu$ can be related to each other by
complex conjugation, $(\theta^\mu)^* = \bar\theta_\mu$, in
accordance with the conjugation rules (\ref{conj}) of the operators
$Q_\alpha$ and $\bar Q_\alpha$.
However, in a $d=3$ superspace with the metric of Lorentzian signature  the
spinor group is $SL(2,{\mathbb R})$ and the spinor index does not change its
position under conjugation. We wish to consider
superfield models on $SU(2|1)/U(1)$ which are related by Wick
rotation to the corresponding models in the $AdS_3$ superspace,
considered, \emph{e.g.}, in \cite{Kuzenko1,Kuzenko2,Kuzenko3,Kuzenko4}. Clearly,
such Wick--rotated models are not necessary real under the conjugation
(\ref{conj}). Therefore, in what follows we will treat the complex coordinates
$\theta^\mu$ and $\bar\theta^\mu$ as independent ones, \emph{i.e.} not related
to each other by the complex conjugation.

The
 $SU(2|1)/U(1)$ supervielbein is given by the set of one--forms,
\be
E^A=dz^M E_M{}^A(z)\,,\qquad
E^A=(E^a,E^\alpha,\bar E^\alpha)\,.
\ee
They are components of the $SU(2|1)$ Cartan form
\be\label{dG}
G^{-1} dG =iE^a M_a
+iE^\alpha Q_\alpha +i\bar E^\alpha \bar Q_\alpha+ i\Omega_{(R)} R
\equiv \omega\,,
\ee
where $G(z^M)$ is a representative of the supercoset $SU(2|1)/U(1)$
and $\Omega_{(R)}$ is the $U(1)$-connection.
In particular, one can consider the following coset representative
\be\label{bf}
G=b(x)f(\theta,\bar\theta)\,, \quad b(x)=e^{i x^m M_m}
\,,\quad f(\theta,\bar\theta)=e^{i\theta^\alpha Q_\alpha}e^{i\bar\theta^\beta \bar
Q_\beta}\,,
\ee
such that
\be\label{dG1}
G^{-1}dG= f^{-1}(d+ie^a(x) M_a)f\,,
\ee
where $e^a(x)=dx^m e^a_m(x)$ is the bosonic vielbein on $S^3\sim SU(2)$. Applying
the algebra (\ref{superalgebra}) we find the components of the
supervielbein in the decomposition (\ref{dG}) explicitly,
\footnote{We use the following conventions for the contractions of spinor indices:
$\theta^2=\theta^\alpha \theta_\alpha$, $\bar\theta^2=\bar\theta^\alpha \bar\theta_\alpha$.
The spinor indices are raised and lowered by the rules
$\theta_\alpha=\varepsilon_{\alpha\beta}\theta^\beta$, $\theta^\alpha=\varepsilon^{\alpha\beta}\theta_{\beta}$,
$\varepsilon_{12}=-\varepsilon^{12}=1$, see Appendix A.}
\bea\label{OR}
E^\alpha&=&{\bf d} \theta^\alpha\,,\nn\\
{\bar E}^\alpha&=&{\bf d} \bar\theta^\alpha - \frac1{r}{\bf d}
\theta^\alpha\, \bar\theta^2\,,\nn\\
E^a&=&e^a-i{\bf d} \theta^\alpha
\gamma^a_{\alpha\beta} \bar\theta^\beta\,,
\eea
where $\bf d$ is the Killing--spinor covariant differential,
\be
{\bf d}  \theta^\alpha= d\theta^\alpha
-\frac i{r}e^a (\gamma_a)_\beta^\alpha \theta^\beta
\,,\qquad
{\bf d}^2=0\,.
\ee
The $U(1)$--connection of the R--symmetry has also very simple form,
\be
\Omega_{(R)}=-\frac i{r}{\bf d} \theta^\alpha\bar\theta_\alpha
=-\frac i{r}E^\alpha \bar\theta_\alpha\,.
\ee

It is easy to see that the $SU(2|1)/U(1)$ supergeometry constructed in this way has a
smooth flat limit at $r\rightarrow \infty$. Note that the
components $E^\alpha$ and $\bar E^\alpha$ enter in (\ref{OR}) asymmetrically.
Therefore we refer to the basis defined by the coset
representative (\ref{bf}) as the chiral basis.

Consider now the inverse supervielbein, \emph{i.e.} the differential operator of the form
\be\label{E_A}
E_A=E_A{}^M\partial_M\,,\qquad E_A{}^M E_M{}^B=\delta_A^B\,.
\ee
In the chiral coordinates corresponding to the choice of the coset
representative (\ref{bf}) its components have the following
explicit form
\bea
E_a&=&\partial_a +\frac i{r}(\gamma_a)^\alpha_\beta \theta^\beta
\partial_\alpha
+\frac i{r}(\gamma_a)^\alpha_\beta \bar\theta^\beta \bar\partial_\alpha
\,,\nn\\
E_\alpha&=&\partial_\alpha+i\gamma^a_{\alpha\beta}\bar\theta^\beta\partial_a
-\frac1{r}\theta^\beta\bar\theta_\beta \partial_\alpha
+\frac1{r}\theta_\alpha\bar\theta^\beta\partial_\beta
-\frac1{2r}\bar\theta^2 \bar\partial_\alpha\,,\nn\\
\bar E_\alpha&=&\bar\partial_\alpha\,.
\label{48}
\eea
Here $\partial_a=e_a^m(x)\partial_m$ is the
differential operator on $S^3$ with the commutation relations
$[\partial_a,\partial_b]=-\frac2r\varepsilon_{abc}\partial_c$.
The differential operators (\ref{48}) obey the following algebra
\bea
\{E_\alpha,\bar E_\beta \}&=&i\gamma^a_{\alpha\beta}E_a
+ \frac1{r}\bar\theta_{\alpha}\bar E_{\beta}\,,
\qquad
\{E_\alpha, E_\beta  \}=-\frac2r \bar\theta_{(\alpha} E_{\beta)}\,,
\nn\\
{}[E_a,\bar E_\alpha]&=&-\frac i{r}(\gamma_a)_\alpha^\beta \bar
E_\beta\,,\qquad
[E_a, E_\alpha]=-\frac i{r}(\gamma_a)_\alpha^\beta
E_\beta\,,\nn\\
{}[E_a,E_b]&=& -\frac 2r\varepsilon_{abc}E_c\,.
\label{DD}
\eea

It is interesting to note that the Berezinian of the supervielbein
is independent of the Grassmann variables,
\be\label{ber}
E\equiv {\rm Ber}E_M{}^A=\det e_m^a(x)=\sqrt{ h(x)}\,,
\ee
where $h(x)=\det h_{mn}(x)$ and $h_{mn}(x)$ is a purely bosonic
metric on $S^3$. The expression \eqref{ber} is obtained for a
particular choice of the coset representative (\ref{bf}),\emph{ i.e.} it
corresponds to the chiral coordinates on $SU(2|1)/U(1)$. However, the
coordinate--independent consequence of (\ref{ber}) is the fact that
the supervolume of the supercoset $SU(2|1)/U(1)$ vanishes
\be
\int d^3x d^2\theta d^2\bar\theta \, E =0\,.
\label{zero-volume}
\ee
In Section\ \ref{flatness} this property will also be checked in a different (superconformally flat) basis.

\subsection{Connection, torsion and curvature}
By construction, the differential form $\omega$ given in
(\ref{dG}) obeys the Maurer--Cartan equation,
\be
d\omega+\frac 12  [\omega, \omega]=0\,.
\ee
The corresponding equations for the components of the
supervielbein $E^A$ and the $U(1)$ connection $\Omega_{(R)}$ are
\bea
dE^a -\frac1{r}\varepsilon^{abc}E^b\wedge E^c
 -i E^\alpha \wedge \bar E^\beta \gamma^a_{\alpha\beta}&=&0\,,\nn\\
dE^\alpha - i\Omega_{(R)}\wedge E^\alpha - \frac i{r} E^a \wedge
E^\beta (\gamma_a)_\beta^\alpha&=&0\,,\nn\\
d\bar E^\alpha+i\Omega_{(R)}\wedge\bar E^\alpha - \frac i{r} E^a \wedge
E^\beta(\gamma_a)^\alpha_\beta &=&0\,,\nn\\
d\Omega_{(R)}-\frac i{r} \varepsilon_{\alpha\beta}E^\alpha\wedge
\bar E^\beta&=&0\,.
\label{24}
\eea

Let us introduce the superconnection $\Omega^{AB}$ with the following non--vanishing
components $\Omega^{ab}$, $\Omega^\alpha_\beta$ and
$\bar\Omega^\alpha_\beta$:
\bea
\Omega^{ab}&=&\frac1{r}\varepsilon^{abc}E^c\,, \nn\\
\Omega^\alpha_{\beta}&=&-\frac i{2r} (\gamma^a)^\alpha_\beta E^a
-i\delta^\alpha_\beta\Omega_{(R)} \,,\nn\\
\bar\Omega^\alpha_{\beta}&=&-\frac i{2r} (\gamma^a)^\alpha_\beta
E^a
+i\delta^\alpha_\beta \Omega_{(R)} \,.
\eea
The superconnestion $\Omega$ appears in the covariant
differential,
\be
{\cal D}=d+\Omega\,.
\label{covdiff}
\ee
In particular, the equations (\ref{24}) take the form
\be \label{DE}
{\cal D}E^A=dE^A +\Omega^{AB}\wedge E^B=T^A\,,
\ee
where the supertorsion $T^A$ has the following components
\bea
T^a &=& i\gamma^a_{\alpha\beta}E^\alpha \wedge \bar E^\beta\,,\nn\\
T^\alpha &=& \frac i{2r}(\gamma_a)^\alpha_\beta E^a\wedge E^\beta\,,\nn\\
\bar T^\alpha &=& \frac i{2r}(\gamma_a)^\alpha_\beta E^a\wedge \bar
E^\beta\,.
\label{torsion}
\eea

Given the superconnection $\Omega^{AB}$ we construct the
supercurvature,
\be
{\cal R}^{AB}=d\Omega^{AB}+\Omega^{AC}\wedge \Omega^{CB}\,,
\ee
or, explicitly,
\bea
{\cal R}^{ab}&=&d\Omega^{ab}+\Omega^{ac}\wedge\Omega^{cb}=\frac1{r^2}E^a\wedge
E^b+\frac i{r} \varepsilon^{abc}\gamma^c_{\alpha\beta}E^\alpha \wedge
\bar E^\beta\,,\nn\\
{\cal R}^\alpha_\beta&=&d\Omega^\alpha_{\beta}+\Omega^\alpha_\gamma\wedge \Omega^\gamma_\beta\
\nn\\&=&-\frac i{4r^2}\varepsilon^{abc}(\gamma^c)^\alpha_\beta
E^a\wedge E^b
-\frac1{2r}(\delta^\alpha_\rho \varepsilon_{\beta\sigma}
+\delta^\alpha_\sigma \varepsilon_{\beta\rho}-2\delta^\alpha_\beta\varepsilon_{\rho\sigma})
E^\rho\wedge \bar E^\sigma\,,\nn\\
\bar{\cal R}^\alpha_\beta&=&d\bar\Omega^\alpha_{\beta}+\bar\Omega^\alpha_\gamma\wedge \bar\Omega^\gamma_\beta
\nn\\&=&-\frac i{4r^2}\varepsilon^{abc}(\gamma^c)^\alpha_\beta
E^a\wedge E^b
-\frac1{2r}(\delta^\alpha_\rho \varepsilon_{\beta\sigma}
+\delta^\alpha_\sigma \varepsilon_{\beta\rho}+2\delta^\alpha_\beta\varepsilon_{\rho\sigma})
E^\rho\wedge \bar E^\sigma\,.
\eea
These equations can be rewritten in one line,
\be
\label{curvature}
{\cal R}=-\frac i{4r} M^{ab} E^a\wedge E^b
+\left(
\frac 1{2} M^{a} \gamma^a_{\alpha\beta} - R \varepsilon_{\alpha\beta} \right) E^\alpha \wedge \bar
E^\beta\,,
\ee
where we assume that the momentum operator $M^{ab}$ acts on the
tangent space vectors $v^a$ and spinors $\psi^\alpha$ by the rule
\be
M^{a}v^b=\frac{2 i}r \varepsilon^{abc} v^c\,,\qquad
M^{a}\psi^\alpha=\frac1{r}(\gamma^a)^\alpha_\beta
\psi^\beta\,.
\ee

The R--symmetry generator acts on a complex superfield $\Phi$ as follows
\be
R\Phi=- q\Phi\,,\qquad
R\bar\Phi=q\bar\Phi\,,
\ee
where $q$ is the R--charge of the field.

\subsection{Covariant derivatives}
Consider the covariant derivatives on the supercoset
$SU(2|1)/U(1)$,
\be
\label{covD}
{\cal D}_A=E_{A}+\Omega_A=({\cal D}_a,{\cal D}_\alpha,\bar{\cal
D}_\alpha)\,.
\ee
They appear in the decomposition of the covariant differential
(\ref{covdiff}) in the tangent--space basis formed by the supervielbein,
\be
\label{D}
\mathcal D=d+\Omega=E^A\mathcal D_{A}=E^a\mathcal D_a +
E^\alpha\mathcal D_\alpha+\bar E^\alpha\bar{\mathcal  D}_\alpha\,.
\ee

To find the algebra of the covariant derivatives we use the fact that
the covariant differential squares to the curvature, $\mathcal D^2={\cal R}$.
This implies that
\be
T^A{\cal D}_A- E^A\wedge E^B{\cal D}_B{\cal D}_A= {\cal R}\,.
\ee
We plug the explicit expressions for the supercurvature
(\ref{curvature}) and supertorsion (\ref{torsion}) into this
equation and obtain the $SU(2|1)$ (anti)commutation relations,
\bea&&
[{\cal D}_a,{\cal D}_b]=-\frac i{2r}M_{ab}\,,\quad
[{\cal D}_a,{\cal D}_\alpha]=-\frac i{2r}(\gamma_a)_\alpha^\beta
{\cal D}_\beta\,,\quad
[{\cal D}_a,\bar{\cal D}_\alpha]=-\frac i{2r}(\gamma_a)_\alpha^\beta
\bar{\cal D}_\beta\,,\nn\\&&
\{{\cal D}_\alpha, \bar{\cal D}_\beta \}=i\gamma^a_{\alpha\beta}{\cal
D}_a
-\frac 1{2} \gamma^a_{\alpha\beta}M_{a}
+\frac 1{r}\varepsilon_{\alpha\beta}R\,,\nn\\&&
\{{\cal D}_\alpha,{\cal D}_\beta \}=\{\bar {\cal D}_\alpha,\bar {\cal D}_\beta
\}=0\,.
\label{deriv}
\eea
The generators $M_{ab}$ and $R$ have the following commutators with
$\mathcal D_A$
\bea&&
[M_{ab},{\cal D}_c]=\frac{2 i}r(\delta_{ac}{\cal D}_b-\delta_{bc}{\cal
D}_a)\,,\nn\\&&
[M_{ab},{\cal D}_\alpha]=-\frac 1{r}\varepsilon_{abc}(\gamma^c)_\alpha^\beta {\cal D}_\beta\,,\quad
[M_{ab},\bar{\cal D}_\alpha]=-\frac 1{r}\varepsilon_{abc}(\gamma^c)_\alpha^\beta\bar {\cal D}_\beta\,,
\nn\\&&
{}[R,{\cal D}_\alpha]={\cal D}_\alpha\,,\qquad
[R,\bar{\cal D}_\alpha]=-\bar{\cal D}_\alpha\,.
\label{deriv1}
\eea

The covariant derivatives ${\cal D}_A=E_A+\Omega_A$ can be written explicitly in the chiral
coordinates corresponding to the coset representative (\ref{bf}).
 To this end, we need to find the form of the superconnection
$\Omega_A$ in these coordinates,
\be
\Omega_A=i\Omega_{(R)A} R+\frac i2\Omega_{ab\, A}M_{ab}\,,
\ee
where the components of $\Omega_{(R)A}$ and $\Omega_{ab\, A}$ read
\bea
\Omega_{(R)a}&=&0\,,\quad \Omega_{(R)\alpha}=-\frac
i{r}\bar\theta_\alpha\,,\quad \bar\Omega_{(R)\alpha}=0\,,\nn\\
\Omega_{ab\,c}&=&-\frac1{2}\varepsilon_{abc}\,,\quad
\Omega_{ab\,\alpha}=\bar\Omega_{ab\,\alpha}=0\,.
\eea
Now recall that the supervielbein in these coordinates is given in
(\ref{48}), so combining the above expressions with (\ref{48}) we get
\bea
{\cal D}_a&=&\partial_a -\frac i2 M_a+\frac i{r}(\gamma_a)^\alpha_\beta \theta^\beta
\partial_\alpha
+\frac i{r}(\gamma_a)^\alpha_\beta \bar\theta^\beta \bar\partial_\alpha
\,,\nn\\
{\cal D}_\alpha&=&\partial_\alpha+i\gamma^a_{\alpha\beta}\bar\theta^\beta\partial_a
-\frac1{r}\theta^\beta\bar\theta_\beta \partial_\alpha
+\frac1{r}\theta_\alpha\bar\theta^\beta\partial_\beta
-\frac1{2r}\bar\theta^2 \bar\partial_\alpha
+\frac1r\bar\theta_\alpha R \,,\nn\\
\bar {\cal D}_\alpha&=&\bar\partial_\alpha\,.
\label{D-explicit}
\eea
One can check that these differential operators obey the algebra (\ref{deriv})
and (\ref{deriv1}). Note that the covariant derivative $\bar{\cal
D}_\alpha$ is short as it should be in the chiral coordinate basis.

\subsection{Superconformal flatness}
\label{flatness}
On general grounds \cite{BILS}, it is natural to expect that the supercoset $SU(2|1)/U(1)$ should
be superconformally flat, since $SU(2|1)/U(1)$  is an Euclidean counterpart of the
$AdS_3$ superspace $\frac{OSp(2|2)\times Sp(2)}{SO(2)\times Sp(2)}$ which was demonstrated to be superconformally flat in
\cite{Kuzenko2,Kuzenko3}. Here we
prove this explicitly by showing that the covariant derivatives on
$SU(2|1)/U(1)$ are related to flat superspace
derivatives by means of a super Weyl transformation.

Let $z^m=(x^m,\theta^\alpha,\bar\theta^\alpha)$ be the coordinates
on the flat Euclidian $\cN=2$, $d=3$ superspace. In the flat case
there is no difference between the indices of the local coordinates
$x^m$ and tangent space, $x^a$, \emph{i.e.}, $\partial_a=\partial_m=\frac{\partial}{\partial
x^m}$.
The flat covariant spinor derivatives in the chiral basis
are given by $D_M=(\partial_m,D_\alpha,\bar D_\alpha)$,
\be
D_\alpha=\partial_\alpha+i\gamma^a_{\alpha\beta}\bar\theta^\beta\partial_a\,,
\quad
\bar D_\alpha=\bar\partial_\alpha\,,\quad
\{D_\alpha ,\bar D_\beta  \}=i\gamma^a_{\alpha\beta}\partial_a\,.
\label{D-flat}
\ee
Following \cite{BILS,Kuzenko2}, we construct the operators
\bea
{\cal D}_\alpha&=&e^{\frac12\rho}(D_\alpha +  \frac r2(D^\beta \rho)\gamma^a_{\alpha\beta}
 M_a - (D_\alpha \rho)R)\,,\nn\\
\bar{\cal D}_\alpha&=&e^{\frac12\rho}(\bar D_\alpha
+ \frac r2(\bar D^\beta \rho)\gamma^a_{\alpha\beta}
 M_a + (\bar D_\alpha \rho)R)\,,\nn\\
{\cal D}_a&=&e^\rho\Big(
\partial_a +i \gamma_a^{\alpha\beta}(D_\alpha \rho) \bar D_\beta
+i\gamma_a^{\alpha\beta} (\bar D_\alpha\rho)D_\beta
\nn\\&&
+\frac{ir}2(D^\alpha \rho)(\bar D_\alpha \rho) M_a
+\frac{ir}2\varepsilon_{abc}\partial^b\rho M^c
+i\gamma_a^{\alpha\beta}(D_{(\alpha}\rho)(\bar D_{\beta)}\rho)R
\Big),
\label{D-conf-flat1}
\eea
with $\rho(x,\theta,\bar\theta)$ being a scalar superfield. These operators happen to
obey the algebra (\ref{deriv}) of covariant derivatives of the supercoset $SU(2|1)/U(1)$
under the condition that the superfield $\rho$ solves for the following equations
\bea
D^2 e^{-\rho}=\bar D^2 e^{-\rho}&=&0\,,
\label{lin}\\
{}[D_{(\alpha},\bar D_{\beta)}]e^\rho&=&0\,,\\
e^{\rho}D^\alpha \bar D_\alpha \rho &=&\frac1{r}\,.
\label{last_}
\label{constr}
\eea
The equation (\ref{lin}) is nothing but the linearity condition
for the superfield $e^{-\rho}$. Note that eq. (\ref{lin}) is not
independent but appears as a differential consequence of
(\ref{last_}).

The equations (\ref{D-conf-flat1}) allow us to expand the differential operator  \eqref{E_A}
in the basis of the covariant derivatives $D_M$ (\ref{D-flat}),
\be
E_A=(E_a, E_\alpha , \bar E_\alpha)=E_A{}^M\partial_M=\tilde E_A{}^M
D_M\,.
\ee
The supermatrix $\tilde E_A{}^M $ has the following explicit form
\be
\tilde E_A{}^M D_M=\left(
\begin{array}{ccc}
e^\rho \delta_a^{m}&
 ie^\rho \gamma_a^{\alpha'\beta}(\bar D_\beta \rho) &
  ie^\rho \gamma_a^{\alpha'\beta}(D_\beta\rho) \\
0 & \delta_\alpha^{\alpha'}e^{\frac12\rho} &0 \\
0 & 0 &\delta_\alpha^{\alpha'}e^{\frac12\rho}
\end{array}
\right)
\left(
\begin{array}c
\partial_{m} \\ D_{\alpha'} \\ \bar D_{\alpha'}
\end{array}
\right).
\ee
The Berezinian of the inverse of this matrix reads
\be
E={\rm Ber}E_M{}^A={\rm Ber}\tilde E_M{}^A=e^{-\rho}\,.
\ee
An important consequence of this equation is the vanishing
of the volume of the $SU(2|1)/U(1)$ superspace (already observed in the chiral basis in Section \ref{sv})
\be
\int d^3x d^2\theta d^2\bar\theta \, E =
\int d^3x d^2\theta d^2\bar\theta \, e^{-\rho} =
-\frac14\int d^3x d^2\theta \,\bar D^2 e^{-\rho} = 0\,.
\ee
The integral is zero owing to the linearity of $e^{-\rho}$, eq.\ (\ref{lin}).

\subsection{Killing supervector}
\label{kil-sect}
Let us consider how the $SU(2|1)$ transformations act on the superfields. These are generated by the Killing supervector defined as follows.

Let us take a local supervector
$\xi^A(z)=(\xi^a,\xi^\alpha,\bar\xi^\alpha)$, and
construct an operator
\be
{\mathbb K}=\xi^a {\cal D}_a +\xi^\alpha {\cal D}_\alpha + \bar\xi^\alpha
\bar{\cal D}_\alpha - i L^{ab}(z)M_{ab}-  i l(z) R\,,
\label{Kil}
\ee
where $L^{ab}(z)$ and $l(z)$ are local $SU(2) \times U(1)$ parameters. $\xi^A$ is said to be a
Killing supervector if the operator $\mathbb K$ associated with $\xi^A$ commutes
with all the covariant derivatives \cite{BKbook}
\be
[{\mathbb K},{\cal D}_A]=0\,.
\label{KD}
\ee
This equation defines the components of the supervector
$\xi^A(z)$ as well as the superfunctions $L^{ab}(z)$ and $l(z)$ in (\ref{Kil}).

The Killing supervector generates the isometries of the superspace and the corresponding symmetries of a dynamical
 system. In the case under consideration, it is
responsible for the supersymmetries generated by
$Q_\alpha$ and $\bar Q_\alpha$,  the $SU(2)$--rotations
 $M_{a}$ and  the R--symmetry of the $SU(2|1)$ algebra
(\ref{superalgebra}).
The $SU(2|1)$ variation of a given superfield $\Phi$ on $SU(2|1)/U(1)$ is
\be
\delta\Phi = {\mathbb K}\Phi\,.
\label{2.49}
\ee
It is worth noticing that the sphere $S^3$ has isometry $SU(2)\times
SU(2)$, but only one of these $SU(2)$'s is taken into account by
$\mathbb K$. To manifestly represent the full isometry group of $S^3$ one
should start with the supercoset (\ref{big-coset}) rather than
(\ref{small-coset}).

The equation (\ref{KD}) leads to a number of differential equations
for the components of $\xi^A$, $L^{ab}(z)$ and
$l(z)$
\\
\\
$[{\cal D}_a,{\mathbb K}]=0$ $\Rightarrow$
\begin{subequations}
\label{2.44}
\begin{align}
&{\cal D}_{(a} \xi_{b)}=0\,,\qquad {\cal D}_{[a}\xi_{b]}=\frac{4}r
L_{ab}\,,
\label{L-xi}
\\
&{\cal D}_a L_{bc}=\frac 1{4r}(\delta_{ac}\xi_b-
\delta_{ab}\xi_c)\,,\\
&{\cal D}_a\xi^\alpha = \frac i{2r}(\gamma_a)^\alpha_\beta\xi^\beta\,,\qquad
{\cal D}_a\bar\xi^\alpha = \frac i{2r}(\gamma_a)^\alpha_\beta\bar\xi^\beta\,,
\label{Ki+}
\\
&{\cal D}_a l=0\,;
\end{align}
\end{subequations}

$[{\cal D}_\alpha,{\mathbb K}]=0$ $\Rightarrow$
\begin{subequations}
\label{2.45}
\begin{align}
&{\cal D}_\alpha \xi^a=i\bar\xi^\beta
\gamma^a_{\alpha\beta}\,,\qquad
{\cal D}_\alpha L^{ab}=-\frac i4
\varepsilon^{abc}(\gamma^c)_{\alpha\beta} \bar\xi^\beta\,,
\label{2.45a}
\\&
{\cal D}_\alpha \xi^\alpha=-2il\,,\qquad
{\cal D}_\alpha l =\frac i{r}\bar\xi_\alpha\,,
\label{l1}
\\&
{\cal D}_{(\alpha} \xi_{\beta)}
=-\frac i {2r}\xi^a (\gamma_a)_{\alpha\beta}
+\frac i{r}L^{ab}\varepsilon_{abc}(\gamma^c)_{\alpha\beta}
\,,\\&
{\cal D}_\alpha \bar\xi^\beta=0\,;
\label{chiral-bxi}
\end{align}
\end{subequations}

$[\bar{\cal D}_\alpha,{\mathbb K}]=0$ $\Rightarrow$
\begin{subequations}
\label{2.46}
\begin{align}
&\bar{\cal D}_\alpha \xi^a=i\xi^\beta
\gamma^a_{\alpha\beta}\,,\qquad
\bar{\cal D}_\alpha L_{ab}=-\frac
i4\varepsilon_{abc}\gamma^c_{\alpha\beta}\xi^\beta\,,
\label{2.46a}
\\&
\bar{\cal D}_\alpha \bar\xi^\alpha=2il\,,\qquad
\bar{\cal D}_\alpha l=-\frac i{r}\xi_\alpha\,,
\label{l2}
\\&
\bar{\cal D}_{(\alpha}\bar\xi_{\beta)}
=-\frac i{2r}\xi^a(\gamma_a)_{\alpha\beta}
+\frac i{r}L^{ab}\varepsilon_{abc}\gamma^c_{\alpha\beta}\,,\\
&
\bar{\cal D}_\alpha\xi^\beta=0\,.
\label{chiral-xi}
\end{align}
\end{subequations}
The analogs of the relations (\ref{2.44})--(\ref{2.46}) for the
(2,0) $AdS_3$ superspace were derived in
\cite{Kuzenko2,Kuzenko3,Kuzenko4}.

The equations (\ref{chiral-bxi}) and (\ref{chiral-xi}) show that
 $\xi^\alpha(z)$ is chiral while $\bar\xi^\alpha(z)$ is
antichiral. All the other paramters are linear as a
consequence  of (\ref{2.45a}), (\ref{l1}), (\ref{2.46a}) and
(\ref{l2}),
\be
{\cal D}^2 \xi^a =\bar {\cal D}^2\xi^a=0\,,\quad
{\cal D}^2 L^{ab}=\bar{\cal D}^2 L^{ab}=0\,,\quad
{\cal D}^2 l=\bar{\cal D}^2 l=0 \,.
\ee

The parameters $L^{ab}$ and $l$ are not independent as they can be
expressed in terms of components of $\xi^A$. Indeed, from
(\ref{l1}) and (\ref{l2}) we have
\be
l=\frac i2 {\cal D}_\alpha \xi^\alpha=-\frac i2 \bar{\cal
D}_\alpha \bar \xi^\alpha\,.
\ee
The second equation in (\ref{L-xi}) implies
\be
L^{ab}=-\frac 14 \varepsilon^{abc}\xi^c\,.
\ee
Hence, the operator (\ref{Kil}) is completely specified by the
components of the supervector $\xi^A$ which obey (\ref{2.44})--(\ref{2.46}).

The general solution of (\ref{2.44})--(\ref{2.46}) is
\bea
\xi^\alpha&=&\bar{\cal D}^2 {\cal D}^\alpha \zeta\,,\qquad
\bar\xi^\alpha = -{\cal D}^2 \bar{\cal D}^\alpha \zeta\,,\nn\\
\xi^a &=&-2i\gamma^a_{\alpha\beta} \bar{\cal D}^\alpha {\cal D}^\beta
\zeta\,,\nn\\
L_{ab} &=& \frac i2 \varepsilon_{abc}\gamma^c_{\alpha\beta}
\bar{\cal D}^\alpha {\cal D}^\beta \zeta\,,\qquad
l = \frac{2i}{r}\bar{\cal D}^\alpha{\cal D}_\alpha \zeta\,,
\label{solution}
\eea
where $\zeta$ is a covariantly constant superparameter with zero R--charge
defined modulo gauge transformations,
\be
{\cal D}_a \zeta =0\,,\quad R \zeta =0\,,\quad
\zeta \sim \zeta - i\Lambda + i
\bar\Lambda\,,
\ee
with $\Lambda$ being a chiral and covariantly constant superfunction,  $\bar{\cal D}_\alpha
\Lambda=0$, ${\cal D}_a \Lambda =0$. In particular,  with the use of $\zeta(z)$
the transformation of a chiral scalar
superfield $\Phi$, $\bar{\cal D}_\alpha \Phi=0$,
can be written as
\be
\delta \Phi = \bar{\cal D}^2[({\cal D}^\alpha\zeta)({\cal
D}_\alpha\Phi)]\,.
\label{varPhi}
\ee
Indeed, using the algebra of the covariant derivatives (\ref{deriv})
the variation (\ref{varPhi}) can be rewritten in the form
(\ref{2.49}) in which the components of the Killing supervector
are given by (\ref{solution}). The Killing vector in the form
(\ref{solution}) and the corresponding superfield transformations
(\ref{varPhi}) derived above will be applied in Sect.\ \ref{Sect4} where
superfield models with extended supersymmetry are considered.

In conclusion of this section we present an explicit expression
for the operator ${\mathbb K}$ in chiral coordinates in which
the covariant derivatives have the form (\ref{D-explicit}):
\be
{\mathbb K}=b^a {\mathbb M}_a + \epsilon^\alpha {\mathbb Q}_\alpha
+\bar\epsilon^\alpha \bar{\mathbb Q}_\alpha +t {\mathbb R}\,,
\label{comb}
\ee
where
\bea
{\mathbb M}_a&=&-i\Lambda_{ab}\partial_b\,,\nn\\
{\mathbb Q}_\alpha&=&-i \Lambda_\alpha{}^\beta
\partial_\beta\,,\nn\\
\bar{\mathbb Q}_\alpha&=&-i\Lambda_\alpha{}^\beta
[\bar\partial_\beta-i \gamma^a_{\beta\gamma}\theta^\gamma
\partial_a+\frac1{2r}\theta^2 \partial_\beta
-\frac1{r}\theta_\beta \bar\theta^\gamma\bar\partial_\gamma
+\frac1{r}\theta_\beta R]\,,\nn\\
{\mathbb R}&=&\bar\theta^\alpha \bar\partial_\alpha
 -\theta^\alpha\partial_\alpha-R\,.
\label{mbbM}
\eea
Here $\Lambda_{a}{}^{b}$ and $\Lambda_\alpha{}^\beta$ are purely
bosonic local $SO(3)\sim SU(2)$ matrices which obey the relations
\be
\partial_d \Lambda_{ab}(x)=\frac2r
\varepsilon_{dcb}\Lambda_{ac}(x)\,,\qquad
\partial_a\Lambda_\alpha{}^\beta
=\frac i{r} \Lambda_\alpha{}^\gamma
(\gamma_a)_\gamma^\beta\,,
\qquad \Lambda_{\alpha}{}^\delta \gamma^{b}_{\delta\rho}\Lambda_{\beta}{}^\rho\Lambda_{b}{}^a= \gamma^{a}_{\alpha\beta}\,.
\ee
Using these properties one can check that each of the operators
(\ref{mbbM}) independently obeys (\ref{KD}). The operator ${\mathbb
M}_a$ corresponds to the $SU(2)$ rotations on the sphere, ${\mathbb
Q}_\alpha$ and $\bar{\mathbb Q}_\alpha$ are the generators of
supersymmetries, and ${\mathbb R}$ is the R-symmetry generator.
The expression (\ref{comb}) is just a linear combination of these
operators with the corresponding constant parameters $b^a$, $\epsilon^\alpha$,
$\bar\epsilon^\alpha$ and $t$.

\section{Superfield actions}
The supergeometry of the $SU(2|1)/U(1)$ supercoset elaborated in
the previous section is characterized by torsion and curvature that
satisfy eqs.\ \eqref{torsion} and \eqref{curvature}.
Comparing these equations with the supergeometry constraints to
be satisfied by the (Euclidean version of)
${\mathcal N}=2$, $d=3$ dynamical supergravity
(see e.g. \cite{Kuzenko1,Kuzenko2,Kuzenko3,K}), one can see that
$SU(2|1)/U(1)$ geometry is a particular (vacuum) solution of
the supergravity constraints. As such, we can bypass the step
of coupling the matter superfields to off--shell supergravity and
construct classical superfield actions directly on
$SU(2|1)/U(1)$ as easy as in flat superspace.

\subsection{Gauge supermultiplet}

Let us take the covariant derivatives ${\cal D}_A= ({\cal D}_a, {\cal D}_\alpha,\bar{\cal
D}_\alpha)$ on $SU(2|1)/U(1)$ and extend them with a gauge superfield connection $V_A$
\be
\nabla_A = {\cal D}_A + V_{A}\,,\qquad
V_{A}=(V_a,V_\alpha,\bar V_\alpha)\,.
\label{3.1}
\ee
$V_A$ take values in the Lie algebra of a gauge group.
Gauge superfield constraints are imposed by requiring that the
gauge--covariant derivatives obey the (anti)commutation relations (\ref{deriv})
deformed by gauge superfield strengths,
\bea&&
\{\nabla_\alpha,\nabla_\beta \}=\{\bar \nabla_\alpha,\bar \nabla_\beta
\}=0\,,\nn\\&&
\{\nabla_\alpha, \bar\nabla_\beta
\}=i\gamma^a_{\alpha\beta}\nabla_a
-\frac 1{2} \gamma^a_{\alpha\beta}M_{a}
+\frac 1{r}\varepsilon_{\alpha\beta}R+ i\varepsilon_{\alpha\beta}G\,,\nn\\&&
[\nabla_a,\nabla_b]=-\frac i{2r}M_{ab}+
i{\bf F}_{ab}\,,\nn\\&&
[\nabla_a,\nabla_\alpha]=-\frac i{2r}(\gamma_a)_\alpha^\beta
\nabla_\beta
- (\gamma_a)_\alpha^\beta \bar W_\beta
\,,\quad
[\nabla_a,\bar\nabla_\alpha]=-\frac i{2r}(\gamma_a)_\alpha^\beta
\bar\nabla_\beta
+(\gamma_a)_\alpha^\beta W_\beta\,,\nn\\&&
[R,\nabla_\alpha]=\nabla_\alpha\,,\quad
[R,\bar\nabla_\alpha]=-\bar\nabla_\alpha\,,\nn\\&&
[M_{ab},\nabla_\alpha]=-\frac 1{r}\varepsilon_{abc}(\gamma^c)_\alpha^\beta \nabla_\beta\,,
\quad
[M_{ab},\bar\nabla_\alpha]=-\frac 1{r}\varepsilon_{abc}(\gamma^c)_\alpha^\beta
\bar\nabla_\beta\,.
\label{c-deriv}
\eea
Here $G$, $W_\alpha$, $\bar W_\alpha$ and ${\bf F}_{ab}$ are gauge superfield
strengths subject to the Bianchi identities. In
particular, $W_\alpha$ is covariantly chiral and $\bar W_\alpha$
is covariantly antichiral,
\be
\bar\nabla_\alpha W_\beta=0\,,\qquad
\nabla_\alpha \bar W_\beta=0\,.
\ee
These superfields obey `standard' Bianchi identity
\be
\nabla^\alpha W_\alpha=\bar\nabla^\alpha \bar W_\alpha \,.
\ee
The spinorial superfield strengths $W_\alpha$ and $\bar W_\alpha$
are expressed in terms of the scalar superfield $G$ as follows
\be
\bar W_\alpha = \nabla_\alpha G\,,\qquad
W_\alpha=\bar\nabla_\alpha G\,.
\ee
The latter is covariantly linear,
\be
\nabla^2 G= \bar \nabla^2 G=0\,.
\ee

The gauge connections $V_A$  in (\ref{3.1}) can be expressed in terms
of a single gauge prepotential $V$. In particular, in the so--called chiral representation
\cite{BKbook,GGRS} the covariant spinor derivatives $\nabla_\alpha$ and
$\bar\nabla_\alpha$ are given by
\be\label{chiralV}
\nabla_\alpha=e^{-V}{\cal D}_\alpha e^V\,,\qquad
\bar \nabla_\alpha = \bar{\cal D}_\alpha\,.
\ee
As a consequence of the constraints (\ref{c-deriv}), the superfield strengths
are expressed in terms of the prepotential $V$ as follows
\be
G= \frac{i}{2}\bar{\cal D}^\alpha(e^{-V}{\cal D}_\alpha
e^V)\,,\quad
W_\alpha=-\frac{i}{4}\bar{\cal D}^2(e^{-V}{\cal D}_\alpha e^V)\,,
\quad
\bar W_\alpha=\frac{i}{2}\nabla_\alpha \bar{\cal D}^\beta
(e^{-V}{\cal D}_\beta e^V)\,.
\label{field-strengths}
\ee
The gauge transformation of $V$ is
\be
e^V\longrightarrow e^{i\bar\Lambda} e^V e^{-i\Lambda}\,,
\label{gauge}
\ee
where $\Lambda$ and $\bar\Lambda$ are covariantly (anti)chiral local gauge parameters
\be
\bar{\cal D}_\alpha \Lambda =0 \,,\qquad
{\cal D}_\alpha \bar\Lambda= 0\,.
\ee
The superfield strengths transform covariantly under the gauge
transformations (\ref{gauge}),
\be
G\to e^{i\Lambda} G e^{-i\Lambda}\,,\qquad
W_\alpha \to e^{i\Lambda} W_\alpha e^{-i\Lambda}\,.
\ee

The super Yang--Mills action can be equivalently written either in
the full $\cN=2$ superspace or in the chiral subspace,
\be
S_{\rm SYM}=-\frac 4{g^2}\tr \int d^3x d^2\theta d^2\bar\theta \,  E\,
G^2=\frac2{g^2}\tr\int d^3x d^2\theta \, {\cal E}\,W^\alpha W_\alpha\,,
\label{S-SYM}
\ee
where $g^2$ is the gauge coupling constant of mass dimension $[g]=1/2$ and $\cal E$
is a chiral density. The variation of the SYM action reads
\be
\delta S_{\rm SYM}=-\frac{4i}{g^2}\tr\int d^3x d^2\theta d^2\bar\theta\,
E \,
\Delta V \nabla^\alpha \bar\nabla_\alpha G\,,
\ee
where $\Delta V$ is the gauge--covariant variation,
\be
\Delta V= e^{-V}\delta e^V = \delta V +\frac12[\delta V,V]+\ldots
\label{DeltaV}
\ee
Hence, the SYM equation of motion is
\be
0=\frac{\delta S_{\rm SYM}}{\Delta V}=-\frac{4i}{g^2}\nabla^\alpha
\bar\nabla_\alpha G=-\frac{4i}{g^2}\nabla^\alpha
W_\alpha \,.
\ee

The Abelian Chern--Simons action is known to be
\be
-\frac k\pi \int d^3x d^2\theta d^2\bar\theta \,E\, VG\,,
\ee
where $k$ is an integer.
The non--Abelian generalization of this action requires the
introduction of an auxiliary parameter $t$ \cite{Ivanov92},
\be
S_{\rm CS}=-\frac{ik}{\pi}\tr\int_0^1 dt\int d^3x d^2\theta
d^2\bar\theta\,E
\, \bar{\cal D}^\alpha(e^{-tV}{\cal D}_\alpha e^{tV})e^{-tV}\partial_t
e^{tV}\,.
\label{CS}
\ee
However, the variation of the Chern--Simons action does not contain
this parameter,
\be
\delta S_{\rm CS}=-\frac{2 k}{\pi}\tr\int d^3x d^2\theta d^2\bar\theta\,E
\,G \Delta V\,.
\label{var-CS}
\ee

Finally, the Fayet--Iliopoulos term is given by
\be\label{FI}
S_{\rm FI}=-4i\xi \int d^3x d^2\theta d^2\bar\theta\,E\, V\,,
\ee
where $\xi$ is the coupling of mass dimension $+1$.

\subsubsection{Component structure}

The vector supermultiplet consists of one scalar field $\sigma(x)$ one
vector $A_a(x)=-i \gamma_a^{\alpha\beta}A_{\alpha\beta}$, spinors
$\lambda_\alpha(x)$ and $\bar\lambda_\alpha(x)$ and one auxiliary field $D(x)$.
In the Wick--rotated (Euclidean) SYM theory under consideration $\lambda_\alpha(x)$ and $\bar\lambda_\alpha(x)$ are regarded as independent fields, not related to each other by complex conjugation, and also the bosonic fields $\sigma$ and $A_a$ are assumed to be complex.

To derive the component structure in supersymmetric gauge theories
it is convenient to impose the Wess--Zumino gauge,
\be
V|=0\,,\quad
{\cal D}_\alpha V|=\bar{\cal D}_\alpha V|=0\,,\quad
{\cal D}^2 V|=\bar{\cal D}^2 V|=0\,,
\ee
where $|$ denotes the component value of the superfields at
$\theta=\bar\theta=0$.
The component fields appear in the following derivatives of the
gauge superfield
\bea
\frac12 [{\cal D}_\alpha,\bar{\cal D}_\beta]V|&=&2iA_{\alpha\beta}
-\varepsilon_{\alpha\beta}i \sigma\,,\nn\\
\frac12 \bar{\cal D}^2 {\cal D}_\alpha V|&=&i\lambda_\alpha\,,
\quad
\frac12 {\cal D}^2 \bar{\cal D}_\alpha
V|=i\bar\lambda_\alpha\,,\nn\\
\frac1{8}\{ {\cal D}^2,\bar {\cal D}^2  \}V|&=&iD\,.
\label{gauge-components}
\eea
Using the algebra of the covariant derivatives (\ref{deriv}) we find the
components of the superfield strengths
(\ref{field-strengths}) and their derivatives to be
\bea
G|&=&\sigma\,,\nn\\
W_\alpha |&=&\frac 12 \lambda_\alpha\,,\qquad
\bar W_\alpha|=\frac12\bar\lambda_\alpha\,,
\nn\\
{\cal D}^\alpha W_\alpha|&=&D+\frac{2\sigma}{r}\,,\nn\\
{\cal D}_{(\alpha} W_{\beta)}|&=&-\frac i4\varepsilon^{abc} \gamma^c_{\alpha\beta}
 F_{ab}
+\frac i2\gamma^a_{\alpha\beta} \hat\nabla_{a} \sigma
\,,\nn\\
{\cal D}^2 W_\alpha |&=&i\gamma^a_{\alpha\beta}
\hat\nabla_a \bar \lambda^\beta
+i[\sigma,\bar\lambda_\alpha]
-\frac 1{2r}\bar\lambda_\alpha\,,
\label{G-comp}
\eea
where
\bea
\hat\nabla_a \bar\lambda^\beta&=&\hat{\cal D}_a\bar\lambda^\beta
+i[A_a,\bar\lambda^\beta]\,,\nn\\
\hat\nabla_a \sigma &=&\hat{\cal D}_a\sigma + i[A_a,\sigma]\,,
\qquad
(\hat\nabla_{\alpha\beta}=-\frac i2\gamma^a_{\alpha\beta}\hat\nabla_a)\,,\nn\\
F_{ab}&=&\hat{\cal D}_a A_b - \hat{\cal D}_b A_a+i[A_a,A_b]\,
\eea
and $\hat{\cal D}_a=\partial_a+\omega_a(x)$ is a covariant derivative on $S^3$.

Consider now the SYM action (\ref{S-SYM}) and replace the
integration over $d^2\theta$ by corresponding spinor covariant
derivatives
\bea
S_{\rm SYM}&=&\frac 2{g^2}  \tr\int d^3x d^2\theta\,{\cal E}\,W^\alpha
W_\alpha
\nn\\&
=&-\frac1{g^2}\tr\int d^3x \sqrt h\left(W^\alpha {\cal D}^2 W_\alpha
-\frac12{\cal D}^\alpha W_\alpha{\cal D}^\beta W_\beta
-{\cal D}_{(\alpha} W_{\beta)}{\cal D}^{(\alpha} W^{\beta)}
\right)\bigg|
\,.
\label{SYM-comp}
\eea
Substituting (\ref{G-comp}) into \eqref{SYM-comp}, we find the component
structure of the classical SYM action
\bea
S_{\rm SYM}&=&\frac1{g^2}\tr\int d^3x\sqrt h\bigg[
\frac14F^{ab}F_{ab}+\frac12\hat\nabla^a \sigma \hat\nabla_a\sigma
+\frac12\left( D+\frac{2\sigma}r\right)^2
\nn\\&&
+\frac i2 \lambda^\alpha(\gamma^a)_\alpha^\beta\hat\nabla_a
\bar\lambda_\beta
-\frac i2 \lambda^\alpha[\sigma,\bar\lambda_\alpha]
+\frac1{4r}\lambda^\alpha\bar\lambda_\alpha
\bigg].
\label{SYMcomp}
\eea
Note that the terms containing the inverse radius of the three--sphere $1/r$ automatically
appear in this procedure and this action is $\cN=2$ supersymmetric
by construction.

In a similar way one recovers the component structure of the
Chern--Simons \eqref{CS} and Fayet--Iliopoulos \eqref{FI} superfield actions,
\bea
S_{\rm CS}&=&\frac{ik}{4\pi}\tr\int d^3x\sqrt h
\left[\varepsilon^{abc}(A_a \hat{\cal D}_b A_c +\frac{2i}{3}A_a A_b A_c)
-\bar\lambda^\alpha \lambda_\alpha
-2\sigma D -\frac{2\sigma^2}{r}\right],
\label{CS-comp}
\\
S_{\rm FI}&=&-4i\xi \int d^3x d^2\theta d^2\bar\theta\,E\, V=\xi \int d^3x\sqrt h\, D\,.
\label{FI-comp}
\eea
The last term in the Chern--Simons action (\ref{CS-comp}) can be eliminated by the
shift of the auxiliary field, $D\to D'=D+\frac{\sigma}{r}$. After
such a shift the Chern--Simons action takes the canonical form.

\subsection{Chiral matter}
Let us now consider a covariantly chiral superfield $\Phi$ and an anti--chiral superfield $\bar\Phi$ \emph{i.e.} the superfields that obey the constraints
\be
\bar{\cal D}_\alpha \Phi =0 \,,\qquad
{\cal D}_\alpha \bar\Phi=0\,.
\label{chrargePhi}
\ee
Again, as for the vector supermultiplet, we do not assume that $\Phi$ and $\bar\Phi$ are related by the complex conjugation.

A general action for the chiral superfields interacting with the
background gauge superfield $V$ is
\be
\label{S-chiral}
S=4\int d^3x d^2\theta d^2\bar\theta \,E \,
\bar\Phi e^{V}\Phi
+2\int d^3x d^2\theta \, {\cal E}\, W(\Phi)
+2\int d^3x d^2\bar\theta \, \bar {\cal E}\, \bar W(\bar\Phi)
\,,
\ee
where $W(\Phi)$ is a superpotential. Here we assume that $\Phi$ transforms under the fundamental
representation of the gauge gorup. In the case of the adjoint representation
 the kinetic term for $\Phi$ includes the trace of the matrix indices
\be
4\,\tr\int d^3x d^2\theta d^2\bar\theta\, E\, e^{-V}\bar\Phi
e^{V}\Phi\,.
\ee
The (anti)chiral superfield may carry an R--charge $q$, \emph{i.e.}
\be
R\Phi=-q \Phi\,,\qquad
R\bar\Phi = q\bar\Phi\,.
\ee
In principle, the R--charge of the chiral superfield can be arbitrary although its canonical
value for the chiral matter is $q=1/2$. Note also that the R--charge of the
superpotential $W(\Phi)$ should be $-2$ since the chiral measure
$d^2\theta$ has the R--charge $+2$. The latter follows form the
fact that $d\theta_\alpha\propto {\cal D}_\alpha$ and from the
commutation relations (\ref{deriv1}).

The (anti)chiral multiplet consists of the complex scalar field  $\phi$
$(\bar\phi)$, the spinor  $\psi_\alpha$  $(\bar\psi_\alpha)$ and the auxiliary
field $F$ $(\bar F)$. These fields appear in the $\theta$--decomposition
of the superfields $\Phi$ and $\bar\Phi$ as follows
\be
\begin{array}{lll}
\phi(x)=\Phi| && \bar\phi(x)=\bar\Phi|\\
\psi_\alpha(x)={\cal D}_\alpha \Phi| && \bar\psi_\alpha(x)=\bar{\cal
D}_\alpha \bar\Phi|\\
F(x)=-\frac12{\cal D}^2 \Phi|&\qquad&
 \bar F(x)=-\frac12\bar{\cal D}^2\bar \Phi|\,.
\end{array}
\label{chiral-components}
\ee
Upon integrating out the Grassmann variables we find the component
structure of the action (\ref{S-chiral}),
\bea
S&=&\int d^3x \sqrt h\bigg[
\hat\nabla^a \bar\phi \hat\nabla_a \phi
+\bar \phi\left(
\sigma^2 + \frac{q(2-q)}{r^2} + \frac{2iq}{r}\sigma +iD
\right)\phi +\bar F F
\nn\\&&
-i\gamma^a_{\alpha\beta} \bar \psi^\alpha \hat\nabla_a \psi^\beta
+\bar \psi^\alpha \left(
i\sigma +\frac{1-2q}{2r}
\right)\psi_\alpha
+i\bar\psi^\beta \bar\lambda_\alpha \phi
+i\bar\phi \lambda^\alpha \psi_\alpha
\bigg]\nn\\&&
+\int d^3x \sqrt h\,
 \left(W'(\phi)F +W'(\bar\phi)\bar F-\frac12W''(\phi)\psi^\alpha\psi_\alpha
  -\frac12 W''(\bar\phi)\bar\psi^\alpha\bar\psi_\alpha\right).
\label{Schiral-comp}
\eea
Here $\hat\nabla_a$ is the gauge covariant derivative on $S^3$ in the
fundamental representation of the gauge group
\be
\hat\nabla_a \phi = \hat{\cal D}_a \phi+i A_a(x)\phi\,,\quad
\hat\nabla_a \bar \phi = \hat{\cal D}_a \bar \phi - i A_a (x)\bar\phi\,,\quad
\hat\nabla_a \psi_\alpha = \hat{\cal D}_a \psi_\alpha + i A_a(x)
\psi_\alpha\,.
\ee
The generalization to any other representation of the gauge group
is straightforward.

\section{Superfield models with extended supersymmetry}
\label{Sect4}
In the previous section we constructed the superfield gauge and matter models on
$S^3$ with minimal ($\cN=2$) supersymmetry\footnote{Recall that since on $S^3$
the spinors are complex, $S^3$ does not admit $\cN=1$ supersymmetry
which would correspond to a single real 2--component spinor.}. This construction was
very similar to the formulation of superfield theories in a
general curved superspace of Lorentz signature  (see, \emph{e.g.} \cite{BKbook} for this
topic in four dimensions or a series of papers
\cite{Kuzenko1,Kuzenko2,Kuzenko3,Kuzenko4,Kuzenko5,Kuzenko6} for relevant three--dimensional
supergravity--matter models in superspace). The classical actions
introduced in this section can be considered as the Wick--rotated
 gauge and matter superfield actions in the (2,0) $AdS_3$
superspace \cite{Kuzenko2,Kuzenko3}.
\if{}
However, we circumvented the
study of the three--dimensional supergravity  and
constructed  the background supergeometry directly with
the supersymmetric Cartan forms on the supercoset $SU(2|1)/U(1)$.
\fi

In this section we will consider models with extended
$\cN>2$ supersymmetry on the three--sphere. In particular, the
classical actions of $\cN=4$ and $\cN=8$ SYM theories, as well as
 the Gaiotto--Witten and ABJM models will be constructed. In
principle, for these models it would be natural to introduce
curved superspaces with extended (${\mathcal N}=4,6,8$) supersymmetry and to construct
the actions directly in these superspaces. However, even in the
flat space the use of the extended superspaces is not always
convenient because it usually employs special methods with
harmonic or projective coordinates which help to achieve
unconstrained superfield formulations. So, we will avoid
introducing extended superspaces and continue to use the $\cN=2$
superspace formalism.

The description of the models with extended supersymmetry in the
$\cN=2$ supercoset $SU(2|1)/U(1)$ mimics the construction of the classical actions of
supersymmetric gauge and matter models in the conventional
component field formulation \cite{KWY}. Let us recall that such a construction is carried out in two steps. First, one couples the flat actions
to the $S^3$ background  geometry and then one finds extra
terms which come with inverse radius of $S^3$ and which are
necessary for the invariance under the supersymmetry on the
sphere generated by $S^3$ Killing spinors. Similarly, in the $\cN=2$ superspace we will
use the chiral matter and gauge superfields for constructing actions with extra supersymmetries and will reveal new terms of order $\frac1r$ required for the action to be invariant under extended supersymmetry for superfields carrying a non--zero R--charge.
As will be shown, the parameters of the extra supersymmetries and their bosonic (R--symmetry) partners are encoded in $\cN=2$ superfield parameters which include, as their
components, $S^3$ Killing spinors corresponding to the extra supersymmetries.

We will consider in detail the construction of the classical
action for an $\cN=4$ SYM model and will shortly discuss
actions for other models ($\cN=8$ SYM, Gaiotto--Witten and
ABJM models).

To have a theory on $S^3$--sphere with $\cN=4$
supersymmetry we need one more copy of the Killing spinors, in addition to
those which have already appeared in $SU(2|1)/U(1)$.
Recall that the Killing spinor equation reads
\be
\hat{\cal D}_a \xi^\alpha(x) \pm
\frac{i}{2r}(\gamma_a)^\alpha_\beta \xi^\beta(x) =0\,,
\label{KilSp_}
\ee
where $\hat{\cal D}_a$ is purely bosonic covariant derivative on
$S^3$. The choice of the sign in (\ref{KilSp_}) can be arbitrary.
In the $\cN=2$ case we should have two
spinors, $\xi_\alpha$ and $\bar\xi_\alpha$, of the same
``chirality''\footnote{We refer to the spinors obeying the equation (\ref{KilSp_})
with different signs as the Killing spinors of different ``chirality''.
We hope that this will not cause the confusion with the conventional
notion of the chiral spinors (which do not exist on $S^3$).}
with respect to the sign in \eqref{KilSp_}, which is required by the $SU(2|1)$
supergroup structure (see eqs. \eqref{Ki+}). In the $\cN=4$ case
we need another copy of Killing spinors associated with extra $\cN=2$ supersymmetries, say $\eta_\alpha$ and
$\bar\eta_\alpha$ the ``chirality'' of which can either coincide with
the one of $\xi_\alpha$, or can be opposite. For instance, in
the case of superfield models in the $AdS_3$ space
\cite{Kuzenko3,Kuz-14}, the corresponding $\cN=4$ superspaces are denoted
as (4,0) and (2,2), respectively. In this paper we
will consider mainly the models with extended supersymmetry
associated with all the $S^3$ Killing spinors of the same
``chirality'' and will shortly describe the models with Killing
spinors of different ``chiralities'' on the example of $\cN=4$ SYM
model.

\subsection{$\cN=4$ SYM with $SU(2)\times SU(2)$ R--symmetry}
\label{Sect4.1}
$\cN=4$ gauge supermultiplet is given by a pair $(V,\Phi)$, where $V(x,\theta,\bar\theta)$
is the $\cN=2$ gauge superfield and $\Phi(x,\theta,\bar\theta)$ is a chiral superfield
in the adjoint representation of the gauge group.
We start the construction by lifting the flat $d=3$,
$\cN=4$ SYM action (written in terms of the $\cN=2$ superfields, see \emph{e.g.} \cite{HKLR})
onto the $SU(2|1)/U(1)$ background
\be
S_{\rm 0}=-\frac4{g^2}\tr\int d^3x d^2\theta
d^2\bar\theta\, E(G^2 - e^{-V}\bar\Phi e^V \Phi)\,.
\label{Sbare}
\ee
The superfields $V$ and $G$ are neutral under the
$U(1)$ R--transformations associated with the manifest $\cN=2$ $SU(2|1)$ supersymmetry, while the R--charge of the chiral
superfield $\Phi$ can be, \emph{a priori}, arbitrary
\be
R \bar{\Phi} = q\bar{\Phi} \,,\qquad
R {\Phi } = -q{\Phi}\,.
\label{RPhi}
\ee

For further convenience, it is useful to introduce the
gauge--covariant chiral superfields,
\be
\overline{\it \Phi}= e^{-V} \bar\Phi e^V\,,\quad {\it \Phi} = \Phi\,,\qquad
\nabla_\alpha \overline{\it \Phi} =0 \,,\quad
\bar\nabla_\alpha{\it \Phi}=0\,.
\label{337}
\ee
In terms of these superfields the gauge transformations are given
by
\be
\Delta V = i\bar\Lambda - i\Lambda\,,\qquad
\delta{\it \Phi} = i[\Lambda,{\it\Phi}]\,,\qquad
\delta\overline{\it\Phi} =
i[\bar{\Lambda},\overline{\it\Phi}]\,,
\ee
with $\Lambda$ being a covariantly chiral gauge superfield
parameter, $\bar\nabla_\alpha\Lambda =0$. Recall that $\Delta V$
is a gauge--covariant variation for the $\cN=2$ gauge superfield
(\ref{DeltaV}).

The general variation of the
action (\ref{Sbare}) reads
\be
\delta S_{\rm 0} = -\frac4{g^2} \tr\int d^3x d^2 \theta
d^2\bar\theta\,E(i\Delta V \bar \nabla^\alpha \nabla_\alpha G
-\delta\overline{\it\Phi}{\it\Phi} -
\overline{\it\Phi}\delta{\it\Phi}
-\Delta V [{\it\Phi},\overline{\it\Phi}])\,.
\label{4.5}
\ee

We now assume that
the hidden $\cN=2$ supersymmetry and its bosonic partners encoded in (anti)chiral superfield parameters $\Upsilon(z)$ and $\bar\Upsilon(z)$ transform the superfields $V$ and $\Phi$ into
each other as follows
\be
\Delta_\Upsilon V = i(\Upsilon\overline{\it \Phi} -\bar \Upsilon{\it\Phi})\,, \quad
\delta_\Upsilon{\it \Phi} =\bar \nabla^\alpha G{\cal D}_\alpha \Upsilon
  + \frac qr G\Upsilon\,,\qquad
\delta_\Upsilon \overline{\it\Phi} = -\nabla^\alpha G \bar {\cal D}_\alpha \bar \Upsilon
  - \frac qr
G\bar \Upsilon\,.
\label{transf-non-Ab}
\ee
In addition to be (anti)chiral
\be
\bar{\cal D}_\alpha \Upsilon=0\,,\qquad
{\cal D}_\alpha \bar\Upsilon =0\,,
\ee
$\Upsilon$ and $\bar\Upsilon$ are also subject to the constraints
\be
{\cal D}_a \Upsilon =0\,,\qquad
{\cal D}_a \bar\Upsilon =0\,.
\label{Dchi=0}
\ee
The above constraints are required for the superfields $\it\Phi$ and $\overline{\it\Phi}$ to remain (anti)chiral upon the extra supersymmetry transformations, \emph{i.e.}
\be\label{nd}
\bar\nabla_\alpha \delta_\Upsilon{\it\Phi}=0\,,\qquad
\nabla_\alpha \delta_\Upsilon \overline{\it\Phi}=0\,.
\ee

Note that $V$ and $\it\Phi$ are non--Abelian superfields in the adjoint
representation of the gauge group, while $\Upsilon$ does not carry the
gauge group indices, so $\nabla_A \Upsilon = {\cal D}_A \Upsilon$.
Note also that $\Upsilon(z)$ should have the same
R--charge as the superfield $\Phi$ (\ref{RPhi}), namely
\be
R\bar\Upsilon=q\bar\Upsilon\,,\qquad
R\Upsilon = -q\Upsilon\,.
\label{Rchi}
\ee
In comparison with the flat case, the transformations
(\ref{transf-non-Ab}) involve additional terms with the inverse
radius of the sphere. These extra terms are necessary to preserve
the covariant chirality of the variation of the chiral superfield \eqref{nd}.

Off the mass shell, the commutator of two transformations (\ref{transf-non-Ab}) closes on the $SU(2|1)$ transformations
considered in Section
\ref{kil-sect}
\bea
{}
[\delta_{\Upsilon_2},\delta_{\Upsilon_1}] {\it\Phi} &=& \bar\nabla^2 [({\cal D}^\alpha\zeta)(\nabla_\alpha{\it\Phi})]\,,\qquad
{}[\delta_{\Upsilon_2},\delta_{\Upsilon_1}] \overline{\it\Phi} = -\nabla^2 [
(\bar {\cal D}^\alpha \zeta)(\bar \nabla_\alpha \overline{\it\Phi})]\,,\nn\\
{}[\delta_{\Upsilon_2} ,\delta_{\Upsilon_1}]G&=&
-2i\gamma^a_{\alpha\beta}(\bar{\cal D}^\alpha {\cal D}^\beta \zeta)
 \nabla_a G
+(\bar{\cal D}^2{\cal D}^\alpha \zeta)\nabla_\alpha G
-({\cal D}^2 \bar{\cal D}^\alpha \zeta)\bar\nabla_\alpha G
\,,
\label{4.12}
\eea
where
\be
\zeta = \frac14(\bar\Upsilon_1 \Upsilon_2 - \bar\Upsilon_2 \Upsilon_1)\,.
\ee
Indeed, the transformation of the chiral superfield in
(\ref{4.12}) has exactly the same form as (\ref{varPhi})
while the transformation of $G$ has the general form (\ref{2.49})
with the parameters given by (\ref{solution}).

Let us consider the commutator of the transformations
(\ref{transf-non-Ab}) with the $SU(2|1)$ transformations
(\ref{2.49}). Using the fact that the operator $\mathbb K$
(\ref{Kil}) commutes with the covariant derivatives (\ref{KD}) we
have
\be
[\delta_\Upsilon,\delta_{\mathbb K}]V = \delta_{\Upsilon'}V\,,\quad
[\delta_\Upsilon,\delta_{\mathbb K}]{\it\Phi} = \delta_{\Upsilon'}{\it\Phi}\,,\quad
[\delta_\Upsilon,\delta_{\mathbb K}]\overline{\it\Phi} =
\delta_{\Upsilon'}\overline{\it\Phi}\,,
\ee
where
\be
\Upsilon ' = {\mathbb K}\Upsilon\,.
\ee
Thus the commutator of (\ref{transf-non-Ab}) and (\ref{2.49}) is
again of the form (\ref{transf-non-Ab}). Therefore, the $SU(2|1)$
transformations and the extra ${\mathcal N}=2$ supertransformations (\ref{transf-non-Ab})
form an ${\mathcal N}=4$ superalgebra. Though we do not have a clear
understanding of algebraic stricture of the transformations
(\ref{transf-non-Ab}) for generic values of $q$, for $q=1$ the form of the
supersymmetry transformations suggests that this superalgebra is
$su(2|2)\times su(2)$.

To show this, consider the component structure
of the chiral superfield parameter $\Upsilon$. In the chiral superspace
coordinates its $\theta$--decomposition is
\be
\Upsilon = a + \theta^\alpha \eta_\alpha + \theta^2 b\,.
\label{chi-comp}
\ee
Using the explicit form of the superspace derivatives
(\ref{D-explicit}) one can easily check that the equation
(\ref{Dchi=0}) implies that the components $a$ and $b$ are
constant
\be
a=const\,,\qquad
b=const\,,
\ee
while $\eta^\alpha$, associated with the extra ${\mathcal N}=2$ supersymmetry, obeys the Killing spinor equation similar to
\eqref{Ki+} satisfied by the supersymmetry parameters of the manifest $SU(2|1)$ supersymmetry
\be
\hat{\cal D}_a \eta^\alpha -
\frac{i}{2r}(\gamma_a)^\alpha_\beta \eta^\beta =0\,.
\label{KilSp}
\ee
The lowest component
$a$ in $\Upsilon$, and its conjugate $\bar a$ appearing in $\bar\Upsilon$,
are the parameters of the coset elements $SU(2)_{R}/U(1)$, where
$SU(2)_{R}$ is part of the R--symmetry group in the $\cN=4$ SYM
theory. This indicates that for $q=1$ the transformations
(\ref{transf-non-Ab})  together with the $SU(2|1)$ symmetry  generate
the supergroup $SU(2|2)$.
One can also verify that the highest component $b$ in
$\Upsilon$, and its conjugate $\bar b$ in $\bar\Upsilon$, are the
parameters of another $SU(2)$ which rotates a triplet of
auxiliary fields in the $\cN=4$ gauge supermultiplet.

One--line computations show that the naive action (\ref{4.5}) is not
invariant under the $\Upsilon$--transformations (\ref{transf-non-Ab})
for $q\ne 0$,
\be
\delta_\Upsilon S_{\rm 0}=-\frac{4q}{rg^2}\tr\int d^3x d^2\theta
d^2\bar\theta\,E\, G(\bar\Upsilon {\it\Phi}-\Upsilon\overline{\it\Phi})\,.
\ee
Surprisingly, the non--invariance of $S_{\rm 0}$ cancels against
the variation of the following Chern--Simons term
\be
S_{\rm CS}=-\frac{2q}{rg^2} \tr\int_0^1 dt
\int d^3x d^2\theta d^2\bar\theta\,E \,\bar{\cal D}^\alpha
(e^{-tV}{\cal D}_\alpha e^{tV})
e^{-tV}\partial_t e^{tV}\,.
\label{CS4}
\ee
This action differs from (\ref{CS}) only by the overall \emph{real}
coefficient in front of the superspace integral. Indeed, using
(\ref{var-CS}) it is easy to find the variation of (\ref{CS4})
under (\ref{transf-non-Ab}),
\be
\delta_\Upsilon S_{\rm CS}=\frac{4iq}{rg^2}\tr\int d^3x d^2\theta
d^2\bar\theta\,E\,G\Delta_\Upsilon V
=\frac{4q}{rg^2}\tr\int d^3x d^2\theta
d^2\bar\theta\,E\,G(\bar\Upsilon{\it\Phi} - \Upsilon\overline{\it\Phi})\,.
\ee

Thus, we conclude that the action of the $\cN=4$ SYM model on the
three--sphere is given by the sum of the action (\ref{Sbare})
and the Chern--Simons term (\ref{CS4}),
\be
S^{\cN=4}_{\rm SYM} =-\frac4{g^2}\tr\int d^3x d^2\theta
d^2\bar\theta\, E\left[G^2 - e^{-V}\bar\Phi e^V \Phi
+\frac q{2r}\int_0^1 dt \bar{\cal D}^\alpha
(e^{-tV}{\cal D}_\alpha e^{tV})
e^{-tV}\partial_t e^{tV}\right]
\,.
\label{N4SYM}
\ee
This action is manifestly invariant under $SU(2|1)$ and under the
hidden $\cN=2$ transformations (\ref{transf-non-Ab}),
\be
\delta_\Upsilon S^{\cN=4}_{\rm SYM}=0\,.
\ee

The requirement to have the Chern--Simons term (for $q\ne 0$) together with the YM term in the
SYM action (\ref{N4SYM}) to make it ${\mathcal N}=4$ supersymmetric
is a somewhat unexpected feature of this
model \footnote{When gauge supermultiplets are part of supergravity
supermultiplets, it is well known that the invariance of the supergravity
action under supersymmetry may require the presence of Chern--Simons terms,
as \emph{e.g.} in the case of $D=11$ supergravity \cite{Cremmer:1978km}
or $\cN=4$, $d=3$ supergravity \cite{Lu:2002uw,Lu:2003yt}.
The necessity to add the Chern--Simons term to the SYM action coupled
to the chiral supermultiplet with the R--charge $q=1$ for getting
the $\cN=4$ SYM theory on $S^3$ was noticed in \cite{HHL}.}. The Chern--Simons
term disappears in the flat limit
as it comes about with the inverse radius of the sphere.
The Chern--Simons term is also absent for
$q=0$, but we stress that the action (\ref{N4SYM}) is
consistent also for $q\ne0$.  However, as we will show in the next section, there is a
natural bound on the values of this parameter $0\leq q\leq 2$ which originates  from the
requirement of the absence of negative energy states in the spectrum
of the model (\ref{N4SYM}). Note that, the choice of $q=1$ is the
most natural since this value of the R--charge coincides with the conformal dimension
of the chiral supermultiplet which has applications in studying
various aspects of dualities of three--dimensional gauge theories \cite{KWY1}.

The term
(\ref{CS4}) comes with a real coefficient in front of the integral, in contrast to the
Chern--Simons action (\ref{CS}) which appears with the imaginary unit factor
since it was obtained by Wick rotating the Chern--Simons action in space--time of Lorentz signature. Hence, the term (\ref{CS4}) results in a negative
topological mass squared for the gauge field and can, in principle,
cause the states with negative energies.
To find the allowed values of $q$ for which these states are absent
we will consider the component form of the action (\ref{N4SYM}).

\subsubsection{Component form of the $\cN=4$ SYM action on $S^3$}
The action (\ref{N4SYM}) consists of the pure $\cN=2$ SYM term,
the (anti)chiral superfield part and the $\cN=2$ Chern--Simons term.
Component structure of all these three terms is given by
(\ref{SYMcomp}), (\ref{Schiral-comp}) and (\ref{CS-comp}),
respectively. Putting these expressions together, we get
\bea
S_{\rm SYM}^{\cN=4}&=&\frac1{g^2}\tr\int d^3x\sqrt h\bigg[
\frac14F^{ab}F_{ab}+\frac12\hat\nabla^a \sigma \hat\nabla_a\sigma
+\frac12\left( D+\frac{2\sigma}r\right)^2
\nn\\&&
+\frac i2 \lambda^\alpha(\gamma^a)_\alpha^\beta\hat\nabla_a
\bar\lambda_\beta
-\frac i2 \lambda^\alpha[\sigma,\bar\lambda_\alpha]
+\frac1{4r}\lambda^\alpha\bar\lambda_\alpha
\nn\\&&
+\hat\nabla^a \bar\phi \hat\nabla_a \phi
+\frac{q(2-q)}{r^2}\bar\phi\phi +\bar\phi[\sigma,[\sigma,\phi]]
+\frac{2iq}r\bar\phi[\sigma,\phi]+i\bar\phi[D,\phi] +\bar F F
\nn\\&&
+i(\gamma^a)_\alpha^\beta \bar \psi^\alpha \hat\nabla_a \psi_\beta
+\frac{1-2q}{2r}\bar\psi^\alpha \psi_\alpha
+i\bar\psi^\alpha[\sigma,\psi_\alpha]
+i\bar\psi^\beta [\bar\lambda_\alpha, \phi]
+i[\bar\phi, \lambda^\alpha] \psi_\alpha
\nn\\&&
+\frac q{2r}\varepsilon^{abc}(A_a \hat{\cal D}_b A_c +\frac{2i}{3}A_a A_b A_c)
-\frac q{2r}\bar\lambda^\alpha \lambda_\alpha
-\frac q{r}\sigma D -\frac{q\sigma^2}{r^2}\bigg]\,.
\label{4.24}
\eea
For $q=1$ the action \eqref{4.24} coincides with that of \cite{HHL} upon a suitable
redefinition of the auxiliary field $D$, $D=F_3+\frac{q-2}r\sigma -i[\phi,\bar\phi]$.
The auxiliary fields
\be
F_3\,,\quad F=\frac1{\sqrt2}(F_1 - iF_2)\,,
\quad\bar F = \frac1{\sqrt2}(F_1+iF_2)
\ee
completely decouple from the physical
sector and form an $SO(3)\sim SU(2)$ triplet contributing to the action (\ref{4.24}) with the term
$\frac12 F^A F_A$, where $F_A=(F_1,F_2,F_3)$.

Analogously, let us decompose the physical scalars $\phi$ and
$\bar\phi$ into their real and imaginary parts
\be \phi  =
\frac1{\sqrt2}(\phi_1- i\phi_2)\,,\qquad \bar\phi =
\frac1{\sqrt2}(\phi_1 +i\phi_2)\,.
\ee
The three scalars
$\phi_1$, $\phi_2$  and $\sigma$ form the triplet of another $SO(3)\sim SU(2)_R$
\be\label{phid}
\phi_I
=(\phi_1,\phi_2,\sigma)\,,\qquad I=1,2,3\,.
\ee
It is
important to note that the physical scalars $\phi_I$ and the
auxiliary fields $F_A$ transform under different $SU(2)$ groups
which together form the $SU(2)\times SU(2)$ R--symmetry of the $\cN=4$ SYM model.

Finally, we introduce the $SU(2)_R$ doublets of spinors
$\psi_{i\alpha}$, $\bar\psi^{i\alpha}$, $i=1,2$. These spinors are
related to the ones in (\ref{4.24}) as follows
\be\label{psid}
\bar\psi^{1\alpha} = \frac1{\sqrt2}\bar\lambda^\alpha \,,\quad
\psi_{1\alpha} = \frac1{\sqrt2}\lambda_\alpha\,,\quad
\psi_{2\alpha}= \bar\psi_\alpha \,,\quad
 \bar\psi^{2\alpha} =\psi^\alpha\,.
\ee
Eliminating the auxiliary fields and using the fields \eqref{phid} and \eqref{psid} we recast the action (\ref{4.24}) in the manifestly $SU(2)_R$
invariant form
\bea
S^{\cN=4}_{\rm SYM}&=&\frac1{g^2}\tr\int d^3x \sqrt h\big(
{\cal L}_{\rm gauge}+{\cal L}_{\rm scalar}+{\cal L}_{\rm spinor}
\big),\label{S4comp}\\
{\cal L}_{\rm gauge}&=&
\frac14 F^{ab}F_{ab}+\frac q{2r}\varepsilon^{abc}(A_a \hat{\cal D}_b A_c +\frac{2i}{3}A_a A_b A_c)\,,
\label{48_}
\\
{\cal L}_{\rm scalar}&=&\frac12\hat\nabla^a \phi^I \hat\nabla_a \phi_I
+\frac{q(2-q)}{2r^2}\phi^I \phi_I
\nn\\&&
-\frac{3 q-2}{6r}\varepsilon^{IJK}\phi_I [\phi_J,\phi_K]
-\frac14[\phi^I,\phi^J][\phi_I,\phi_J]\,,
\label{4.31_}\\
{\cal L}_{\rm spinor}&=&
i(\gamma^a)_\alpha^\beta \bar\psi^{i\alpha} \hat\nabla_a
\psi_{i\beta}
+\frac{1-2q}{2r}\bar\psi^{i\alpha} \psi_{i\alpha}
+i\bar\psi^{i\alpha}(\gamma^I)_i^j[\phi_I,\psi_{j\alpha}]\,.
\eea
Here $(\gamma^I)_i^j$ are $SO(3)\sim SU(2)_R$ gamma--matrices similar to
(\ref{gamma-matrices}). It is straightforward to check that
(\ref{S4comp}) is invariant under the following $\cN=4$ supersymmetry
transformations
\bea
\delta A_a &=& i (\gamma_a)_\alpha^\beta (
\bar \eta^{i\alpha} \psi_{i\beta}
+\eta_{i\beta} \bar \psi^{i\alpha})
\,,\nn\\
\delta\phi^I &=&(\gamma^I)^i_j(\bar\eta^{j\alpha}\psi_{i\alpha}
+\eta_{i\alpha}\bar\psi^{j\alpha})\,,\nn\\
\delta\bar\psi^{i\alpha}&=&\frac i2 \varepsilon^{abc}
(\gamma_c)^\alpha_\beta \bar\eta^{i\beta}F_{ab}
+i (\gamma^a)^\alpha_\beta (\gamma^I)^i_j \hat\nabla_a \phi^I
\bar\eta^{j\beta}
\nn\\&&
-\frac qr\phi^I (\gamma_I)^i_j\bar\eta^{j\alpha}
-\frac12\varepsilon^{IJK}(\gamma_K)^i_j [\phi_I,\phi_J]
\bar\eta^{j\alpha}
\,,\nn\\
\delta\psi_{i\alpha}&=&-\frac i2 \varepsilon^{abc}
(\gamma_c)_\alpha^\beta \eta_{i\beta}F_{ab}
+i (\gamma^a)_\alpha^\beta (\gamma^I)_i^j \hat\nabla_a \phi^I
\eta_{j\beta}
\nn\\&&
+\frac qr\phi^I (\gamma_I)_i^j\eta_{j\alpha}
+\frac12\varepsilon^{IJK}(\gamma_K)_i^j [\phi_I,\phi_J]
\eta_{j\alpha}
\,,\label{N4susy-comp}
\eea
where $\eta_{i\alpha}$ and $\bar\eta^i_\alpha$ are $SU(2)$--doublets
of Killing spinors obeying standard equation
(\ref{KilSp}).
For $q=0$ and $q=1$ these transformations close according to
the (anti)commutation relations in the $su(2|2)$ superalgebra. The
algebraic properties of these transformations for generic values
of $q$ should still be understood.%
\footnote{Note that an $\cN=4$ superfield description of
a similar model in an $AdS_3$ superspace with
$OSp(4|2)\times SL(2,\mathbb{R})$ as its symmetry group
was developed in a recent paper \cite{Kuz-14}.}

Let us consider the  gauge field equations of motion which follow from the Lagrangian (\ref{48_}),
for simplicity, in the Abelian case
\be
\frac{\delta }{\delta A_a}\int d^3x \sqrt h \,{\cal L}_{\rm gauge}=0
\quad \Rightarrow \quad
\hat{\cal D}^b F_{ab} +\frac{q}{2r}\varepsilon_{abc} F^{bc}=0\,.
\label{4.33_}
\ee
It is convenient to introduce the dual field strength
\be
\tilde F_a = \frac12 \varepsilon_{abc}F^{bc}\,,
\label{eqF}
\ee
which obeys the Bianchi identity
\be
\hat{\cal D}^a \tilde F_a =0\,.
\ee
{}From equation (\ref{4.33_}) it follows that the dual field strength
satisfies the massive ``Klein--Gordon" equation
\be
(-\hat{\cal D}^a \hat{\cal D}_a+\frac 2{r^2} )\tilde F_b -\frac{q^2}{r^2}\tilde
F_b=0\,.
\label{eq-tilde-F}
\ee

Note that the Laplacian operator acting in the space of
divergenceless vector fields on $S^3$ is given by
\be
\Delta =-\hat{\cal D}^a \hat{\cal D}_a +\frac 2{r^2}\,.
\ee
Its spectrum is given in (\ref{vec-spec}). In particular, its
lowest eigenvalue is $\frac{4}{r^2}$. Hence, to avoid negative
energy states in the solution of eq. (\ref{eq-tilde-F})
we should impose the bound $q\leq 2$. Similarly, the
absence of negative energy states for the scalar field in
(\ref{4.31_}) requires $q\geq 0$. Hence, the allowed values of the
parameter $q$ are
\be
0\leq q \leq  2\,.
\label{bound}
\ee
The situation here  is analogous to the Breitelohner--Freedman bound \cite{BF} on the negative mass
square of fields in $AdS$.

\subsection{$\cN=4$ SYM with $U(1)\times U(1)$ R--supersymmetry}
\label{N4add}
The authors of \cite{KWY1} considered an $\cN=4$ SYM model which
consists of one $\cN=2$ gauge multiplet and one chiral multiplet
with the unite R--charge  $q=1$. In contrast with (\ref{N4SYM}) the
classical action of this model is given simply by (\ref{Sbare}) with no extra Chern--Simons
term. In this section we demonstrate that this model is invariant under
the $\cN=4$ supersymmetry which has two Killing spinors with
positive ``chirality'' and other two with the negative one.

Recall that the isometries of the $SU(2|1)/U(1)$ supercoset are
generated by the operator $\mathbb K$ given by (\ref{Kil}) which
includes the Killing spinors of the positive ``chirality'', see
eq.\ (\ref{Ki+}). Consider now the Killing spinors on $S^3$ of the opposite
``chirality'',
\be
\hat{\cal D}_a \eta^\alpha +
\frac{i}{2r}(\gamma_a)^\alpha_\beta \eta^\beta =0\,.
\label{KilSp+}
\ee
On $S^3$ one can choose such a gauge for the Lorentz connection in
which the covariant derivative acts on spinors as
\be
\hat{\cal D}_a \eta^\alpha
=\partial_a \eta^\alpha -\frac i{2r} (\gamma_a)^\alpha_\beta
\eta^\beta\,,
\ee
where $\partial_a= e_a^m(x) \partial_m$ is purely bosonic. Hence,
in this gauge the Killing spinor equation (\ref{KilSp+}) is
simply
\be
\partial_a \eta^\alpha =0\,.
\label{4.42}
\ee
Moreover, we require that $\eta^\alpha$ is neutral under the action of the
$U(1)$ R--symmetry of the manifest $SU(2|1)$ supersymmetry
\be
R \eta^\alpha =0\,.
\label{R=0}
\ee
In this case, using (\ref{D-explicit}), it is straightforward to
check that $\eta^\alpha$ is annihilated by the covariant spinor
derivatives,
\be
{\cal D}_\alpha \eta^\beta= \bar{\cal D}_\alpha \eta^\beta =0\,.
\label{4.44}
\ee

Given a pair of Killing spinors $\eta^\alpha$ and
$\bar\eta^\alpha$ with the properties described above
one can construct an analog of superfield
transformations (\ref{transf-non-Ab})
\bea
\Delta_\eta V&=&\Theta^\alpha\eta_\alpha \overline{\it \Phi}
-\bar\theta^\alpha \bar\eta_\alpha{\it \Phi}\,,\nn\\
\delta_\eta{\it \Phi}&=&-i \eta^\alpha \bar\nabla_\alpha G\,,\qquad
\delta_\eta\overline{\it \Phi}=i\bar\eta^\alpha\nabla_\alpha G\,,
\label{hiddenN2}
\eea
where we have introduced the object
\be
\Theta^\alpha=\theta^\alpha -\frac1r \theta^2\bar\theta^\alpha\,,
\label{Theta}
\ee
which has an important property
\be
{\cal D}_\alpha \Theta^\beta =\delta_\alpha^\beta\,.
\ee
Using this property one can also find the transformation of the
superfield strength $G$,
\be
\delta_\eta G = \frac i2\eta^\alpha\bar\nabla_\alpha \overline{\it\Phi}
-\frac i2 \bar\eta^\alpha \nabla_\alpha{\it\Phi}
-\bar\theta^\alpha\bar\eta_\alpha [G,{\it\Phi}]\,.
\label{susyG}
\ee
Note that, because of (\ref{R=0}), the R--charge of the chiral
superfield is fixed as
\be
R{\it\Phi}  = -{\it\Phi}\,,\qquad
R\overline{\it\Phi} = \overline{\it\Phi}\,.
\ee

The general variation of the action (\ref{Sbare}) is given by
(\ref{4.5}). It is a simple exercise to check that this variation
vanishes for the transformations of the fields (\ref{hiddenN2}),
$\delta_\eta S_0=0$. So, this action is manifestly invariant under
the $\cN=2$ supersymmetry and respects also the hidden $\cN=2$ supersymmetries
(\ref{hiddenN2}). By construction, these supersymmetries are
generated by the Killing spinors of opposite ``chiralities''.

It is instructive to find the closure of the transformations
(\ref{hiddenN2}). For instance, with the use of (\ref{susyG})
one can easily find the commutator of two transformations
(\ref{hiddenN2}) for the chiral superfield
\bea
[\delta_{\eta_2},\delta_{\eta_1}]{\it \Phi}&=&i
\zeta^{\alpha\beta}\gamma^a_{\alpha\beta}\nabla_a {\it\Phi}
+\frac\zeta{r} {\it\Phi}
+ i \zeta[G,{\it\Phi}]
+2i\zeta^{\alpha\beta}\bar\theta_\beta[W_\alpha,{\it\Phi}]\,,
\label{4.45}
\eea
where
\be
\zeta^{\alpha\beta}=\frac12(\eta_1^\alpha\bar \eta_2^\beta -
\eta_2^\alpha\bar\eta_1^\beta)\,,\qquad
\zeta=\zeta^\alpha_\alpha\,.
\ee
The first term in the r.h.s.\ of (\ref{4.45}) is the bosonic
translation while the second one is a $U(1)$ transformation. The
terms with commutators in (\ref{4.45}) provide the covariant
chirality of the superfield expression in the r.h.s. These terms are
required in the non--Abelian case only.

The relation (\ref{4.45}) suggests that the transformations
(\ref{hiddenN2}) close according to the commutation relations of
the super Lie algebra of the group $SU(2|1)$. However, this
$SU(2|1)$ group is different from the one generated by the Killing
supervector considered in Sect.\ \ref{kil-sect}. Indeed,
using the equations (\ref{4.42})--(\ref{4.44}) one can verify that
the Killing spinor $\eta^\alpha$ (as well as $\bar\eta^\alpha$)
defined by (\ref{KilSp+}) is
annihilated by the operator (\ref{Kil}),
\be
\mathbb{K} \eta^\alpha = \mathbb{K} \bar\eta^\alpha =0\,.
\ee
As a consequence, the transformations (\ref{hiddenN2}) commute
with the $SU(2|1)$ ones, up to a field dependent gauge
transformation,
\be
[\delta_\eta,\delta_{\mathbb K}]{\it\Phi}=0\,,\quad
[\delta_\eta,\delta_{\mathbb K}]\overline{\it\Phi}=0\,,\quad
[\delta_\eta,\delta_{\mathbb K}]G=[({\mathbb
K}\bar\theta^\alpha\bar\eta_\alpha) {\it\Phi},G]\,.
\label{closure-eta-K}
\ee
Note that the expression $({\mathbb
K}\bar\theta^\alpha\bar\eta_\alpha)$ in the last commutator is
chiral, $\bar{\cal D}_\alpha({\mathbb
K}\bar\theta^\alpha\bar\eta_\alpha)=0$. This can be verified,
\emph{e.g.} using the explicit form of the operator $\mathbb K$ in
the chiral coordinates given in (\ref{comb}) and (\ref{mbbM}).
In the Abelian case the superfield strength is gauge invariant and
the last commutator (\ref{closure-eta-K}) vanishes identically.
So, the full symmetry group of the model
(\ref{Sbare}) is $SU(2|1)\times SU(2|1)$. A similar model in the
$AdS_3$ space was considered recently  in
the Abelian case \cite{Kuz-14}.

For completeness, in this section we present the component
structure of the action (\ref{Sbare}),
\bea
S_0&=&\frac1{g^2}\tr\int d^3x\sqrt h\bigg[
\frac14F^{ab}F_{ab}+\frac12\hat\nabla^a \sigma \hat\nabla_a\sigma
+\hat\nabla^a \bar\phi \hat\nabla_a \phi
+\frac{1}{r^2}\bar\phi\phi
\nn\\&&
-[\sigma,\phi][\sigma,\bar\phi]
+\frac12[\phi,\bar\phi]^2 +\bar F
F+\frac12\left(D+\frac{2\sigma}{r}+i[\phi,\bar\phi]\right)^2
\nn\\&&
+\frac i2 \lambda^\alpha(\gamma^a)_\alpha^\beta\hat\nabla_a
\bar\lambda_\beta
-\frac i2 \lambda^\alpha[\sigma,\bar\lambda_\alpha]
+\frac1{4r}\lambda^\alpha\bar\lambda_\alpha
\nn\\&&
+i(\gamma^a)_\alpha^\beta \bar \psi^\alpha \hat\nabla_a \psi_\beta
-\frac{1}{2r}\bar\psi^\alpha \psi_\alpha
+i\bar\psi^\alpha[\sigma,\psi_\alpha]
+i\bar\psi^\beta [\bar\lambda_\alpha, \phi]
+i[\bar\phi, \lambda^\alpha] \psi_\alpha
\bigg]\,.
\eea
In contrast with (\ref{S4comp}), the scalars $\phi$, $\bar\phi$
and $\sigma$ have different masses
and do not form an $SU(2)$ triplet. The full R--symmetry of
this model is $U(1)\times U(1)$ because the complex scalars
$\phi$, $\bar\phi$ and the auxiliary fields $F$, $\bar F$
transform independently under two differen $U(1)$ groups.

\subsection{$\cN=8$ SYM}

In the $\cN=2$ superfield description of $\cN=8$, $d=3$ SYM theory its multiplet consists of the gauge
superfield $V$ and an $SU(3)$--triplet of chiral superfields $\Phi_i$, $i=1,2,3$ in the adjoint
representation. The generalization of the flat $\cN=8$, $d=3$ SYM classical action \cite{BPS2}
to the supercoset $SU(2|1)$ is
\bea
\label{N8SYM}
S^{\cN=8}_{\rm SYM}&=&S_{\rm YM}+S_{\rm CS}+S_{\rm pot}\,,\\
S_{\rm YM}&=&-\frac 4{g^2}\tr\int d^3x d^2\theta
d^2\bar\theta\,E(G^2- e^{-V}\bar\Phi^i e^V \Phi_i)\,,\\
S_{\rm CS}&=&-\frac{4}{3rg^2} \tr\int_0^1 dt
\int d^3x d^2\theta d^2\bar\theta\,E \,\bar{\cal D}^\alpha
(e^{-tV}{\cal D}_\alpha e^{tV})
e^{-tV}\partial_t e^{tV}\,,\\
S_{\rm pot}&=&-\frac{i}{3g^2}\tr\int  d^3x d^2 \theta\, {\cal E}\,
\varepsilon^{ijk}\Phi_i[\Phi_j , \Phi_k]
+\frac{i}{3g^2}\tr\int  d^3x d^2 \bar\theta\,
\bar{\cal E}\,
\varepsilon_{ijk}\bar\Phi^i[\bar\Phi^j , \bar\Phi^k].
\label{superpot}
\eea
This action is invariant under the following transformations which
include hidden $\cN=6$ supersymmetry,
\bea
\delta_\Upsilon V &=& i\Upsilon_i\overline{\it \Phi}^i - i \bar \Upsilon^i{\it\Phi}_i\,, \nn\\
\delta_\Upsilon{\it \Phi}_i &=&\bar \nabla^\alpha G{\cal D}_\alpha \Upsilon_i
+ \frac2{3r} G\Upsilon_i+\frac12\varepsilon_{ijk}\bar\nabla^2(\bar \Upsilon^j
\overline{\it\Phi}^k)\,,\nn\\
\delta_\Upsilon \overline{\it\Phi}^i &=& -\nabla^\alpha G \bar {\cal D}_\alpha \bar \Upsilon^i -
\frac2{3r}
G\bar \Upsilon^i -\frac12 \varepsilon^{ijk}\nabla^2(\Upsilon_j
{\it\Phi}_k)\,,
\label{N8susy}
\eea
where $\Upsilon_i$ is a triplet of chiral superfield parameters, $\bar{\cal
D}_\alpha\Upsilon_i=0$, subject to
\be
{\cal D}_a \Upsilon_i =0\,.
\label{constr-Ups}
\ee
Similarly to (\ref{chi-comp}), the superparameters $\Upsilon_i$ contain
three Killing spinors $\eta_{i\alpha}$, each of
which obeys (\ref{KilSp}). In (\ref{N8susy}) we use covariantly
chiral superfields $\it\Phi_i$, $\overline{\it\Phi}^i$ defined
as in (\ref{337}).

The form of the superpotential (\ref{superpot}) fixes the
R--charges of the chiral superfields to be
\be
R \Phi_i = -\frac23\Phi_i\,,\qquad
R\bar\Phi^i = \frac23 \bar\Phi^i\,.
\label{RN8}
\ee
This R--charge differs from the scaling dimension of the chiral
superfields. As a consequence, the localization methods cannot be
directly applied to the $\cN=8$ SYM theory, see \cite{KWY1} for
a discussion of this issue.

Clearly, the superparameters $\Upsilon_i$ should have the same charges as $\Phi_i$
\be
R \Upsilon_i = -\frac23 \Upsilon_i\,,\qquad
R \bar \Upsilon^i = \frac 23\bar \Upsilon^i\,.
\ee
Taking these values of the R--charges into account, it is
straightforward to check that the transformations of the (anti)chiral
superfields in (\ref{N8susy}) preserve the chirality
\be
\bar\nabla_\alpha \delta_\Upsilon{\it\Phi}_i=0\,,\qquad
\nabla_\alpha \delta_\Upsilon \overline{\it\Phi}^i=0\,.
\ee

It is also rather straightforward but a bit lengthy to check, using the identities
\be
\bar\nabla^2 \nabla_\alpha {\it\Phi}_i=4i[W_\alpha,{\it\Phi}_i]\,,\qquad
\nabla^2\bar\nabla_\alpha \overline{\it\Phi}^i =- 4i[\bar W_\alpha,
\overline{\it\Phi}^i]\,,
\ee
that (\ref{N8SYM}) is invariant under (\ref{N8susy}),
$\delta_\Upsilon S_{\rm SYM}^{\cN=8}=0$. In this
procedure, the cancelation of some terms becomes evident only
after passing to the (anti)chiral subspace.

The action (\ref{N8SYM}) contains the real Chern--Simons term
similar to that in the $\cN=4$ SYM model (\ref{N4SYM}). However, in
contrast to the $\cN=4$ case the value of the R--charge of the
chiral superfields is now fixed (\ref{RN8}) by the presence of the superpotential. This value is within
the bound (\ref{bound}), hence, although the Chern--Simons term in
(\ref{N8SYM}) gives a negative topological mass squared, there are
no negative energy states in the theory.

In this section we considered the $\cN=8$ SYM model with Killing
spinors of the same ``chirality'' (obeying the equation (\ref{KilSp_}) with
minus sign). Similar to Section \ref{N4add}, it is
straightforward to construct an $\cN=8$ SYM action invariant under
supersymmetry with Killing spinors of different ``chiralities''.
For instance, one can check that the action (\ref{N8SYM}) with
vanishing Chern--Simons term is still invariant under hidden
$\cN=6$ supersymmetry, but which is associated with six Killing spinors
obeying (\ref{KilSp+}) rather than (\ref{KilSp}). The
transformations of these hidden supersymmetries are a simple
generalization of (\ref{hiddenN2}). There is also a
possibility of constructing an $\cN=8$ SYM model with four Killing
spinors of positive ``chirality'' and four extra ones of the
negative ``chirality''. It would be of interest to study all
these cases in detail and determine corresponding underlying supergroup structures.

\subsection{Gaiotto--Witten theory}
In this section we construct a classical action of the
Gaiotto--Witten model \cite{GWth} on $S^3$. This is a
superconformal Chern--Simons--matter model with $\cN=4$
supersymmetry which consists of two
$\cN=2$ gauge superfields $V$ and $\tilde V$ corresponding
to two different gauge groups and two chiral superfields (a hypermultiplet),
$X_+$ and $X_-$, in the bi--fundamental representation. We find the
classical action of this model in the form
\bea
S_{\rm GW}&=&S_{\rm CS}[V] - S_{\rm CS}[\tilde V]+S_X\,,
\label{GW}\\
S_X&=&4\,\tr\int d^3x d^2\theta d^2\bar\theta\, E(
\bar X_+ e^{V} X_+ e^{-\tilde V}
+ X_- e^{-V}\bar X_- e^{\tilde V}
)\,,
\label{GW-FI}
\eea
where $S_{\rm CS}[V]$ and $S_{\rm CS}[\tilde V]$ are two
Chern--Simons terms for left and right gauge superfields each of
which has the form (\ref{CS}). The action $S_X$ is the standard
action for the chiral superfields minimally interacting with gauge
superfields in the bi--fundamental representation and carrying
R--charge $q$. It is straightforward to check that the action
(\ref{GW}) is invariant under the following superfield
transformation
\bea
\Delta V &=& \bar \Sigma {\cal X}_+{\cal X}_- + \Sigma \bar{\cal X}_- \bar{\cal
X_+}\,,\quad
\Delta \tilde V = \bar \Sigma {\cal X}_- {\cal X}_+ + \Sigma \bar{\cal X}_+ \bar{\cal
X}_-\,,\nn\\
\delta {\cal X}_\pm&=&\pm \bar\nabla^2(\bar \Upsilon \bar{\cal
X}_\mp)\,,\qquad\qquad
\delta\bar{\cal X}_\pm=\pm\nabla^2 (\Upsilon{\cal X}_\mp)\,.
\label{susy-GW}
\eea
Here ${\cal X}_\pm$ and $\bar{\cal X}_\pm$ are covariantly
(anti)chiral superfields,
\be
\bar{\cal X}_+=e^{-\tilde V}\bar X_+ e^{V}\,,\quad
{\cal X}_+= X_+\,,\quad
\bar{\cal X}_- = e^{-V}\bar X_- e^{\tilde V}\,,\quad
{\cal X}_- = X_-\,,
\label{covQ}
\ee
and $\Upsilon$ is a chiral superfield parameter subject to the
constraint (\ref{Dchi=0}). As is shown in (\ref{chi-comp}), in
components it contains the Killing spinor $\eta_\alpha$ and a
parameter of the $SU(2)_R$ symmetry group. Hence, the
variations (\ref{susy-GW}) include transformations of the hidden
$\cN=2$ supersymmetry as well as part of the $SU(2)_R$ R--symmetry.

The superfields
$\Upsilon$ and $\Sigma$ possess the following $U(1)$ R--charges
associated with the manifest $\cN=2$ $SU(2|1)$ supersymmetry
\be
R \Upsilon = 2(q-1)\Upsilon \,,\quad
R \Sigma = -2q\Sigma\,,\quad
R \bar \Upsilon =2(1-q) \bar \Upsilon\,,\quad
R \bar \Sigma =2q \bar \Sigma\,.
\ee
We stress that the (anti)chiral superfields $\Sigma$ and
$\bar\Sigma$ in (\ref{susy-GW}) are not independent. They are related to $\Upsilon$
and $\bar\Upsilon$ as follows
\be
{\cal D}_\alpha \Sigma = - \frac{8i\pi}{k}\bar{\cal D}_\alpha \bar
\Upsilon\,,\qquad
\bar{\cal D}_\alpha \bar \Sigma =- \frac{8i\pi}{k}
 {\cal D}_\alpha \Upsilon\,.
\ee
These equations define $\Sigma$ and $\bar\Sigma$ in terms of
$\Upsilon$ and $\bar\Upsilon$ uniquely. For instance, for the chiral
superfield parameter $\Upsilon$ in the form (\ref{chi-comp}) we find the
following component field decomposition of $\bar\Sigma$ in the chiral basis
\be
\bar \Sigma = -\frac{8i\pi}{k}\left(
\frac{q-1}{r}\bar\theta^2 b
+\bar\theta^\alpha \eta_\alpha
+\frac{q-1}{r}\bar\theta^2 \theta^\alpha \eta_\alpha
+2\theta^\alpha \bar\theta_\alpha a + \frac{q-1}{r}\theta^2
\bar\theta^2 a - \frac rqa
\right)\,.
\label{A}
\ee

\subsubsection{Component form of the Gaiotto--Witten action on $S^3$}
Let us denote the components of the $\cN=2$ superfieds in the Gaiotto--Witten model as follows
\be
V: \{ \sigma, A_a, \lambda_\alpha,\bar\lambda_\alpha, D
\}\,,\quad
\tilde V: \{ \tilde\sigma,\tilde A_a,\tilde \lambda_\alpha,\tilde{\bar\lambda}_\alpha
,\tilde D\}\,,\quad
X_{\pm}: \{\phi_\pm, \psi^\alpha_\pm, F_\pm \}\,.
\ee
These components are defined in accordance with the rules (\ref{gauge-components})
and (\ref{chiral-components}).

The action (\ref{GW}) contains Chern--Simons terms for the gauge
superfields $V$ and $\tilde V$ each of which has the component
structure (\ref{CS-comp}) as well as the matter superfields part
$S_X$ the component structure of which can be read from
(\ref{Schiral-comp}). We thus get
\bea
S_{\rm GW}&=&\tr\int d^3x \, \sqrt h ({\cal L}_{\rm CS}+{\cal L}_{X_+}+{\cal L}_{X_-})\,,\\
{\cal L}_{\rm CS}&=&
\frac{ik}{4\pi}\left[\varepsilon^{abc}(A_a \hat{\cal D}_b A_c +\frac{2i}{3}A_a A_b A_c)
-\bar\lambda^\alpha \lambda_{\alpha}
-2\sigma D -\frac{2\sigma^2}{r}\right]
\nn\\&&
-\frac{ik}{4\pi}\left[\varepsilon^{abc}(\tilde A_a \hat{\cal D}_b \tilde A_c
  +\frac{2i}{3}\tilde A_a \tilde A_b \tilde A_c)
-\tilde{\bar\lambda}^\alpha \tilde\lambda_{\alpha}
-2\tilde\sigma \tilde D -\frac{2\tilde\sigma^2}{r}\right]\,,\\
{\cal L}_{X_+}&=&
\bigg[
\hat\nabla^a \bar\phi_+ \hat\nabla_a \phi_+
+\frac{q(2-q)}{r^2}\bar\phi_+\phi_+
-(\sigma \phi_+ - \phi_+ \tilde\sigma)(\tilde\sigma \bar\phi_+ - \bar\phi_+
\sigma)\nn\\&&
+\frac{2iq}{r}\bar\phi_+(\sigma \phi_+ - \phi_+ \tilde\sigma)
+i\bar\phi_+(D\phi_+ -\phi_+ \tilde D) + F_+ \bar F_+\nn\\&&
+i(\gamma^a)_\alpha^\beta\bar\psi_+^\alpha \hat\nabla_a \psi_{+\beta}
+\frac{1-2q}{2r}\bar\psi_+^\alpha \psi_{+\alpha}
+i\bar\psi_+^\alpha(\sigma \psi_{+\alpha} - \psi_{+\alpha}\tilde\sigma)
\nn\\&&
+i\bar\psi_+^\alpha (\bar\lambda_{\alpha}\phi_+ - \phi_+
\tilde{\bar\lambda}_{\alpha})
-i(\tilde\lambda^\alpha \bar\phi_+ - \bar\phi_+
\lambda^\alpha)\psi_{+\alpha}
\bigg]\,,\\
{\cal L}_{X_-}&=&\bigg[
\hat\nabla^a \bar\phi_- \hat\nabla_a \phi_-
+\frac{q(2-q)}{r^2}\bar\phi_-\phi_-
-(\tilde\sigma \phi_- - \phi_- \sigma)(\sigma \bar\phi_- -
\bar\phi_-\tilde\sigma)\nn\\&&
+\frac{2iq}{r}\bar\phi_-(\tilde\sigma \phi_- - \phi_-\tilde\sigma)
+i\bar\phi_-(\tilde D \phi_- -\phi_- D) + F_- \bar F_-\nn\\&&
+i(\gamma^a)_\alpha^\beta\bar\psi_-^\alpha \hat\nabla_a \psi_{-\beta}
+\frac{1-2q}{2r}\bar\psi_-^\alpha \psi_{-\alpha}
+i\bar\psi_-^\alpha(\tilde\sigma \psi_{-\alpha} - \psi_{-\alpha}\sigma)
\nn\\&&
+i\bar\psi_-^\alpha (\tilde{\bar\lambda}_{\alpha}\phi_- - \phi_-
\bar\lambda_{\alpha})
-i(\lambda^\alpha \bar\phi_- - \bar\phi_-
\tilde\lambda^\alpha)\psi_{-\alpha}
\bigg]\,.
\eea

The component fields $F_\pm$, $\bar F_\pm$, $D$, $\tilde D$, $\sigma$, $\tilde
\sigma$, $\lambda_\alpha$, $\tilde\lambda_\alpha$,
$\bar\lambda_\alpha$, $\tilde{\bar\lambda}_\alpha$ enter the
action $S_{\rm GW}$ algebraically. They can be eliminated using
their equations of motion,
\bea
&&
F_\pm=\bar F _\pm =0\,,\nn\\
&&\sigma = \frac{2\pi}{k}(\phi_+ \bar \phi_+ - \bar\phi_-
\phi_-)\,,\qquad
\tilde\sigma = \frac{2\pi}{k}(\bar\phi_+ \phi_+ - \phi_- \bar\phi_-)\,,\nn\\
&&\lambda_{\alpha}=\frac{4\pi}{k}(\phi_+ \bar\psi_{+\alpha}
 -\bar\psi_{-\alpha}\phi_-)\,,\qquad
\bar\lambda_{\alpha}=\frac{4\pi}{k}(\psi_{+\alpha}\bar\phi_+ - \bar\phi_-
\psi_{-\alpha})\,,\nn\\&&
\tilde\lambda_{\alpha}=\frac{4\pi}{k}(\bar\psi_{+\alpha}\phi_+ - \phi_-\bar\psi_{-\alpha})
\,,\qquad
\tilde{\bar\lambda}_{\alpha}=\frac{4\pi}{k}(\bar\phi_+\psi_{+\alpha}
 - \psi_{-\alpha}\bar\phi_-)\,.
\eea
Next, we combine the scalar and spinor fields into $SU(2)$
doublets as follows
\bea
&&
\phi^i = (\phi^1,\phi^2) = (\phi_+, \bar\phi_-)\,,\qquad
\bar\phi_i = (\bar\phi_1, \bar\phi_2) =
(\bar\phi_+,\phi_-)\,,\nn\\
&&
\psi^i_\alpha=(\psi^1_\alpha,\psi^2_\alpha)=
(\psi_{+\alpha},\bar\psi_{-\alpha})\,,\qquad
\bar\psi_{i\alpha}=(\bar\psi_{1\alpha},\bar\psi_{2\alpha})
=(\bar\psi_{+\alpha},\psi_{-\alpha})\,.
\eea
As a result, we get the component form of the Gaiotto--Witten
action on $S^3$ in the form
\bea
S_{\rm GW}&=&S_{\rm CS}+S_2 +S_{\rm int}\,,
\label{SGW-comp}\\
S_{\rm CS}&=&\frac{ik}{4\pi}\tr\int d^3x\sqrt h\,
\varepsilon^{abc}(A_a \hat{\cal D}_b A_c +\frac{2i}{3}A_a A_b A_c-
\tilde A_a \hat{\cal D}_b \tilde A_c -\frac{2i}{3}\tilde A_a \tilde A_b \tilde A_c)\,,\\
S_2&=&\tr\int d^3x \sqrt h\Big[
\hat\nabla^a \phi^i \hat\nabla_a \bar\phi_i
+\frac{q(2-q)}{r^2}\phi^i\bar\phi_i
+i(\gamma^a)_\alpha^\beta\bar\psi^\alpha_i \hat\nabla_a
\psi^i_\beta+\frac{1-2q}{2r}\bar\psi_i^\alpha \psi^i_{\alpha}
\Big]\,,\label{S2-GW}\\
S_{\rm int}&=&\frac{2\pi}{k}\tr\int d^3x \sqrt h\Big[
\frac ir(1-2q)\phi^i\bar\phi^j \phi_i \bar\phi_j
+\frac{2\pi}{k}
(\phi^i \bar\phi_i \phi^j\bar\phi^k \phi_j \bar\phi_k
+ \bar\phi_i \phi^i \bar\phi^j \phi^k \bar\phi_j \phi_k)
\nn\\&&
-i\psi^{i\alpha}\bar\psi_{i\alpha}\phi^j\bar\phi_j
+i\bar\psi_i^\alpha\psi^i_\alpha\bar\phi_j\phi^j
+i\phi^i\bar\psi_j^\alpha \phi_i \bar\psi^j_\alpha
-i\bar\phi^i \psi_j^\alpha \bar\phi_i \psi^j_\alpha
\Big]\,.
\eea
Here the gauge--covariant derivative $\hat\nabla$ acts on the
matter fields in the bi--fundamental representation by the rule
$\hat\nabla \phi^i = \hat{\cal D}_a\phi^i + i A_a\phi^i - i\phi^i \tilde
A_a$.

Although the canonical value of the $\cN=2$ supersymmetry R--charge of the chiral matter
is $q=\frac12$, the action (\ref{SGW-comp}) is explicitly $SU(2)$ invariant
for arbitrary value of the R--charge, thus manifesting the presence of the extended $\cN=4$ supersymmetry. The natural bound for this
parameter $q$ is (\ref{bound}) for which the mass square of the
scalar fields is positive.

\subsection{ABJ(M) model}
Finally, let us construct the classical action of the ABJ(M) theory \cite{ABJM,Benna,ABJ}
on $S^3$. This model can be considered as an $\cN=6$ supersymmetric generalization of the
Gaoitto--Witten theory \cite{GWth} which involves two
hypermultiplets, $(X_{+i}, X_-^i)$, $i=1,2$, where $X_{+i}$  and
$X_-^i$ are chiral superfields in the bi--fundamental representation
of the gauge group. Each of these two chiral superfields can be
rotated independently by its own $SU(2)$ group which make part
of the full $SU(4)$ R--symmetry group of the ABJM model.
The action of the ABJM model involves a
superpotential which is consistent with this symmetry.
We find the following generalization of this action on $S^3$:
\bea
\label{S-ABJM}
S_{\rm ABJM}&=&S_{\rm CS}[V]-S_{\rm CS}[\tilde V]+S_X +
S_{\rm pot}\,,\\
S_X&=&4\tr\int d^3x d^2\theta d^2\bar\theta \,E \left(
\bar X_+^i e^{V}X_{+i} e^{-\tilde V} +X_-^i e^{-V}\bar X_{-i} e^{\tilde V}
\right)\,,\\
S_{\rm pot}&=&-\frac{4\pi i}{k} \tr\int d^3x d^2\theta\,{\cal E}\left(X_{+i} X_-^i  X_{+j} X_-^j - X_-^i X_{+i}X_-^j X_{+j}\right)
\nn\\&&
-\frac{4\pi i }{k} \tr\int d^3x d^2\bar\theta\,\bar{\cal E}
\left(\bar X_{-i}\bar X_+^i \bar X_{-j}\bar X_+^j - \bar X_+^i\bar X_{-i} \bar X_+^j\bar X_{-j}
\right)\,.
\eea
Similar to the Gaiotto--Witten model (\ref{GW}), this action has two
Chern--Simons terms $S_{\rm CS}[V]$ and $S_{\rm CS}[\tilde V]$ for
the two gauge superfields and the standard kinetic term $S_X$ for the
chiral superfields minimally interacting with the gauge
superfields. The superpotential $S_{\rm pot}$ has the standard ABJM
form which is fixed by the requirement that the
action (\ref{S-ABJM}) be invariant under the following superfield
transformations
\bea
\Delta V &=&-\frac{8i\pi}{k}( \bar \Upsilon^i{}_j {\cal X}_{+i}{\cal X}_-^j
+ \Upsilon_i{}^j \bar{\cal X}_{-j}\bar{\cal X}_+^i )\,,\nn\\
\Delta \tilde V&=&-\frac{8i\pi}{k}(\bar \Upsilon^j{}_i{\cal X}_-^i {\cal X}_{+j}
 + \Upsilon_i{}^j\bar{\cal X}_+^i \bar{\cal X}_{-j})\,,\\
\delta{\cal X}_{+i}&=& \bar\nabla^2(\bar \Upsilon_i{}^j \bar{\cal X}_{-j})\,,\qquad
\delta{\cal X}_-^j = -\bar\nabla^2(\bar \Upsilon_i{}^j \bar{\cal
X}_+^i)\,,\\
\delta\bar{\cal X}_+^i&=&\nabla^2(\Upsilon^i{}_j{\cal X}_-^j)\,,\qquad
\delta\bar{\cal X}_{-j}=-\nabla^2(\Upsilon^i{}_j{\cal X}_{+i})\,.
\label{transfABJM}
\eea
Here ${\cal X}_{\pm i}$ and $\bar{\cal X}_{\pm i }$ are
covariantly (anti)chiral superfields defined similarly to
(\ref{covQ}) and $\Upsilon^i{}_j$ is a quartet of chiral
superfield parameters each of which is constrained by
(\ref{Dchi=0}). In components, it involves four Killing spinors
$(\eta^i{}_j)_\alpha$ (their conjugate are present in
$\bar\Upsilon_i{}^j$) which, together with the manifest supersymmetry,
form the $\cN=6$ supersymmetry of the ABJ(M) model.

To summarize, in this section we  have constructed $\cN=2$ superfield
actions for the models with extended supersymmetry, namely, for $\cN=4$ and
$\cN=8$ SYM, Gaiotto--Witten and ABJ(M) theories. For these models
we have derived the transformations of $\cN=2$ superfields under the hidden supersymmetries.
Although these transformations are the generalization to $SU(2|1)/U(1)$ superspace of the corresponding flat--space
supersymmetries,
to the best of our knowledge, their explicit form has not been given in the literature
before. The extended $\cN=4$ and $\cN=8$ supersymmetry, associated with the $S^3$ Killing spinors of the same ``chirality", requires the extension
of the SYM actions on $S^3$ with the Chern--Simons terms. It would be of interest to
understand the nature of these terms from the point of view of
$\cN=4$ superfield formulations of these theories and
coupling these models to the extended three--dimensional supergravities
considered \emph{e.g.} in \cite{Kuzenko1,Kuzenko3,Kuzenko6,Kuz-14}.

\section{One--loop partition functions}
We will now compute one--loop effective actions and corresponding  partition functions for superfield theories on $SU(2|1)/U(1)$ discussed in the previous section.
\label{quant-sect}
\subsection{Chiral superfield on the gauge superfield background}
Let us consider a pair of chiral superfields $\Phi$ and $\widetilde\Phi$
interacting with an Abelian external background gauge superfield $V$
\be
\label{S-2chiral}
S=4\int d^3x d^2\theta d^2\bar\theta \,E \,
(\overline\Phi e^{V}\Phi + \widetilde{\overline\Phi}
e^{-V}\widetilde\Phi)\,.
\ee
A reason why we consider the pair of the chiral fields is because they carry opposite
charges with respect to the $U(1)$ gauge group.
Hence, there is no
parity anomaly and the Chern--Simons term is not generated at one
loop \cite{Niemi,Redlich1,Redlich2,Aharony1997}.

The problem of computing
the partition function of the chiral supermultiplet on $S^3$ with
an arbitrary R--charge was considered in \cite{KWY,KWY1,HHL,KW,Jafferis} using component
field calculations. Here we will derive similar results using
superfield methods. Note also that the problem of low--energy
effective action of the model (\ref{S-2chiral}) in flat space--time
was considered in \cite{BPS1}.

As we have already done in the previous Section, it is convenient to introduce gauge--covariant (anti)chiral
superfields
\be
\overline{\it \Phi}=\overline\Phi e^V\,,\quad
{\it \Phi}= \Phi\,,\quad
\widetilde{\overline{\it\Phi}} = \widetilde{\overline\Phi} e^{-V}\,,\quad
\widetilde{\it\Phi}=\widetilde\Phi\,,
\ee
such that $\nabla_\alpha\overline{\it\Phi}=0$ and
$\bar\nabla_\alpha{\it\Phi}=0$.
In terms of these superfields the classical action (\ref{S-2chiral}) is simply
\be
\label{S-2chiral1}
S=4\int d^3x d^2\theta d^2\bar\theta \,E \,
(\overline{\it\Phi}{\it\Phi} + \widetilde{\overline{\it\Phi}} \widetilde{\it\Phi})\,.
\ee

Since the background gauge field is non--propagating, the effective action in this model is one--loop
exact,
\be\label{ef1}
\Gamma=-\frac12 \Tr\ln H-\frac12 \Tr\ln \widetilde H\,,
\ee
where $H$ and $\widetilde H$ are the operators acting in the space of the
superfields $({\it\Phi,\overline{\it\Phi}})$ and
$(\widetilde{\it\Phi},\widetilde{\overline{\it\Phi}})$, respectively, \emph{i.e.}
\be
H=\left(
\begin{array}{cc}
0 & -\bar\nabla^2 \\
-\nabla^2 & 0
\end{array}
\right).
\label{H}
\ee
The operator $\widetilde H$ differs from $H$ only in the sign of the
background gauge superfield $V$ due to the opposite $U(1)$ charges of $\it \Phi$ and $\widetilde{\it\Phi}$. The standard procedure of
computing the effective action in the chiral superfield model is
based on squaring the operators $H$ and $\widetilde H$ \cite{BKbook} and rewriting \eqref{ef1} as follows
\be
\Gamma=-\frac14 \Tr\ln H^2-\frac14 \Tr\ln \widetilde H^2\,.
\label{4.6}
\ee
However, one should be careful with this squaring because some part of the
effective action can be lost.\footnote{This is similar to the case of
the Dirac operator on $S^3$ which has both positive and negative
eigenvalues. So, if one naively takes its square, the negative eigenvalues will not be counted.}
Therefore, we will avoid naive squaring like (\ref{4.6}) and consider instead
 the variation of the effective action with respect to the
background gauge superfield $V$,
\be
\label{varGamma}
\delta\Gamma= \int d^3x d^2 \theta d^2 \bar\theta\, E\,\delta V
\langle J \rangle\,,
\ee
where $\langle J\rangle$  is an effective current which is expressed in
terms of the Green's functions of the chiral superfields as follows
\be
\langle J \rangle= \langle \frac{\delta S}{\delta V}\rangle=4\langle
\overline{\it\Phi}{\it \Phi} \rangle-4\langle
\widetilde{\overline{\it\Phi}} \widetilde{\it\Phi} \rangle\,.
\label{J}
\ee
Once the variation \eqref{varGamma} is computed, its integration will give us the value of the effective action.

To compute \eqref{J}, consider the Green's function $\langle
\overline{\it\Phi}(z){\it\Phi}(z')\rangle\equiv{\rm G}_{-+}(z,z')$ which obeys the equation
\be
\bar\nabla^2{\rm G}_{-+}(z,z')=\delta_+(z,z')\,,
\ee
where $\delta_+(z,z')$ is a chiral delta-function ($\bar\nabla_\alpha \delta_+(z,z')=0$),
\be
 \delta_+(z,z')=-\frac14\bar\nabla^2 \delta^7(z,z')\,,\qquad
\delta^7(z,z')=\frac1E
\delta^3(x-x')\delta^2(\theta-\theta')\delta^2(\bar\theta-\bar\theta')\,.
\ee
As a result, to obtain the variation of the effective action
(\ref{varGamma}) we should find the Green's function ${\rm G}_{-+}$ at
coincident superspace points.

As in the flat superspace \cite{BKbook},
the Green's function ${\rm G}_{-+}$ is related to the covariantly chiral Green's
function ${\rm G}_+$,
\be
{\rm G}_{-+}(z,z')=-\frac14 \nabla^2{\rm G}_+(z,z')\,,
\label{G+-}
\ee
where ${\rm G}_+$ obeys
\be
\square_+{\rm G}_+(z,z')=-\delta_+(z,z')\,,\qquad
\square_+ \equiv \frac14\bar\nabla^2 \nabla^2\,.
\label{G+eq}
\ee
Using the algebra (\ref{c-deriv}), the operator $\square_+$
can be represented as
\be
\square_+=-\nabla^a\,\nabla_a+(G-\frac
i{r}R)^2
+i(\bar\nabla^\alpha\bar W_\alpha)
+2i W^\alpha \nabla_\alpha
+\frac1{r}[\nabla^\alpha,\bar\nabla_\alpha]\,.
\label{box+}
\ee

Let us take a very particular background gauge superfield $V=V_0$ such
that its superfield strength $G=G_0$ is constant,
\be
G_0= \frac i2 \bar{\cal D}^\alpha {\cal D}_\alpha V_0 = \sigma_0
=cosnt\,,\qquad
W_{0\,\alpha}=W_{0\,\alpha}=0\,.
\label{const-back}
\ee
As will be discussed in the next section, exactly the background of this kind is
interesting from the point of view of the localization technique.

In the chiral coordinates, the background gauge superfield $V_0$
corresponding to (\ref{const-back}) is
\be
V_0=i\sigma_0(\theta\bar\theta-\frac1{2r}\theta^2\bar\theta^2)\,,
\label{V0}
\ee
and from (\ref{G-comp}) we see that the background values of the component fields
are
\be
\sigma=\sigma_0\,,\quad
D=-\frac{2\sigma_0}{r}\,,\quad
F_{ab}=0\,,\quad
\lambda_\alpha=\bar\lambda_\alpha =0\,.
\label{back-comp}
\ee
For this background the spinorial components of the superfield strengths
vanish (\ref{const-back}), and the form of the operator
(\ref{box+}) simplifies to
\be
\square_+=-\nabla^a \nabla_a +m^2\,,\qquad m^2
 \equiv G_0^2 +\frac{2 i}rG_0(q-1)+\frac{q(2-q)}{r^2}\,,
\label{square+}
\ee
where $m$ is the effective mass. Here we have assumed that $\square_+$ acts on the covariantly chiral
scalar superfields of R--charge $q$.

For the gauge superfield background described above
the chiral Green's function ${\rm G}_+$ (\ref{G+eq}) can be written as
\footnote{Four--dimensional analogs of the relations (\ref{G+Gv}) and (\ref{GG+-}) were first derived
in \cite{KM1,KM2}.}
\be
{\rm G}_+(z,z')=-\frac14\bar \nabla^2{\rm G_o}(z,z')=-\frac14 \bar\nabla'^2{\rm G_o}(z,z')\,,
\label{G+Gv}
\ee
where $\bar\nabla'$ acts on $z'$ and ${\rm G_o}(z,z')$ solves for
\be
\square_{\rm o}{\rm G_o}(z,z')=-\delta^7(z,z')\,,
\qquad
\square_{\rm o}= -\nabla^a \nabla_a + m^2\,.
\ee
The operator $\square_{\rm o}$ has the same expression as
$\square_+$, eq.\
(\ref{square+}), but it acts on the superfields defined in the
full superspace rather than on the chiral superfields.
To check that (\ref{G+Gv}) obeys (\ref{G+eq}) one should use the identities
\be
[\nabla^2 , \square_{\rm o}]=[\bar\nabla^2,\square_{\rm o}]=0\,,
\label{nabla-square}
\ee
which hold for the considered gauge superfield background.

Combining (\ref{G+-}) with (\ref{G+Gv}) we find
\be\label{GG+-}
{\rm G}_{-+}(z,z')=\frac1{16}\nabla^2 \bar\nabla'^2{\rm G_o}(z,z')
=-\frac1{16}\nabla^2 \bar\nabla'^2 \frac1{-\nabla^a \nabla_a
+m^2}\delta^7(z,z')\,.
\ee
Next, using (\ref{nabla-square}) we commute the operators
$\nabla^2$ and $\bar\nabla'^2$ with $(-\nabla^a \nabla_a
+m^2)^{-1}$ and consider the Green's function \eqref{GG+-} at coincident
superspace points
\be
{\rm G}_{-+}(z,z)=-\frac1{-\nabla^a \nabla_a +m^2}\frac1{16}\nabla^2 \bar\nabla'^2
\delta^7(z,z')|_{z=z'}=-\frac1{\Delta_{S^3}
+m^2}\delta^3(x,x')|_{x=x'}\,.
\ee
Note that all the fermionic components of the superspace
delta--function $\delta^7(z,z')$ should be differentiated out by
the operators $\nabla^2$ and $\bar\nabla'^2$ to get the non--vanishing
result. The remaining expression is nothing but the trace of the
inverse of the purely bosonic Laplace--Beltrami operator $\Delta_{S^3}$
acting on scalar fields on the $S^3$--sphere
\be
-\tr\frac1{\Delta_{S^3}
+m^2}\propto -\sum_{j=0}^\infty
\frac{d_j}{\lambda_j+m^2}\,,
\label{sum}
\ee
where $\lambda_j$ are the eigenvalues of the Laplace--Beltrami
operator and $d_j$ are their degeneracies
\be
\lambda_j=\frac 1{r^2}j\left(j+2\right)\,,\qquad
d_j=(j+1)^2\,,\quad j=0,1,2,\ldots
\ee

The sum (\ref{sum}) is divergent. Regularizing it in a standard
way, $\sum 1 = \zeta(0)=-\frac12$, we find
\bea
{\rm G}_{-+}(z,z)&=&\frac{c\pi r^2}2 \sqrt{1- m^2 r^2}\cot \left(
\pi \sqrt{1- m^2 r^2}
\right)\nn\\
&=&\frac{c\pi r^2}{2}(irG_0 +1-q)\cot\big(\pi\left(irG_0+1-q\right) \big).
\label{G-+=cot}
\eea
Here we used the explicit expression for the effective mass $m^2$
given in (\ref{square+}).
The constant $c$ can be fixed from the flat space limit which
was studied in \cite{BPS1}, namely
\be
\lim_{r\to\infty}{\rm G}_{-+}=\frac1{4\pi}G_0 \quad\Rightarrow\quad
c=\frac1{2\pi^2 r^3}\,.
\ee

The formula (\ref{G-+=cot}) is valid for the arbitrary value of the
R--charge $q$. Let us consider several particular values of $q$.
$q=\frac12$ corresponds to the chiral matter fields with canonical
R--charge, $q=0$ and $q=2$ are carried by ghost superfields in
the SYM theory (see next subsection), and $q=1$ is the value of
R--charge of the adjoint chiral multiplet in the $\cN=4$ SYM action
(\ref{N4SYM}) which is singled out by its equality with the scale
dimension of the chiral superfield. For these particular cases the formula
(\ref{G-+=cot}) reduces to
\bea
{\rm G}_{-+}|_{q=\frac12}&=&\frac1{4\pi} G_0 \tanh \pi rG_0
- \frac i{8\pi r} \tanh \pi r G_0\,,
\label{307}
\\
{\rm G}_{-+}|_{q=0}&=&\frac1{4\pi} G_0 \coth \pi r G_0
-\frac i{4\pi r}\coth \pi rG_0
\,,\label{4.28}\\
{\rm G}_{-+}|_{q=1}&=&\frac1{4\pi} G_0 \coth \pi r G_0\,,
\label{q=1}
\\
{\rm G}_{-+}|_{q=2}&=&\frac1{4\pi} G_0 \coth \pi r G_0
+\frac i{4\pi r}\coth \pi rG_0
\,.
\label{4.29}
\eea

Let us now consider in detail the computation of the effective action for the
chiral superfield with the R--charge $q=\frac12$.
Recall that the Green's function
$\langle\widetilde{\overline{\it\Phi}}\widetilde{\it\Phi}\rangle$ is obtained from
$\langle\overline{\it\Phi}{\it\Phi}\rangle$ by changing the sign of the
gauge superfield
\be
\langle\overline{\it \Phi}{\it\Phi}\rangle \xrightarrow{G\to -G}
\langle\widetilde{\overline{\it\Phi}}\widetilde{\it\Phi}\rangle\,.
\ee
As a result, the real part of the Green's functions (\ref{307})
cancel in the effective current (\ref{J})
\be
\langle J \rangle=-\frac{i}{\pi r} \tanh \pi rG_0=-\frac{i}{\pi r} \tanh \pi r\sigma_0\,.
\ee
Now, we substitute this expression for the effective current into
(\ref{varGamma}) and compute the superspace integral similarly to the Fayet--Iliopoulos term (\ref{FI-comp}),
\be
\delta\Gamma=-\frac{i}{\pi r} \tanh(\pi r\sigma_0)\int d^3 x d^2\theta
d^2\bar\theta \,E\,\delta V=\frac1{4\pi r} \tanh(\pi r\sigma_0) \int d^3x\sqrt
h\, \delta D\,.
\ee
Recall that for the considered background (\ref{back-comp})
the auxiliary field $D$ is proportional to the scalar $\sigma$,
$\delta D=-\frac2 r\delta\sigma_0$. Taking into account that
$\sigma_0$ is a constant parameter, we obtain
\be
\delta\Gamma=-\frac{1}{2\pi r^2} \delta \sigma_0 \tanh (\pi r\sigma_0) {\rm
Vol}S^3=-\pi r \delta \sigma_0 \tanh(\pi r \sigma_0)\,.
\ee
Hence,
\be
\Gamma=- \ln\left( c_1 \cosh(\pi r\sigma_0)\right)\,,
\ee
where $c_1$ is an integration constant. The corresponding
partition function is
\be
Z=e^\Gamma=\frac1{c_1 \cosh(\pi r\sigma_0) }\,.
\label{4.36}
\ee

For $\sigma_0=0$ the expression (\ref{4.36}) should reproduce the partition
function of a free chiral supermultiplet on $S^3$, \cite{DMP}. This fixes
the value of the integration constant $c_1$,
\be
c_1=2\,.
\ee

Using the Green's functions (\ref{4.28})--(\ref{4.29}) in a
similar way we find that the partition functions of the chiral superfields
with R--charges $q=0$, $q=1$ and $q=2$ have the following form
\bea
q=0: &&
Z=\frac{1}{(2\sinh \pi r\sigma_0)^2}\,,
\label{Zq=1}\\
q=1: && Z=1\,, \label{Zq=11}\\
q=2: &&Z=(2\sinh \pi r\sigma_0)^2 \,.
\label{Zq=2}
\eea
The partition function of the chiral superfield with the R--charge
$q=1$ is equal to one because the propagator (\ref{q=1}) has no
imaginary part which could contribute to the effective current
(\ref{J}). The fact that this partition function is trivial
was first noticed in \cite{KWY1}.

\subsection{$\cN=2$ super Yang--Mills partition function}
\label{Zsym}
Let us now consider the $\cN=2$ super Yang--Mills theory (\ref{S-SYM}) with the gauge
group $SU(N)$. We are interested in the one--loop partition
function $Z$ which is related to the one--loop effective action
$\Gamma$ as
\be
Z^{\cN=2}_{\rm SYM}=e^{\Gamma[V]}\,.
\ee

To derive the effective action $\Gamma$ we perform the standard
background--quantum splitting \cite{GGRS} $V\to(V_0,v)$ such that
\be
e^V = e^{\Omega^\dag} e^{gv}e^{\Omega}\,,
\label{Vv}
\ee
where $v$ is the Hermitian \emph{quantum} gauge superfield and
$\Omega$ is a complex unconstrained prepotential which defines the Hermitian
\emph{background} gauge superfield $V_0$ as follows
\be
e^{V_0} = e^{\Omega^\dag}e^\Omega\,.
\label{Vv1}
\ee
With this splitting we acquire extra gauge
symmetry which leaves eqs.\ \eqref{Vv} and (\ref{Vv1}) invariant
\be
e^{\Omega}\to e^{i\tau} e^{\Omega} \,,\qquad
e^{gv}\to e^{i\tau }e^{gv}e^{-i\tau}\,,
\label{bg-tr}
\ee
where $\tau(z)$ is a real (Hermitian) superfield parameter. These transformations are called the `background' gauge transformations.

The so--called `quantum' form of the original gauge transformation (\ref{gauge}) is
\be
e^{\Omega}\to e^{i\lambda}e^{\Omega}e^{-i\lambda}\,,\qquad
e^{gv}\to e^{i\bar\lambda}e^{gv}e^{-i\lambda}\,,
\label{q-tr}
\ee
 where
$\lambda(z)$ is a chiral superfield parameter. The basic idea of the background field
method is to fix the gauge symmetry corresponding to the parameter
$\lambda$ such that the effective action remains invariant under the
background gauge transformations (\ref{bg-tr}) with arbitrary
$\tau$.

In general, it is a difficult problem to find the effective action
$\Gamma[V_0]$ for an arbitrary unconstrained background gauge superfield $V_0$. To
simplify the problem, we restrict ourself to the consideration of the
low--energy effective action for
$V_0$ taking vales in the Cartan subalgebra of $su(N)$,
\be
V_0={\rm diag}(V_1,V_2,\ldots V_N)\,,\qquad \sum_{I=1}^N V_I =0\,.
\label{V-backg}
\ee
Moreover, we assume that each of the superfields $V_I$ in (\ref{V-backg})
has a constant superfield strength,
$G_I=\frac i2 {\cal D}^\alpha {\cal D}_\alpha V_I=\sigma_I =
const$, $I=1,\ldots,N$. In components, such a background is given in (\ref{back-comp}).
Although these restrictions may look too strong, as we will show in the next
section, they will allow
us to compute the $\cN=2$ Chern--Simons partition function with the
localization method applied to the superfield action.

One--loop partition function is defined by quadratic fluctuations
of the quantum superfield $v$ around the classical gauge
superfield background $V_0$,\footnote{The details of the
background--quantum expansion of the SYM action in $\cN=1$, $d=4$ superspace
can be found in \cite{GGRS}. This procedure is also directly applied to the
$\cN=2$, $d=3$ SYM model under consideration.}
\be
S_2 =-\frac12\tr\int d^7z\, E\, v(\nabla^\alpha \bar\nabla^2 \nabla_\alpha -4iW^\alpha
\nabla_\alpha)v\,,
\label{S2}
\ee
where the superfield strength $W_\alpha$ and gauge--covariant derivatives
$\nabla_\alpha$ and $\bar\nabla_\alpha$ are constructed with the use of the
background gauge superfield $V_0$ by the rules
(\ref{chiralV}) and (\ref{field-strengths}). These derivatives obey the (anti)commutation
relations similar to (\ref{c-deriv}).
Note that $V_0$ in (\ref{V-backg}) has a constant superfield strength $G_0$.
Hence, the superfield $W_\alpha$ vanishes, $W_\alpha=0$, and the action for the quadratic
fluctuations simplifies to
\be
S_2 =-\frac12\tr\int d^7z\, E\, v\nabla^\alpha \bar\nabla^2 \nabla_\alpha v\,.
\label{S2_}
\ee

The operator $\nabla^\alpha \bar\nabla^2 \nabla_\alpha$ in (\ref{S2_})
is degenerate and requires gauge fixing.
Following the conventional background field method in the $\cN=2$, $d=3$
superspace \cite{BPS2,BPS3}, we fix the gauge
freedom for the quantum transformations (\ref{q-tr}) by imposing the conditions
\be
i\bar\nabla^2 v=f\,,\qquad
i\nabla^2 v=\bar f\,,
\label{gauge-f}
\ee
where $f$ is a fixed covariantly chiral superfunction, $\bar\nabla_\alpha f=0$. This gauge is manifestly supersymmetric.

The corresponding ghost superfield action has the form
\be
S_{\rm FP}=\tr\int d^7z \, E(b+\bar b)L_{gv}
[c+\bar c+\coth(L_{gv})(c-\bar c)]
=\tr\int d^7z\,E(\bar b c-b \bar c)+O(g)\,,
\label{FP}
\ee
where $b$ and $c$ are two covariantly chiral anticommuting ghost
superfields and $L_{gv}X$ denotes the commutator, $L_{gv}X=[g\,v,X]$.
As a result, the one--loop partition function in the SYM theory
is given by the following functional integral
\be
Z^{\cN=2}_{\rm SYM}=\int {\cal D}v {\cal D}b{\cal D}c\,
\delta(f-i\bar\nabla^2 v)
\delta(\bar f- i\nabla^2 v)
e^{-S_2 -S_{\rm FP}}\,.
\label{Z1}
\ee

To represent the delta--functions in (\ref{Z1}) in the Gaussian form, we average
this functional integral with the weight
\be
1=\int {\cal D}f{\cal D}\varphi
e^{\alpha\tr \int d^7z \, E [\bar f f+\bar \varphi \varphi]}\,,
\ee
where $\alpha$ is a real parameter and $\varphi$ is the Grassmann--odd
Nielsen--Kallosh ghost. This yields the following gauge--fixing and
Nielsen--Kallosh ghost actions
\be
S_{\rm gf}=-\frac \alpha2\tr\int d^7z \, E\, v\{\nabla^2,\bar\nabla^2
\}v\,,\qquad
S_{\varphi}=\alpha\, \tr\int d^7z\, E\, \bar\varphi\varphi\,.
\ee
For $\alpha=1/2$ we have
\be
S_2+S_{\rm gf}=-\tr\int d^7z \, E\, v\square_{\rm v} v\,,
\ee
where
\be
\square_{\rm v}=\frac1{4}\{\nabla^2,\bar\nabla^2 \}-\frac12
\nabla^\alpha \bar\nabla^2 \nabla_\alpha
+2i W^\alpha\nabla_\alpha
\ee
is a covariant d'Alembertian operator in the space of real
superfields $v$. With the use of the algebra of the covariant derivatives
(\ref{c-deriv}), this operator can be represented as
\bea
\square_{\rm v}&=&-\nabla^a \nabla_a
+(G_0-\frac i{r}R)^2
+\frac1{r}[\nabla^\alpha,\bar\nabla_\alpha]
\nn\\&&
+2iW^\alpha \nabla_\alpha
-2i \bar  W^\alpha \bar\nabla_\alpha
-i(\nabla^\alpha  W_\alpha)\,.
\label{box-v}
\eea
Since we consider the constant gauge superfield background
$G_0=const$ for which $W_\alpha=0$ and the gauge superfield $v$ has
vanishing R--charge,  the form of the operator (\ref{box-v})
gets simplified to
\be
\square_{\rm v}=-\nabla^a \nabla_a
+G_0^2
+\frac1{r}[\nabla^\alpha,\bar\nabla_\alpha]\,.
\label{4.57}
\ee

In the one--loop approximation the functional integrals in (\ref{Z1})
for the gauge and ghost superfields factorize and the partition function takes the form
\be
Z^{\cN=2}_{\rm SYM}={\rm Det}^{-1/2}\square_{\rm v} \cdot Z_\varphi \cdot
Z_{b,c}\,.
\label{Z-det}
\ee
Here $Z_\varphi$ and $Z_{b,c}$ are one--loop partition functions
corresponding to the chiral ghost superfields $\varphi$ and $(b,c)$, repsectively.
It is important to note that, as is seen from the action
(\ref{FP}), the $b,c$ ghosts have vanishing R--charge while
the Nielsen--Kallosh ghost $\varphi$ has R--charge $+2$ as a
consequence of the gauge--fixing (\ref{gauge-f}),
\be
q_{(b,c)}=0\,,\qquad q_{(\varphi)}=2\,.
\ee

Let us consider the operator $\square_{\rm v}$ in (\ref{Z-det}). In
general, as a consequence of the gauge invariance of the effective action,
the trace of the logarithm of this operator is given by a
functional of the gauge superfield strength $G$
\be
-\frac12\Tr\ln \square_{\rm v} = \int d^7z\, E \, {\cal L}(G_0)\,,
\label{5.60}
\ee
with some effective Lagrangian ${\cal L}(G_0)$. We stress that $\cal
L$ explicitly depends on the superfield strength $G_0$, but not
on the gauge field potential $V_0$, since the Chern--Simons like terms can be
produced by chiral field loops only. So, since we consider
the constant superfield background, $G_0=const$, ${\cal L}(G_0)$
is also a constant. Therefore, the full superspace integral over
this effective Lagrangian vanishes owing to (\ref{zero-volume}).
We conclude that\footnote{A direct proof of (\ref{unit}) based
on the analysis of the spectrum of the operator $\square_{\rm v}$ is
given in Appendix~\ref{AppB2}.}
\be
{\rm Det}^{-1/2}\square_{\rm v}=1\,,
\label{unit}
\ee
\emph{i.e.} there are no contributions from the quantum superfield $v$ to the partition function
(\ref{Z-det}).

At first glance the result (\ref{unit}) might look strange, because
the component field computations of the $\cN=2$ SYM partition
function \cite{KWY} show that the fields of the gauge multiplet
contribute non--trivially. In our case,
the $\cN=2$ SYM partition function is entirely due to the
chiral ghost superfields, while the gauge multiplet itself brings
only trivial contribution (\ref{unit}). In fact, this mismatch is
not so surprising, since we use the supersymmetric gauge (\ref{gauge-f})
while in the component field computation \cite{KWY} one imposes the standard
Lorentz gauge which is obviously non--supersymmetric. In different gauges the modes
giving non--trivial contributions to the partition function can be
distributed differently among the gauge multiplet and ghosts, however
 the final result should be the same, since the partition function is
a gauge invariant object.

Consider now the contributions to the partition function (\ref{Z-det}) of the chiral ghost superfields. For simplicity, let us look at the
Nielsen--Kallosh ghost $\varphi$, the contributions from
$b,c$--ghosts can be analyzed in a similar way. Recall that
$\varphi$ is a covariantly chiral superfield,
$\bar\nabla_\alpha\varphi=0$, with the action
\be
S_{\varphi}=\int d^3x d^2\theta
d^2\bar\theta\,E\,\bar\varphi\varphi\,.
\label{Svarphi}
\ee
These superfields are in the adjoint representation of $SU(N)$.
They can be expanded in the basis elements $e_{IJ}$
\footnote{Here we exclude the diagonal (Cartan) elements form the
sum, $\sum_{I=1}^N e_{II}\varphi_{II}$, as they do not interact
with the background gauge superfield (\ref{V-backg}).}
\be
\varphi=\sum_{I\ne J}^N e_{IJ}\varphi_{IJ}\,,\qquad
\bar\varphi=\sum_{I\ne J}^N e_{JI}\bar\varphi_{IJ}\,.
\ee
where $e_{IJ}$ are $N\times N$ matrices in $gl(N)$ with the
following matrix elements
\be
(e_{IJ})_{KL}=\delta_{IK}\delta_{JL}\,.
\ee
Thus, the action (\ref{Svarphi}) is given by the sum of actions
for covariantly chiral superfields $\varphi_{IJ}$ which do not
interact with each other
\be
S_{\varphi}=\sum_{I\ne J}^N \int d^3x d^2\theta d^2\bar\theta\,E\,
\bar\varphi_{IJ}\varphi_{IJ}\,.
\label{S-sum}
\ee
Each of the superfields $\bar\varphi_{IJ}$ is covariantly
antichiral,
\be
e^{-V_{IJ}}{\cal D}_\alpha e^{V_{IJ}}
\bar\varphi_{IJ}=0
\mbox{ for } I<J\, ,\qquad
e^{V_{IJ}}{\cal D}_\alpha e^{-V_{IJ}}
\bar\varphi_{IJ}=0
\mbox{ for } I>J\,,
\label{cov-chir-varphi}
\ee
where
\be
V_{IJ}=V_I-V_J\,.
\ee
The equations (\ref{cov-chir-varphi}) show that the superfields
$\varphi_{IJ}$ appear in the action (\ref{S-sum}) in pairs in which the two fields have
opposite charges associated with the gauge superfield
$V_{IJ}$. Hence, each of the terms in the sum (\ref{S-sum}) is equivalent to
the chiral superfield action (\ref{S-2chiral1}) for which
the partition function was given in (\ref{Zq=2}). There are
$N(N-1)/2$ pairs of the superfields $\varphi_{IJ}$, hence
\be
Z_\varphi = \prod_{I<J}^N\frac1{(2\sinh \pi r\sigma_{IJ})^{2}}\,.
\label{Zvarphi1}
\ee
Note that the ghost $\varphi$ has Grassmann--odd statistics and contributes
as in (\ref{Zq=2}), but in the inverse power.

The ghost superfields $b$ and $c$ can be considered analogously, keeping in mind that they have vanishing R--charges. So one uses the expression given in
eq. (\ref{Zq=1}), but in the inverse power since the ghosts are Grassmann--odd,
\be
Z_{b,c} = \prod_{I<J}(2\sinh \pi r\sigma_{IJ})^{4}\,.
\label{Zbc1}
\ee

We plug the equations (\ref{Zvarphi1}) and (\ref{Zbc1}) into
(\ref{Z-det}) and obtain the one--loop partition function of the
SYM theory,
\be
Z^{\cN=2}_{\rm SYM}=\prod_{I<J}4\sinh^2( \pi r\sigma_{IJ})\,.
\label{ZSYM}
\ee
This partition function differs from the one computed in \cite{KWY} by the
factor $\prod_{I<J}(\sigma_I-\sigma_J)^2$. As we prove in Appendix
\ref{appC}, this mismatch is due to the fact that in (\ref{Z1}) we perform the functional
integration over the unconstrained superfield $v$ while in the calculations of
\cite{KWY} the zero modes of the scalar field $\sigma$ in the $\cN=2$, $d=3$ gauge
multiplet are effectively removed from the corresponding functional integration.
In Section \ref{localization} we will demonstrate that
the partition function (\ref{ZSYM}) gives the correct result for the
$\cN=2$ Chern--Simons partition function calculated with the use of the superfield version of the
localization method.

\subsection{$\cN=4$ SYM partition function}
In comparison to the $\cN=2$ case (\ref{S-SYM}), the classical action of
$\cN=4$ SYM theory on $S^3$ (\ref{N4SYM}) has one extra chiral superfield
and a Chern--Simons term which comes about with a real
parameter $q$. It is natural, from the point of view of the $SU(2|2)$
group structure of $\cN=4$ supersymmetry, to consider two cases, $q=0$
and $q=1$, both of which are within the bound
(\ref{bound}). For $q=0$ the action (\ref{N4SYM}) has no
Chern--Simons term and resembles the $\cN=4$ SYM action in flat
space. The value $q=1$ is interesting from the point of view of
applications of localization methods \cite{KWY1} because it
coincides with the scaling dimension of the chiral superfield $\Phi$
which constitutes part of the $\cN=4$ gauge multiplet.
Consider one--loop partition functions in the model
(\ref{N4SYM}) for these two values of $q$ separately.

For $q=0$ the one--loop partition function in
the $\cN=4$ SYM model can be represented as
\be
Z^{\cN=4}_{\rm SYM}={\rm Det}^{-1/2}(\square_{\rm v})\cdot Z_\varphi \cdot
Z_{b,c}\cdot Z_\Phi
=Z_{\rm SYM}^{\cN=2}\cdot Z_\Phi\,,
\label{Z-det-N4-q=0}
\ee
where ${\rm Det}^{-1/2}(\square_{\rm v})$ corresponds to the
one--loop determinant for the gauge superfield, $Z_{b,c}$ and
$Z_\varphi$ are contributions from the ghost superfields which are
the same as for the $\cN=2$ SYM while $Z_\Phi$ takes into
account the contribution from the chiral superfield $\Phi$. For
$q=0$ the latter was computed in (\ref{Zq=1}), namely
\be
Z_\Phi=\prod_{I<J}\frac1{4\sinh^2( \pi r\sigma_{IJ})}\,.
\ee
This expression is the inverse for (\ref{ZSYM}). Thus, we conclude
that for $q=0$ the $\cN=4$ SYM one--loop partition function is
\be
Z^{\cN=4}_{\rm SYM}=1\,.
\ee

Consider now the $\cN=4$ SYM partition function for $q=1$,
\be
Z^{\cN=4}_{\rm SYM}={\rm Det}^{-1/2}(\square_{\rm v}
-\frac1{2r}[\nabla^\alpha,\bar\nabla_\alpha])\cdot Z_\varphi \cdot
Z_{b,c}\cdot Z_\Phi\,.
\label{Z-det-N4}
\ee
In contrast to the previous case, the quadratic operator for the
quantum gauge superfield $\square_{\rm v}$ gets shifted by the term
$-\frac1{2r}[\nabla^\alpha,\bar\nabla_\alpha]$ which
originates from the second variational derivative of the Chern--Simons
term in the $\cN=4$ SYM action (\ref{N4SYM}). The same arguments
as in eqs.\ (\ref{5.60}) and (\ref{unit}) can be employed to show
that
\be
{\rm Det}^{-1/2}(\square_{\rm v}
-\frac1{2r}[\nabla^\alpha,\bar\nabla_\alpha]) =1\,.
\label{5.57}
\ee
One can check this identity by analyzing the spectrum of this
operator by the methods of Appendix \ref{appC} and to verify that
this operator has equal numbers of bosonic and fermionic states with the same eigenvalue. Note that for $q=1$ the one--loop partition
function of the chiral superfield is trivial, (\ref{Zq=11}).

The identities (\ref{Zq=11}) and (\ref{5.57}) show that the $\cN=4$
SYM partition function (\ref{Z-det-N4}) receives non--trivial
contributions only from the ghost superfields which have the same
structure as in the $\cN=2$ SYM. Thus, we conclude that
the partition functions of the $\cN=4$ and $\cN=2$ SYM theories
coincide
\be
Z^{\cN=4}_{\rm SYM}=Z^{\cN=2}_{\rm SYM}=\prod_{I<J}4\sinh^2( \pi r\sigma_{IJ})\,.
\label{ZSYM-N4}
\ee
This fact was first noticed in \cite{KWY1}.

Naively, it is straightforward to extend the formula
(\ref{Z-det-N4}) to the case of the $\cN=8$ SYM model
(\ref{N8SYM}), just by taking the factor $Z_\Phi$ in eq.\
(\ref{Z-det-N4}) three times, since there are three chiral
superfields in the game. However, in contrast to the $\cN=4$ case,
each of these factors becomes non--trivial as soon as
the R--charges of the chiral superfields are fractional, eq.\
(\ref{RN8}), and do not coincide with the scaling dimensions of
these fields. As is argued in \cite{KWY1}, the naive computation
of the partition function with the chiral superfields having the
fractional R--charge (\ref{RN8}) does not give the partition
function corresponding to an infrared fixed point of the $\cN=8$
supersymmetric gauge theory. The authors of \cite{KWY1} showed
that to get the relevant partition function one should consider a
`mirror' version of the $\cN=8$ SYM theory which consists of
$\cN=4$ SYM action supplemented with one adjoint and one
fundamental hypermultiplet. The partition function in the latter
model describes the $\cN=8$ SYM theory in the infrared regime and
agrees with the partition function in the ABJM theory.

\section{On localization in $\cN=2$ Chern--Simons theory}
\label{localization}
Before gauge fixing, the path integral for the $\cN=2$
Chern--Simons partition function is given by
\be
Z_{\rm CS}=\int {\cal D}V e^{-S_{\rm CS}}\,.
\ee
According to the localization method \cite{Pestun}, one deforms this
partition function by an operator $X$,
\be
Z_{\rm CS}(t)=\int {\cal D}V e^{-S_{\rm CS}- t X}\,,
\label{Zt}
\ee
which should be $Q$--exact, $X=Q Y$,
with respect to a supersymmetry generator $Q$. This guarantees that
the partition function does not depend on the deformation parameter $t$
\be
\frac{dZ(t)}{dt}=0\,.
\ee

The quantity $X$ should obey some reasonable constraints. Namely, it
should be given by a local gauge--invariant functional of the gauge
superfield $V$ with a `good' kinetic term. The conventional choice
of this operator is just the SYM action \cite{KWY}
\be
X=S_{\rm SYM}\,.
\ee
The Lagrangian of the $\cN=2$ SYM action is known to be $Q$--exact
\cite{KWY}.

In the functional integral (\ref{Zt}) one performs the
background--quantum splitting similar to eq.\ (\ref{Vv}), but with
the parameter $1/\sqrt{t}$ instead of the gauge coupling constant $g$
\be
e^V = e^{\Omega^\dag} e^{\frac1{\sqrt{t}}v'}e^{\Omega}\,,\qquad
e^{V_0}=e^{\Omega^\dag}e^{\Omega}\,.
\label{Vv2}
\ee
Note that at this stage the background gauge field $V_0$ is not restricted to be constant yet.

This generic background--quantum splitting should satisfy the
following natural property
\be
\{ V\}= \{{V_0}\}\oplus \{ v' \}\,,
\ee
\emph{i.e.} the space of all the fields (trajectories) $\{ V \}$ is a direct sum
of the spaces of the fields
$\{{V_0}\}$ and $\{ v' \}$. Then, the integration measure
factorizes
\be
{\cal D}V = {\cal D}{ V_0} {\cal D} v'\,.
\label{decomposition}
\ee
For instance, when all the fields are represented as series in
spherical harmonics on $S^3$ (modes) the decomposition
(\ref{decomposition}) assumes that some of these modes (in particular zero modes) are in ${\cal
D}{V_0}$ and the others are accounted by ${\cal D}v'$. For different choices of
$V_0$ the corresponding redistributions of the modes between the background
and the quantum fields are different. This will be important for the comparison
of the superfield computations with the component field ones.

The functional integration in (\ref{Zt}) requires gauge fixing. We
use the same gauge fixing procedure as in Section \ref{Zsym}, by
taking the  gauge--fixing functions (\ref{gauge-f}) and inserting them
into the functional integral in a standard way
\be
Z_{\rm CS}(t)=\int {\cal D}{V_0} {\cal D}v' {\cal D}b {\cal D}c \,
\delta(f-i\bar\nabla^2 v')
\delta(\bar f- i\nabla^2 v')\,
e^{-S_{\rm CS}[{V_0},\frac1{\sqrt t}v']-tS_{\rm SYM}[{ V_0},\frac1{\sqrt t}v']
-S_{\rm FP}}\,.
\label{5.8}
\ee
The Faddeev--Popov ghost action $S_{\rm FP}$ has the form of
eq.\ (\ref{FP}) but with the gauge coupling constant $g$ replaced with
$\frac1{\sqrt t}$.

The main idea of the localization method is to compute the
functional integral (\ref{Zt}) at  $t\to\infty$. In this
limit the contribution to the functional integral \eqref{Zt} is dominated by
quadratic fluctuations around the so--called critical points, \emph{i.e.} the
points for which $X=0$. In the case under consideration these are
the values of the gauge superfield $V$ for which the classical SYM
action vanishes
\be
S_{\rm SYM}[V_0]=0 \,.
\ee
According to the equation (\ref{SYM-comp}),
the SYM action is equal to zero for the vanishing superfield strength
$W_\alpha$,
\be
S_{\rm SYM}=0 \quad\Leftrightarrow\quad  W_\alpha=0\,,\quad
G=G_0=\mbox{constant matrix}\,.
\label{bg}
\ee
So, the functional integral over ${\cal D}V_0$ in (\ref{Zt}) is localized to
such gauge superfield configurations $V_0$
which have a constant gauge superfield strength $G_0$. Recall that
the lowest component of $G$ is the scalar $\sigma(x)$ which takes its
values in the Lie algebra $\mathfrak g$ of the gauge group, hence,
\be
G_0=\sigma_0 \in {\mathfrak g}\,, \qquad V_0=i\sigma_0(\theta\bar\theta-\frac1{2r}\theta^2\bar\theta^2)\,.
\label{Gg}
\ee
As a result, the integration measure ${\cal D}V_0$ exactly
corresponds to the integration over the zero modes of the
Lie--algebra--valued scalar $\sigma(x)$. Therefore, according to
(\ref{decomposition}), these zero modes should be removed from the
measure ${\cal D}v'$.

The actions $S_{\rm CS}[V_0,\frac1{\sqrt t}v']$ and $S_{\rm SYM}[V_0,\frac1{\sqrt
t}v']$ in (\ref{5.8}) should be expanded in series with respect to $v'$ around the
background field $V_0$. It is easy to see that for large $t$
only classical part in the Chern--Simons action remains
\be
S_{\rm CS}[V_0,g\,v'/\sqrt t]=S_{\rm CS}[V_0]+O(1/\sqrt t)\,,
\ee
while in the SYM action only the quadratic fluctuations survive,
\be
-tS_{\rm SYM}[V_0,g\,v'/\sqrt t]=-S_2[V_0,v']+O(1/\sqrt t)\,,
\ee
where $S_2[V_0,v']$ is given by (\ref{S2_}).
Thus, the path integral defining the partition
function in the Chern--Simons theory takes the following form
\be
Z_{\rm CS} = \int {\cal D}V_0\,  e^{-S_{\rm CS}[V_0]}\cdot
Z'_{\rm SYM}[V_0]\,,
\label{Zcs}
\ee
where
\be
Z'_{\rm SYM}[V_0] = \int' {\cal D}v' {\cal D}b{\cal D}c\,
\delta(f-i\bar\nabla^2 v')
\delta(\bar f- i\nabla^2 v')\,
e^{-S_2[V_0,v']
-S_{\rm FP}}
\label{Z'}
\ee
is the functional integral which has one important difference
from the $\cN=2$ SYM partition function (\ref{Z1}).
In (\ref{Z'}) the integration is over the fields $v'$ excluding their zero modes because they
are already taken into account by the measure ${\cal D}V_0$ in
(\ref{Zcs}) while in (\ref{Z1}) there are no restrictions on the
field $v$. The reason for this is that in (\ref{Z1}) we computed the partition function for the particular background in which the gauge field has vacuum expectation values only in the Cartan subalgebra of the gauge group.

Within the superfield methods considered in the previous section
the computation of the functional integral (\ref{Z'}) in a
generic Lie--algebra valued background $V_0$ is much more
subtle as compared with (\ref{Z1}) because it requires
the separation of zero modes from non--zero ones \emph{within} superfields.
Fortunately, it is possible to
rearrange the integration measures in (\ref{Zcs}) such that the $\cN=2$
SYM one--loop partition function (\ref{Z1}) can be used instead of
(\ref{Z'}). To this end, let us separate the Cartan subalgebra directions
of the Lie--algebra--valued $V_0$ from the rest,
\be
V_0 =V_0^{\mathfrak h} + V_0^{\mathfrak x}\,,\qquad
V_0^{\mathfrak h}\in {\mathfrak h}\,,\quad
V_0^{\mathfrak x}\in {\mathfrak x}\,.
\ee
Here ${\mathfrak h}$ stands for the Cartan subalgebra of $\mathfrak g$ and
$\mathfrak x$ labels the root space directions, ${\mathfrak g}
={\mathfrak h}\oplus {\mathfrak x}$. The integration measure ${\cal
D}V_0$ decomposes as
\be
{\cal D}V_0 = {\cal D}V_0^{\mathfrak h}{\cal D}V_0^{\mathfrak
x}\,.
\ee
Now, let us combine the measure ${\cal
D}V_0^{\mathfrak x}$ with ${\cal D}v'$
\be
{\cal D}v = {\cal D}V_0^{\mathfrak x} {\cal D}v'\,.
\ee
This new integration measure ${\cal D}v$ includes zero modes
(as well as all the non--zero ones)
which were missing in (\ref{Z'}). As a result, we get
\be
Z_{\rm CS} = \int {\cal D}V_0^{\mathfrak h}\,  e^{-S_{\rm CS}[V_0^{\mathfrak h}]}\cdot
Z_{\rm SYM}[V_0^{\mathfrak h}]\,,
\label{Zcsh}
\ee
where $Z_{\rm SYM}[V_0^{\mathfrak h}]$ is exactly the $\cN=2$ SYM partition function
(\ref{Z1}).

Let us consider now the gauge group $U(N)$ with the Lie algebra ${\mathfrak
g}=u(N)$. In this case $V_0^{\mathfrak h}$ is a diagonal
matrix
\be
V_0^{\mathfrak h} = {\rm diag}(V_1,V_2,\ldots V_N)\,,
\label{Vh}
\ee
where each of $V_I$ is as in (\ref{V0}),
$V_I=i\sigma_I(\theta\bar\theta-\frac1{2r}\theta^2\bar\theta^2)$,
$\sigma_I=const$. Hence, the integration measure ${\cal D}V_0^{\mathfrak
h}$ reduces to
\be
{\cal D}V_0^{\mathfrak h} = \prod_{I=1}^N d\sigma_I\,.
\label{measure+}
\ee

It is easy to compute the value of the
Chern--Simons action (\ref{CS-comp}) for the constant gauge superfield background
(\ref{back-comp}). One gets
\be
S_{\rm CS}[V_0^{\mathfrak h}]=i\pi k r^2\,\tr\, \sigma_0^2
=i\pi k r^2 \sum_{L=1}^N \sigma_L^2\,.
\label{SSCS}
\ee
Finally, we substitute (\ref{Z1}), (\ref{SSCS}) and
(\ref{measure+}) into (\ref{Zcsh}) and
arrive at the well--known expression for the
partition function of the Chern--Simons theory
\cite{KWY},
\be
Z_{\rm CS}=\int \prod_{L=1}^N d\sigma_L \, e^{-i\pi k r^2
\sigma^2_L} \prod_{I<J}^N(2\sinh \pi r(\sigma_{I}-\sigma_J))^{2}\,.
\label{result}
\ee

We point out that the expression (\ref{result}) of the partition function  is exactly the same as in \cite{KWY},
but the procedure of arriving at this result is different. Let
us discuss this difference in more detail.

The authors of \cite{KWY} notice that the functional integral
(\ref{Zcs}) has a residual symmetry \eqref{bg-tr} which can be used to reduce the integration over the
Lie--algebra--valued field $V_0$ to the integration over its Cartan subalgebra values
\be
{\cal D}V_0 \to \prod_{I=1}^N d  V_I
\prod_{K<L}(V_K-V_L)^2
=\prod_{I=1}^N d\sigma_I
\prod_{K<L}(\sigma_K-\sigma_L)^2\,,
\label{Vand}
\ee
where $V_I$ parametrize the Cartan subalgebra as in
(\ref{Vh}). Here $\prod_{K<L}(V_K-V_L)^2$, which appears in the reduced measure, is the
so--called Vandermonde (or Weyl) determinant
see, \emph{e.g.} \cite{BT}. Next, one evaluates the factor $Z'_{\rm
SYM}[V_0]$ in (\ref{Zcs}) by computing one--loop determinants for
all the component fields in the $\cN=2$ gauge multiplet with the
following outcome
\be
Z'_{\rm SYM}[V_0]=\prod_{I<J}^N\left(\frac
{2\sinh \pi
r(\sigma_{I}-\sigma_J)}{\sigma_I-\sigma_J}\right)^{2}.
\label{Z''}
\ee
The denominator in (\ref{Z''}) exactly cancels
the Vandermonde factor in (\ref{Vand}) and one gets the same
result for the partition functions as the one obtained by the superfield computations,
\emph{i.e.} eq.\ (\ref{result}).

In the superfield approach for computing the partition function we have effectively
imposed an additional constraint that the critical points around which
the theory is localized are not generic constant scalars valued in the Lie algebra
(\ref{Gg}), but take values only in the Cartan subalgebra.
In this case there is no residual symmetry and the
Vandermonde factor does not appear. The `non--Cartan'
degrees of freedom are taken into account in the factor
$Z_{\rm SYM}[V_0^{\mathfrak h}]$ in (\ref{Zcsh}).

Comparing (\ref{Z''}) with (\ref{ZSYM}) one can see that the
one--loop partition function in the $\cN=2$ SYM differs from the
one computed in \cite{KWY} by the Vandermonde factor
\be
Z_{\rm SYM} = Z'_{\rm SYM} \cdot \prod_{I<J}(\sigma_I -
\sigma_J)^2\,.
\label{ZZ'}
\ee
This identity is proved explicitly in Appendix \ref{appC} by comparing
one--loop determinants contributing to $Z_{\rm SYM}$ and $Z'_{\rm
SYM}$ within the component field approach. In Appendix \ref{appC} we show
that the factor $\prod_{I<J}(\sigma_I -
\sigma_J)^2$ in (\ref{ZZ'}) appears due to the zero modes of the scalar
field which were systematically removed from the SYM partition function
considered in \cite{KWY}.

To summarize, these
two ways of computing the partition function are equivalent since
they differ only in the place where the zero modes of the scalars $\sigma$ are
accommodated, \emph{i.e.} either in the measure ${\cal D}V_0$ or in ${\cal D}v$.
The latter option has turned out to be more convenient in the superfield approach
because it is easier to compute the
one--loop superfield partition function $Z_{\rm SYM}[V_0^{\mathfrak h}]$ with
no restrictions on the integration measure (\emph{i.e.} without separating the
zero modes).

In this section we considered the Coulomb branch localization formula
only for the pure $\cN=2$ Chern--Simons theory.\footnote{Higgs branch localization of various $\cN=2$ gauge theories on $S^3$
have been considered recently in \cite{Fujitsuka-Higgs,Benini-Higgs}.}
It is straightforward to generalize this procedure to models of major interest,
such as the Gaiotto--Witten or ABJM theories. To this end one should include into the consideration
additional chiral matter fields taking values in appropriate representations of the gauge group.
Then the localization formula
(\ref{Zcsh}) just acquires extra factors with one--loop partition
functions of these additional matter superfields. In the component field formulation many
such examples were studied in \cite{KWY,KWY1}.

\section{Discussion}
In this paper, we have constructed the $\cN=2$ superfield formulations of
gauge and matter field theories with rigid $\cN=2$ supersymmetry on three--sphere
$S^3$. Our construction is based on the supercoset $SU(2|1)/U(1)$
which has $S^3$ as its bosonic body. For this coset we have derived an
explicit form of the supervielbein, covariant derivatives and curvature
and used these objects to construct superfield actions for gauge and
matter $\cN=2$ supermultiplets. Upon the integration over the Grassmann--odd coordinates these actions
reduce to the known component field actions which contain terms with $S^3$ curvature \cite{KWY}. The $\cN=2$ superfield actions on $\frac{SU(2|1)}{U(1)}\sim \frac{SU(2|1)_L\times SU(2)_R}{U(1)\times SU(2)}$ are Euclidean counterparts of actions on the $AdS_3$ supercoset $\frac{OSp(2|2)_L\times Sp(2)_R}{SO(2)\times Sp(2)}$  constructed in \cite{Kuzenko1,Kuzenko2,Kuzenko3} within the study of three--dimensional
superfield supergravities.

Using $\cN=2$ superfields on $SU(2|1)/U(1)$ we have also constructed
superfield actions with extended supersymmetry for
$\cN=4$ SYM and Gaiotto--Witten theories, $\cN=8$ SYM, and
$\cN=6$ ABJM theory. An interesting new feature of the $\cN=4$ SYM
action is that it respects the $\cN=4$ supersymmetry and $SU(2)$
R--symmetry for arbitrary value of the charge $q$ of the chiral superfield $\Phi$ under
the $U(1)_R$ subgroup of $SU(2|1)$. This parameter $q$ appears
explicitly both in the action and in the supersymmetry
transformations. The value
$q=1$ corresponds to the canonical scaling dimension of this
superfield. To understand the nature of generic
values of $q$ from the point of view of $\cN=4$ superalgebra
it would be interesting to develop an $\cN=4$
superfield formulation of this model.
Analogously, the extended supersymmetry does not
impose constraints on the values of the $U(1)_R$ charge of the chiral
superfields in the Gaiotto--Witten theory and they may be, in
principle, different from the canonical one $q=\frac12$.

As a further extension and application of the superfield methods it will be interesting to
consider superfield theories on the supercoset $SU(2|1)/[U(1)\times U(1)]$ which contains
the sphere $S^2$ as its bosonic body. Gauge and matter
multiplets on $S^2$ were considered in components in \cite{Gomes12,Benini}
where their partition functions were studied with the localization technique. It would be also of interest to
develop a superfield formulation for five--dimensional gauge
theories on curved backgrounds considered, \emph{e.g.}  in
\cite{Zabzine1,Zabzine2,Hosomichi5,Jafferis5}.

The localization method in supersymmetric field theories effectively reduces the computation of the
full partition functions  to
the calculation of one--loop partition functions for quadratic
fluctuations around critical points \cite{BT,Marino11}. As a rule, in the process of the computation of these
one--loop determinants many cancelations happen among bosonic and
fermionic eigenvalues due to supersymmetry. In superspace, these cancelations occur automatically  in the supersymmetric gauge in which the operators of the quadratic fluctuations of the superfields in gauge theory are manifestly
supersymmetric. In particular, in $\cN=2$, $d=3$ superspace the SYM
partition function is represented as a product of one--loop
determinants for the gauge superfield $v$ itself and the ghost
superfield contributions. Simple superspace arguments allowed us to
conclude that the one--loop determinant of the Laplace--like
operator $\square_{\rm v}$ for the superfield $v$ is equal to one and only the ghost
superfields contribute to the SYM partition function. The
cancelation of the bosonic and fermionic eigenvalues of this operator
is verified also by explicit computations of its spectrum given in
Appendix B.

In superspace, the problem of computing the one--loop partition functions of chiral
superfields reduces to finding the chiral superfield propagator at
coincident superspace points. We have obtained the result by analyzing the
formal superspace expression for this propagator and reducing the
problem to the eigenvalue problem of usual bosonic Laplace
operator acting on scalar fields. However, it will be useful to
derive exact expressions for the chiral and gauge superfield
propagators on curved supersymmetric backgrounds such as $AdS$
space or a sphere. Having at hand exact
superfield propagators one could compute one--loop partition
functions in supersymmetric theories without appealing to the
eigenvalue problem for the component fields. Note that
propagators of some superfields on $AdS_5\times S^5$ superspace were studied in
\cite{Siegel}. It would be useful to extend
these results to chiral and gauge superfields considered in the
present paper.

The one--loop partition function in  $\cN=2$, $d=3$ SYM theory computed
in Section \ref{Zsym} differs from the one obtained in \cite{KWY} by
the factor $\prod_{I<J}(\sigma_I-\sigma_J)^2$, where $\sigma_I$
are vacuum expectation values of the scalar $\sigma(x)$ in the $\cN=2$ gauge supermultiplet.
This mismatch is due to the fact that when computing the one--loop SYM partition
function in the superfield formulation we performed functional integration over
unconstrained superfields, while in \cite{KWY} the zero
modes of component fields are effectively removed from the functional integrals. Such a partition
function with removed zero modes appeared in the localization
formula for the Chern--Simons partition function.
In Section \ref{localization} we have shown
that the $\cN=2$ SYM partition function (\ref{ZSYM}) which
includes contributions of all the modes is equally good for the
localization formula of the Chern--Simons partition function.
To this end, one should take care that the scalar zero modes are
not counted twice. With the use of superfields, it is more natural
to exclude the scalar zero modes from the measure in the
localization formula for the Chern--Simons partition function rather
than from the one--loop SYM partition function.

To conclude, we have demonstrated that the superfield methods
not only simplify the problem of the construction of classical actions for supersymmetric field theories on curved backgrounds, but are also useful for studying their quantum
aspects with the use of the localization method. Although we have restricted
ourselves to three--dimensional gauge and matter theories, it is straightforward
to extend these results to models in other space--time dimensions in which
superspace description is applicable.

\vspace{5mm} {\bf Acknowledgements.} The authors wish to thank
Jaume Gomis for the suggestion to look at the superfield
description of field theories on curved supermanifolds and their
localization. We are grateful to N. Berkovits, I. Buchbinder, J.
Gomis, E. Ivanov, S. Kuzenko, P. Lavrov, O. Lechtenfeld, M.
Mari\~no and P. Sorba
for useful discussions and comments. This work was
partially supported by the Padova University Project CPDA119349
and by the MIUR-PRIN contract 2009-KHZKRX. Work of I.B.S. was also
supported by the Marie Curie research fellowship Nr.\ 909231
``QuantumSupersymmetry'', by the RFBR grants
Nr.\ 12-02-00121, 13-02-90430 and 13-02-91330 and by the LRSS grant Nr.\
88.2014.2.

\appendix
\section{Euclidian $d=3$ gamma--matrices}
The three--dimensional gamma--matrices, taken to be those of Pauli
\be (\gamma_1)_\alpha{}^\beta=\left(
\begin{array}{cc}
0 & 1 \\ 1 &0
\end{array}
\right)\,,\quad (\gamma_2)_\alpha{}^\beta =\left(
\begin{array}{cc}
0 & -i \\ i &0
\end{array}
\right)\,,\quad  \gamma_3=-i\gamma_1\gamma_2=\left(
\begin{array}{cc}
1 &0 \\ 0 &-1
\end{array}
\right)
\label{gamma-matrices}
\ee
obey the Clifford algebra
\be
\{\gamma_a,\gamma_b \}=2\delta_{ab}\,,\qquad
a,b=1,2,3\,,
\ee
and generate the spinor representation of $SU(2)$
\be\label{ssu2}
[\gamma_a,\gamma_b]=i\varepsilon_{abc}\gamma_c\,.
\ee
Basic gamma--matrix relations are
\be
(\gamma_a)_{\alpha\beta} (\gamma_a)^{\gamma\delta}
=-(\delta_\alpha^\gamma \delta_\beta^\delta
 +\delta_\alpha^\delta \delta_\beta^\gamma)
\,,\qquad \tr\, \gamma_a \gamma_b=2\delta_{ab}\,,
\ee
where the spinorial indices are raised and lowered with the
antisymmetric tensors $\varepsilon^{\alpha\beta}$ and $\varepsilon_{\alpha\beta}$
$\varepsilon_{12}=-\varepsilon^{12}=1$. Useful formulae for
products of gamma--matrices:
\be
\gamma^a \gamma^b =
i\varepsilon^{abc}\gamma^c + \delta^{ab}{\bf 1}\,,\qquad
\gamma^a
\gamma^b \gamma^c = i\varepsilon^{abc} {\bf 1} +
\delta^{ab}\gamma^c +\delta^{bc}\gamma^a - \delta^{ac}\gamma^b\,.
\ee

An antisymmetric tensor $\omega^{ab}$ can be converted to a vector
$\omega^c$ and vice versa with the help of Levi--Civita symbol,
\be
\omega^{ab}=\varepsilon^{abc}\omega_c\,,\qquad
\omega_c=\frac12 \varepsilon_{abc}\omega^{ab}\,.
\ee
We use the following conventions for converting the vector and
spinorial indices into each other,
\bea
\omega^a=-i\gamma^a_{\alpha\beta}\omega^{\alpha\beta}\,,&\qquad&
\omega^{\alpha\beta} = -\frac i2
\gamma_a^{\alpha\beta}\omega_a\,,\nn\\
\omega^{ab}=-i\varepsilon^{abc}\gamma^c_{\alpha\beta}\omega^{\alpha\beta}\,,&\qquad&
\omega^{\alpha\beta}=-\frac
i4\varepsilon_{abc}\gamma_c^{\alpha\beta}\omega^{ab}\,.
\eea
In particular, for the bosonic derivative we have
\be
\partial_{\alpha\beta}=-\frac i2\gamma^a_{\alpha\beta}\partial_a\,,\quad
\partial_a=-i
\gamma_a^{\alpha\beta}\partial_{\alpha\beta}\,,\quad
\partial^{\alpha\beta}\partial_{\alpha\beta}=\frac12\partial_a
\partial_a\,.
\ee

\section{Spectra of supersymmetric operators on $S^3$}
The supersymmetric Laplacian operator on the sphere has the form
 (\ref{box+}) or (\ref{box-v}) depending on whether it acts
in the space of covariantly chiral $\Phi$ or vector superfields $V$. As we
will show below, the supersymmetric eigenvalue problems of these
operators are always reduced to the eigenvalue problems of the
component fields in $\Phi$ and $V$. Therefore, before we start considering
supersymmetric operators we summarize the result about the spectra
of conventional Laplacian and Dirac operators on $S^3$. All these
results are well known and can be found \emph{e.g.} in the appendices
of \cite{Marino11}.
\begin{itemize}
\item Laplacian operator $ -\partial^a \partial_a $ acting
on scalar fields $\phi$ has the following eigenvales
\be
-\partial^a \partial_a \phi^{(n)} = \lambda_n \phi^{(n)}\,,\quad
\lambda_n = \frac1{r^2}n(n+2)\,,\quad
d_n = (n+1)^2\,,\quad
n=0,1,2,\ldots
\label{scalar-spectrum}
\ee
Here (and further) $d_n$ means the degeneracy of the
corresponding eigenvalue.
\item Dirac operator $-i \gamma^a \hat{\cal D}_a$ on $S^3$ has the spectrum
\be
\lambda_n^\pm = \pm \frac1r(n+\frac12)\,,\qquad
d_n^\pm = n(n+1)\,,\qquad
n=1,2,3,\ldots
\label{spec-Dirac}
\ee
\item The operator of square of the full angular momentum ${\bf J}^2=-(\partial_a+\frac ir
\gamma_a)^2$ acting on spinors $\psi_\alpha$ has the spectrum
\be
-(\partial_a+\frac ir
\gamma_a)^2\psi^{(n)}_\alpha = \lambda_n \psi^{(n)}_\alpha
\,,\quad
\lambda_n = \frac1{r^2} n(n+2)\,,\quad
d_n = 2(n+1)^2\,,\quad
n=0,1,2,\ldots
\label{spec-ferm}
\ee
This spectrum coincides with the scalar spectrum
(\ref{scalar-spectrum}), but the number of states is doubled
because the spinor $\psi_\alpha$ has two independent components.
Indeed, the operator $\bf J$ of the total angular momentum is given by the sum of orbital
and spin parts,
\be
{\bf J}={\bf L} + {\bf S}\,,\qquad
{\bf L}_a= -\frac i2 \partial_a \,,\qquad
{\bf S}_a = \frac12 \gamma_a\,.
\ee
All these three operators $\bf J$, $\bf L$ and $\bf S$ obey the
commutation relations of the $su(2)$ algebra,
\be
[{\bf J}_a,{\bf J}_b]=i\varepsilon_{abc}{\bf J}_c\,,\qquad
[{\bf L}_a,{\bf L}_b]=i\varepsilon_{abc}{\bf L}_c\,,\qquad
[{\bf S}_a,{\bf S}_b] = i\varepsilon_{abc}{\bf S}_c\,.
\ee
Hence, the spectrum of ${\bf L}^2$ is $\frac1{r^2}n(n+2)$ and the
spectrum of ${\bf J}^2$ is similar, but with shifted values of $n$ as $n\to n\pm
1$,
\be
\lambda_n = \left\{
\begin{array}l
\frac1{r^2}(n+1)(n+3)\,,\qquad
 d_n = (n+2)(n+1)\\
\frac1{r^2}(n-1) (n+1)\,,\qquad
 d_n = n(n+1)\,,\qquad
 n=0,1,2,\ldots
\end{array}
  \right.
\ee
This spectrum is equivalent to (\ref{spec-ferm}).
\item The covariant Laplacian operator $\Delta = -\hat{\cal D}^a \hat{\cal D}_a + \frac
2{r^2}$ acting in the space of divergenceless one--forms $B_a$ on $S^3$, $\partial^a
B_a=0$, has the spectrum
\be
\lambda_n = \frac1{r^2} (n+1)^2\,,\qquad d_n = 2n(n+2)\,.
\label{vec-spec}
\ee

\end{itemize}
\subsection{Chiral superfield Laplacian}
Consider the eigenvalue problem for the operator $H$ (\ref{H}) in
the case of vanishing gauge superfield background,
\be
\frac12\left(
\begin{array}{cc}
0 & -\bar{\cal D}^2 \\
-{\cal D}^2 & 0
\end{array}
\right)
\left(
\begin{array}c
\Phi \\ \bar\Phi
\end{array}
\right)=
\lambda
\left(
\begin{array}c
\Phi \\ \bar\Phi
\end{array}
\right).
\ee
Here $\Phi$ is a chiral superfield, $\bar {\cal D}_\alpha \Phi=0$.
For any $\lambda\ne 0 $ this equation implies
\be
\frac14 \bar{\cal D}^2 {\cal D}^2 \Phi = \lambda^2 \Phi\,,\qquad
\frac14 {\cal D}^2 \bar {\cal D}^2 \bar\Phi = \lambda^2
\bar\Phi\,,
\ee
or
\be
\left(-{\cal D}^a {\cal D}_a+M^2\right) \Phi = \lambda ^2 \Phi \,,\qquad
\left(-{\cal D}^a {\cal D}_a  +M^2\right)\bar\Phi = \lambda^2
\bar\Phi\,,
\label{B3}
\ee
where
\be
M^2 = \frac{q(2-q)}{r^2}\qquad (R\Phi = -q \Phi)\,.
\ee
The equations (\ref{B3}) allow one to find the eigenvalues
$\lambda$ up to signs.

Using the explicit expression (\ref{D-explicit}) for the derivative ${\cal D}_a$
we find
\be
{\cal D}^a {\cal D}_a \Phi =(\partial_a +\frac ir (\gamma_a)^\alpha_\beta
\theta^\beta \partial_\alpha)^2\Phi\,.
\ee
Recall that component field decomposition for the chiral
superfield reads
\be
\Phi = \varphi+\theta^\alpha \psi_\alpha +\frac12\theta^2 F\,.
\label{Phi-comp}
\ee
As a result, we get the following equations for the component
fields
\bea
-\partial^a \partial_a \varphi +M^2\varphi &=& \lambda_{(\varphi)}^2  \varphi\,,\qquad
-\partial^a \partial_a F +M^2 F = \lambda_{(F)}^2 F\,,\label{B6}\\
-(\partial_a + \frac ir \gamma_a)^2 \psi
+M^2\psi&=& \lambda_{(\psi)}^2
\psi\,.
\label{B7}
\eea
The bosonic spectrum for the fields $\varphi$ and $F$ in
(\ref{B6}) can be found from (\ref{scalar-spectrum}),
\be
\lambda_{(\varphi)n}^2=\lambda_{(F)n}^2  = \frac 1{r^2} n (n+2)+M^2\,,\qquad
 d_n = (n+1)^2\,,\qquad n=0,1,2,\ldots
\label{spec-bos}
\ee
Owing to (\ref{spec-ferm}), the fermions spectrum for the fields
$\psi_\alpha$ in (\ref{B7}) appears to be exactly the same,
\be
\lambda_{(\psi)n}^2=\frac 1{r^2} n (n+2)+M^2\,,\qquad
 d_n = 2(n+1)^2\,,\qquad n=0,1,2,\ldots
\ee
Hence, these eigenvalues cancel among each other and the
determinant of the operator (\ref{B3}) is equal to one,
\be
\det\left(-{\cal D}^a {\cal D}_a+M^2\right)
=\frac{\prod_m (\lambda^2_{(\varphi)m})^{d_m}\prod_n (\lambda^2_{(F)n})^{d_n}}{
\prod_k (\lambda^2_{(\psi)k})^{d_k}} =
1\,.
\ee

We point out that thr operator $-{\cal D}^a {\cal D}_a+M^2$ appears
by squaring the operator $H$ in (\ref{H}). However, this
squaring is possible for every $\lambda\ne 0$ while the zero modes
require special considerations. Indeed, the zero modes obey the
equations
\be
{\cal D}^2 \Phi =0 \,, \qquad
\bar{\cal D}^2 \bar\Phi = 0\,,
\ee
instead of (\ref{B3}). These two equations are equivalent and we
consider the first of them. Using the explicit form of the
covariant spinor derivatives (\ref{D-explicit}) for the components of
the chiral superfield (\ref{Phi-comp}) we find
\bea
F&=&0\,,\\
-\partial^a \partial_a \varphi + \frac{q(2-q)}{r^2} \varphi &=&0\,,
\label{412}\\
-i(\gamma^a)^\beta_\alpha \hat{\cal D}_a \psi_\beta +\frac{2q-1}{2r}
\psi_\alpha&=&0\,,\label{413}\\
-(\partial_a + \frac ir\gamma^a)^2\psi_\alpha
+\frac{q(2-q)}{r^2}\psi_\alpha &=& 0\,.
\label{414}
\eea
Here $\hat{\cal D}_a=\partial_a -\frac i2 M_a$ is purely bosonic
covariant derivative acting on the spinor field.
Note that (\ref{414}) is a differential consequence of
(\ref{413}), hence, it does not require separate treatment.

Using (\ref{scalar-spectrum}) and (\ref{spec-Dirac}) we find the
eigenvalues of the operators in the equations (\ref{412}) and (\ref{413}):
\bea
\lambda_{(\varphi)n}&=&\frac1{r^2}n(n+2)+\frac{q(2-q)}{r^2}\,,\qquad
d_n=(n+1)^2\,,\quad n=0,1,2,\ldots\\
\lambda_{(\psi)n}&=&\pm \frac1r(n+\frac12)+\frac{2q-1}{2r}\,,\qquad
d_n^\pm = n(n+1)\,,\qquad
n=1,2,3,\ldots
\eea
These eigenvalues can vanish for some particular values of the
charge $q$. In particular, the values
$q=0$ and $q=2$ should be investigated.

For $q=0$ the equation (\ref{412}) has one zero mode
$\varphi=const$ while the fermionic equation (\ref{413}) has no
zero modes. Hence, for $q=0$ the operator $H$ has two bosonic zero
modes in its spectrum corresponding to $\varphi=const$ and
$\bar\varphi=const$ (the latter appears in the antichiral superfield
$\bar\Phi$).

For $q=2$ the equation (\ref{412}) has one bosonic zero mode, but
there are also two fermionic zero modes in (\ref{413}) as follows
from (\ref{spec-Dirac}). Hence, for $q=2$ the operator $H$ has two
bosonic and four fermionic zero modes (the doubling is because the
antichiral superfield $\bar\Phi$ contributes similarly as $\Phi$).

We point out that for $q=\frac12$ the equations (\ref{412}) and
(\ref{413}) do not have zero modes and $\lambda =0$ only for $\Phi=0$.
Hence, for the chiral matter superfields with canonical R--charge
the operator $H$ has no zero modes.

\subsection{Vector superfield Laplacian on gauge superfield background}
\label{AppB2}
In this section we perform direct computation of the determinant
of the vector superfield Laplacian by calculating its spectrum. We
will use chiral coordinates in which the covariant derivatives are
given by (\ref{D-explicit}) and the background gauge superfield
has the form (\ref{V0}). Then, the operator (\ref{4.57}) can be
written as
\be
\square_{\rm v} =
-{\cal D}^a {\cal D}_a +\frac1r[{\cal D}^\alpha,\bar{\cal D}_\alpha]
+\frac{2i}r \sigma_0 \bar\theta^\alpha \bar{\cal D}_\alpha + \sigma_0^2 -\frac
{2i}r\sigma_0\,,
\ee
where $\sigma_0$ is a constant. We consider the eigenvalue problem
\be
\square_{\rm v} V =\lambda V\,,
\label{eqV}
\ee
where $V$ is a chargeless superfield without any further
constraints. It has the following expansion over Grassmann
coordinates
\be
V(x,\theta,\bar\theta)=w(x,\theta)+\bar\theta^\alpha \Psi_\alpha(x,\theta)
+\bar\theta^2 F(x,\theta)\,,
\label{Vcomp1}
\ee
where $w$, $\Psi_\alpha$ and $F$ are chiral superfields,
\bea
w&=&w_0(x)+\theta^\alpha w_\alpha(x) + \theta^2 {\bf w}(x)\,,\nn\\
\Psi_\alpha&=&\psi_\alpha(x)+\theta_\alpha \varphi(x) + \theta^\beta
A_{(\alpha\beta)}(x) + \theta^2 {\bf \Psi}_\alpha(x)\,,\nn\\
F&=& F_0(x) + \theta^\alpha F_\alpha(x) +\theta^2 {\bf F}(x)\,.
\label{Vcomp2}
\eea

Substituting (\ref{Vcomp1}) into (\ref{eqV}) we get the following eigenvalue problems
for the chiral superfields $w$, $\Psi_\alpha$ and $F$,
\bea
&&(-{\cal D}^a {\cal D}_a +\sigma_0^2 -\frac{2i}{r}\sigma_0)w -\frac 2r \partial_\alpha \Psi^\alpha =
\lambda w\,,\label{344}\\
&&-({\cal D}_a +\frac i{2r}\gamma_a)^2 \Psi_\alpha
+\sigma_0^2 \Psi_\alpha\nn\\&&
\qquad\quad +\frac{2i}r(\gamma^a)_\alpha^\beta \partial_a \Psi_\beta
 +\frac 2{r^2} \theta_\alpha \partial_\beta \Psi^\beta
 +\frac 2{r^2} \theta_\beta\partial_\alpha \Psi^\beta
 -\frac2{r^2}\Psi_\alpha
 +\frac 4r\partial_\alpha F = \lambda\Psi_\alpha\,,\label{345}\\
&&(-{\cal D}_a {\cal D}^a +\sigma_0^2 +\frac{2i}{r}\sigma_0)F = \lambda F\,.\label{346}
\eea
Here we used the fact that the R--charges of the chiral superfields
are $R\Psi_\alpha = -\Psi_\alpha$, $R F= -2F$. Next, we expand
remaining derivatives in (\ref{344}), (\ref{345}) and (\ref{346})
and arrive at the following set of equations for the component
fields
\begin{subequations}
\label{B26}
\bea
-\partial^a \partial_a w_0 -\frac 4r\varphi +(\sigma_0^2 -\frac{2i}{r}\sigma_0)w_0 &=& \lambda w_0 \,,\label{351}\\
-(\partial_a + \frac ir\gamma_a)^2 w_\alpha +\frac 4r
{\bf\Psi}_\alpha+(\sigma_0^2 -\frac{2i}{r}\sigma_0)w_\alpha &=& \lambda w_\alpha\,,\label{B26b}\\
-\partial^a \partial_a {\bf w} +(\sigma_0^2 -\frac{2i}{r}\sigma_0){\bf w}&=& \lambda {\bf w}\,;
\label{B26c}
\eea
\end{subequations}
\begin{subequations}
\label{B27}
\bea
-\partial^a \partial_a F_0 +(\sigma_0^2 +\frac{2i}{r}\sigma_0)F_0 &=& \lambda F_0\,,\label{B27a}\\
-\partial^a \partial_a {\bf F} + (\sigma_0^2 +\frac{2i}{r}\sigma_0){\bf F} &=& \lambda{\bf F}\,,\label{B27b}\\
-(\partial_a + \frac ir \gamma_a)^2 F_\alpha +(\sigma_0^2 +\frac{2i}{r}\sigma_0)F_\alpha &=& \lambda
F_\alpha\,;
\label{B27c}
\eea
\end{subequations}
\begin{subequations}
\label{B28}
\bea
(-\partial_a^2 +\frac1{r^2} +\sigma_0^2)\psi_\alpha + \frac 4{r^2}  F_\alpha
&=&\lambda \psi_\alpha\,,
\label{B28a}\\
(-\partial_a^2 +\frac1{r^2}+\sigma_0^2){\bf \Psi}_\alpha &=&\lambda{\bf
\Psi}_\alpha\,,
\label{B28b}\\
(-\partial_a^2 +\frac4{r^2}+\sigma_0^2)\varphi +\frac ir(\gamma_a)^{\alpha\beta
}\partial_a A_{\alpha\beta}  +\frac8r{\bf F} &=&\lambda\varphi\,,
\label{B28c}
\\
(-\partial_a^2 +\frac4{r^2}+\sigma_0^2)A_{\alpha\beta}
 -\frac{2i}r \gamma^a_{\alpha\beta}\partial_a \varphi
 -\frac{2i}r (\gamma^a)^\gamma_{(\beta}\partial_a A_{\alpha)\gamma}
 &=&\lambda A_{\alpha\beta}\,.
 \label{last}
\eea
\end{subequations}

The bispinor $A_{\alpha\beta}$ is equivalent to a vector,
$A_a = -i \gamma_a^{\alpha\beta}A_{\alpha\beta}$.
Hence, the equation (\ref{last}) can be rewritten as
\be
(-\hat{\cal D}^b \hat{\cal D}_b +\frac 2{r^2}+\sigma_0^2)A_a +\frac 4r \partial_a \varphi
= \lambda A_a\,.
\label{360}
\ee
where we used the fact that the covariant derivative
acts on the vector by the rule
$\hat{\cal D}_a A_b= \partial_a A_b
+\frac1r\varepsilon_{abc}A_c$. Next, we decompose this vector
into the divergenceless $B_a$ and gradient parts,
\be
A_a = B_a + \partial_a b\,,\qquad \partial^a B_a =0\,,\quad
b\ne const\,.
\ee
The equation (\ref{360}) leads to two independent equations for
these components,
\bea
(-\hat{\cal D}_b^2 +\frac 2{r^2 }+\sigma_0^2)  B_a  &=& \lambda B_a\,,\label{362}\\
(-\hat{\cal D}_b^2 +\sigma_0^2)b +\frac 4r \varphi&=& \lambda b\,.\label{363}
\eea
Note also that eq.\ (\ref{B28c}) is equivalent to
\be
(-\partial_a^2 +\frac4{r^2}+\sigma_0^2)\varphi - \frac1r\partial_a^2 b +\frac 8r {\bf F} = \lambda \varphi\,.\label{367}
\ee

Our purpose now is to find the eigenvalues $\lambda$ from the system of equations
(\ref{351})--(\ref{B28b}) and (\ref{362})--(\ref{367}). Some of
these equations are entangled because of the fact that we work in
the chiral coordinates. We start with the case when the equations
(\ref{B27}) have trivial solution, $F_0=F_\alpha={\bf F}=0$. In
this case (\ref{B28}) can be rewritten as
\bea
(-\partial_a^2 +\frac1{r^2}+\sigma_0^2)\psi_\alpha&=&\lambda \psi_\alpha\,,
\label{B34}
\\
(-\partial_a^2 +\frac1{r^2}+\sigma_0^2){\bf \Psi}_\alpha &=&\lambda{\bf
\Psi}_\alpha\,,
\label{B35}
\\
(-\hat{\cal D}_b^2 +\frac 2{r^2 }+\sigma_0^2)  B_a  &=& \lambda B_a\,,
\label{B36}
\\
(-\partial_a^2 +\sigma_0^2)b +\frac 4r \varphi&=& \lambda b\,,\\
(-\partial_a^2 +\frac4{r^2}+\sigma_0^2)\varphi - \frac1r\partial_a^2 b &=& \lambda \varphi\,.
\label{B37}
\eea

The equations (\ref{B34}) and (\ref{B35}) for the spinors
$\psi_\alpha$ and ${\bf\Psi}_\alpha$ have the form of the bosonic
equation (\ref{scalar-spectrum}), but with shifted value of
$\lambda$. Hence, we find the spectrum,
\be
\lambda_n = \frac1{r^2} (n+1)^2+\sigma_0^2\,,\qquad
n=0,1,2,\ldots,
\label{B39}
\ee
with altogether $d_n= 4(n+1)^2$ fermionic states on the corresponding level.

The operator $\Delta=-\hat{\cal D}_a^2 +\frac 2{r^2 }$ in (\ref{B36}) is
nothing but the Laplacian operator acting in the space of
divergenceless one--forms. Its spectrum is given (\ref{vec-spec}).
Thus, the equation for the vector $B_a$ gives eigenvalues
$\lambda_n=\frac1{r^2}(n+1)^2+\sigma_0^2$, with degeneracies
$d_n=2n(n+2)$.

The equations (\ref{B36}) and (\ref{B37}) also have the spectrum
(\ref{B39}) with $d_n = 2n^2 + 4n + 4$ states on the corresponding level.
Thus, the equations (\ref{B34})--(\ref{B37}) have non--trivial
solutions for the values of $\lambda$ given by (\ref{B39}) with
$4(n+1)^2$ bosonic and $4(n+1)^2$ states on the $n$--th level. Finally,
we point out that for every non--trivial solution of these
equations the system (\ref{B26}) has the unique solution of the form
\be
{\bf w}=0\,,\quad
w_\alpha = w_\alpha ({\bf \Psi}_\alpha)\,,\quad
w_0 = w_0(\varphi)\,.
\ee
with some functions $w_\alpha ({\bf \Psi}_\alpha)$ and
$w_0(\varphi)$.
Therefore, no new independent degrees of freedom appear from
(\ref{B26}).

Let us turn to the case when the system (\ref{B27}) has
non--trivial solutions. Equations (\ref{B27}) are similar to (\ref{B6})
and (\ref{B7}) which correspond to the supersymmetric Laplacian
operator acting on the chiral superfield (\ref{B3}) in the
case of vanishing R--charge $q$. Therefore the
equations (\ref{B27}) give the spectrum
\be
\lambda_n = \frac1{r^2}n(n+2)+\sigma_0^2 +\frac{2i}{r}\sigma_0\,,
\qquad n=0,1,2,\ldots,
\label{chir-spec}
\ee
with $d_n = 2(n+1)^2$ bosonic and $d_n = 2(n+1)^2$ fermionic
states on $n$-th level. One can easily see that for
every non--trivial solution of (\ref{B27}) it is possible to find
unique solution of the remaining equations (\ref{B26}) and
(\ref{B28}). Hence, these equations do not give any new degrees of
freedom corresponding to the eigenvalues (\ref{chir-spec}).

The last case to consider is when both systems (\ref{B27}) and
(\ref{B28}) have trivial solutions, $F_0={\bf F}=F_\alpha=0$,
$\psi_\alpha={\bf \Psi}_\alpha=A_{\alpha\beta}=\varphi=0$.
In this case the set of equations (\ref{B26}) is simply
\bea
-\partial^a \partial_a w_0 +(\sigma_0^2 -\frac{2i}{r}\sigma_0)w_0 &=& \lambda w_0 \,,\label{3511}\\
-(\partial_a + \frac ir\gamma_a)^2 w_\alpha +(\sigma_0^2 -\frac{2i}{r}\sigma_0)w_\alpha &=& \lambda w_\alpha\,,\label{B26bb}\\
-\partial^a \partial_a {\bf w} +(\sigma_0^2 -\frac{2i}{r}\sigma_0){\bf w}&=& \lambda {\bf
w}\,.
\eea
These equations are identical to the ones (\ref{B6}), (\ref{B7})
arising from the chiral superfield eigenvalue problem. Hence,
using (\ref{spec-bos}), we can immediately write down the
spectrum,
\be
\lambda_n=\frac1{r^2}n(n+2) +\sigma_0^2 -\frac{2i}{r}\sigma_0\,,
\qquad
n=0,1,2,\ldots.
\label{antichir-spec}
\ee
For any given eigenvalue there are $d_n=2(n+1)^2$ bosonic and
fermionic modes.

To summarize, the system of equations (\ref{B26})--(\ref{B28}) has
the spectrum (\ref{B39}), (\ref{chir-spec}) and
(\ref{antichir-spec}). The numbers of states (degeneracies) for these eigenvalues are
given in Table \ref{Tab}.
\begin{table}[tbh]
\begin{center}
\begin{tabular}{c||c|c|c}
 & $\lambda = \frac1{r^2}(n+1)^2 +\sigma_0^2$ & $\lambda= \frac1{r^2}n(n+2)
 +\sigma_0^2 +\frac{2i}r\sigma_0 $
  & $\lambda= \frac1{r^2}n(n+2)
 +\sigma_0^2 -\frac{2i}r\sigma_0 $
 \\ \hline\hline
$w_0$ & 0 & 0 & $(n+1)^2$ \\ \hline
${\bf w}$ & 0 & 0 & $(n+1)^2$ \\ \hline
$w_\alpha$ & 0 & 0 & $2(n+1)^2$ \\ \hline
$F_0$ & 0 & $(n+1)^2$ & 0 \\ \hline
${\bf F}$ & 0 & $(n+1)^2$ & 0 \\ \hline
$F_\alpha$ & 0 & $2(n+1)^2$ & 0 \\ \hline
$\psi_\alpha$ & $2(n+1)^2$ & 0 & 0 \\ \hline
${\bf \Psi}_\alpha$ & $2(n+1)^2$ & 0 & 0\\ \hline
$B_a$ & $2n(n+2)$ & 0 &0  \\ \hline
$\varphi$, $b$ & $2n^2 + 4n +4$ & 0 & 0 \\ \hline
\end{tabular}
\end{center}
\caption[b]{Degeneracies of eigenvalues of the operator
$\square_{\rm v}$ acting on general superfield $V$.
\label{Tab}}
\end{table}
This table shows that for every eigenvalue $\lambda_n$ there are equal
numbers of bosonic and fermionic eigenstates.  Hence, they exactly
cancel against each other in the determinant of the operator $\square_{\rm v}$,
\be
\det \square_{\rm v} = \frac{\prod \lambda_{\rm bos}}{\prod
\lambda_{\rm ferm}}=1\,.
\ee
This result was used in sect.\ \ref{Zsym} when computing the SYM
partition function.

\section{Component field calculation of the $\cN=2$ SYM one--loop partition
function revisited}
\label{appC}

The one--loop partition function in the $\cN=2$ SYM theory was
computed in \cite{KWY} by considering the spectra of operators of
quadratic fluctuations for bosonic and fermionic fields of the
$\cN=2$ gauge multiplet. Here we revisit these computations with a
special attention to zero modes of scalar fields. In contrast to
\cite{KWY} we use a modified Lorentz gauge which has no zero
modes and gives a mass term to the Laplacian operators of the
Faddeev--Popov ghosts and physical scalar $\sigma$ making these
operators invertible.

Consider the $\cN=2$ super Yang--Mills action in the component form
(\ref{SYMcomp}) and make background--quantum splitting for the
scalar field $\sigma$,
\be
\sigma \to \sigma_0 + g\,\sigma \,,\quad
A_a \to gA_a\,,\quad
\lambda_\alpha \to  g \lambda_\alpha\,,\quad
D\to  g D\,,
\ee
were $g$ is the gauge coupling and $\sigma_0$ is a constant background
field which is chosen to belong to the Cartan subalgebra of the
gauge algebra. For computing the one--loop
partition function it is sufficient to consider the part of the
action (\ref{SYMcomp}) which describes quadratic fluctuations
around this background,
\bea
S_2&=&\tr\int d^3x\, \sqrt h ({\cal L}_{\rm bos} +{\cal L}_{\rm
ferm})\,,\label{C2}\\
{\cal L}_{\rm bos}&=&
\frac12 \hat{\cal D}_a A_b \hat{\cal D}^a A^b - \frac12 \hat{\cal D}_a A_b
\hat{\cal D}^b A^a
+\frac12 \partial^a \sigma \partial_a \sigma
+i\partial_a \sigma [A^a ,\sigma_0]
-\frac12[A_a, \sigma_0]^2
\nn\\ &&+\frac12\left( D+ \frac{2\sigma}{r} \right)^2 \,, \label{C3}\\
{\cal L}_{\rm ferm}&=&\frac i2 \lambda^\alpha
(\gamma^a)_\alpha^\beta \hat{\cal D}_a \bar\lambda_\beta
-\frac i2\lambda^\alpha [\sigma_0,\bar\lambda_\alpha]
+\frac1{4r}\lambda^\alpha \bar\lambda_\alpha\,.
\label{C4}
\eea
Here $\hat{\cal D}_a $
is purely bosonic covariant derivative on $S^3$ with standard
commutation rule, $
[\hat{\cal D}_a ,\hat{\cal D}_b ] = -\frac i{4r} M_{ab}
$.

The one--loop partition function
\be
Z_{\rm SYM}[\sigma_0] = \int {\cal D}A_a \,{\cal D}\sigma\,
 {\cal D}\lambda_\alpha \,{\cal D}D\, e^{-S_2}
\label{C7}
\ee
requires gauge fixing since the SYM action is gauge invariant. The
standard Lorentz gauge
\be
\hat{\cal D}^a A_a =0
\label{Lorentz-gauge}
\ee
(although admissible) is not convenient
 here because there is the cross--term $i\partial_a \sigma [A^a
,\sigma_0]$ in (\ref{C3}). It is desirable to have a propagator in
the diagonal form, without mixing of the fields $A_a$ and
$\sigma$. The simplest way to eliminate this crossing term from the
action is to impose the modified Lorentz gauge,
\be
f= \hat{\cal D}^a A_a + i [\sigma_0,\sigma]\,,
\label{f}
\ee
where $f(x)$ is some fixed function. In principle, one can put
this function to zero, but we keep it to represent the gauge--fixing condition
in the functional integral in Gaussian form. Indeed, the functional
delta--function $\delta(\hat{\cal D}^a A_a + i
[\sigma_0,\sigma]-f)$, after averaging
over $f$ with a suitable weight,
leads to the gauge--fixing term
\be
S_{\rm gf} = \tr\int d^3x\,\sqrt h \, {\cal L}_{\rm
gf}\,,\qquad
{\cal L}_{\rm gf}=\frac12f^2\,.
\ee
Adding this action to (\ref{C2})
we find\footnote{We omit the term $\frac12\left( D+ \frac{2\sigma}{r} \right)^2$
in (\ref{C3}) since the functional integration over the auxiliary field
$D$ gives trivial contribution to the partition function.}
\be
{\cal L}_{\rm bos}+{\cal L}_{\rm gf} =
\frac12 A_a \Delta A^a - \frac12 [A_a,\sigma_0]^2
-\frac12 \sigma \partial^2 \sigma -\frac12
[\sigma_0,\sigma]^2\,,
\label{C10}
\ee
where
\be
\Delta = -\hat{\cal D}^2+ \frac2{r^2}
\ee
is the covariant Laplacian operator in the space on one--forms on
$S^3$. As a result, the gauge fixed version of the functional
integral (\ref{C7}) reads
\be
Z_{\rm SYM}[\sigma_0] = \int {\cal D}A_a \,{\cal D}\sigma\,
{\cal D}\lambda_\alpha \,
 \Delta_{\rm FP}\, e^{-\int d^3x\, \sqrt h({\cal L}_{\rm bos}
+{\cal L}_{\rm ferm}+{\cal L}_{\rm gf})}\,,
\label{C12}
\ee
where $\Delta_{\rm FP}$ is the Faddeev--Popov determinant.
We stress that the functional integration $\int{\cal
D}\sigma$ in (\ref{C12}) runs over all configurations of the scalar field
$\sigma$, including its zero mode (\emph{i.e.}, the zero mode of the operator
$\partial^2$).

Consider the variation of the gauge--fixing function (\ref{f})
under gauge transformations with local gauge parameter
$\lambda=\lambda(x)$,
\be
\delta f = i(\partial^a \partial_a \lambda
-[\sigma_0,[\sigma_0,\lambda]]
+ig\partial^a[A_a,\lambda]
-g[\sigma_0,[\sigma,\lambda]])\,.
\label{C13}
\ee
The last two terms in (\ref{C13}) are not essential for one--loop computations as
they are responsible for interactions of the ghost fields with
the vector $A_a$ and scalar $\sigma$. The quadratic term for the ghost fields
corresponds to the operator
\be
{\cal O}=-\partial^a \partial_a +[\sigma_0,[\sigma_0,\cdot]]\,.
\label{O}
\ee
Hence, the one--loop Faddeev--Popov determinant $\Delta_{\rm FP}
={\rm Det}\,{\cal O}$ is represented by the functional integral
over anticommuting Faddeev-Popov ghosts $b$ and $c$,
\be
\Delta_{\rm FP} = \int {\cal D}b {\cal D}c \,
e^{-S_{\rm FP}}\,,\qquad
S_{\rm FP} = \tr\int d^3x \, \sqrt h\, b (-\partial^2 c
+[\sigma_0,[\sigma_0,c]])\,.
\label{C15}
\ee

Note that the functional integration in (\ref{C15}) is taken over
unrestricted ghost fields $b$ and $c$, including their zero modes.
Indeed, the operator (\ref{O}) is non--degenerate owing to the last
term which is nothing but the mass parameter. This term can be
also interpreted as the interaction of the ghost fields with
the background field $\sigma_0$. This is the crucial difference of our
computation from the one given in \cite{KWY} where the Lorentz
gauge (\ref{Lorentz-gauge}) was imposed and the zero modes of
$\sigma$ did not enter the functional integral over ${\cal
D}\sigma$ in (\ref{C7}) (we will comment on this case in the end
of this Section).

In what follows we concentrate on the gauge group $SU(N)$. In this
case all the fields are given by Hermitian matrices. Consider, for
instance, the gauge field $A_a$ and expand it over the basis in
the Lie algebra $gl(N)$,
\be
A_a = \sum_{I,J=1}^N e_{IJ} A^{IJ}_a\,,\quad
\bar A_a^{IJ} = A_a^{JI}\,,\quad
\sum_{I=1}^N A_a^{II}= 0\,,
\ee
where the basis elements $e_{IJ}$ are given by the matrices
\be
(e_{IJ})_{KL} = \delta_{IK}\delta_{JL}
\ee
with the orthogonality property
\be
\tr\, e_{IJ} e_{KL} = \delta_{IL} \delta_{JK}\,.
\ee
The field $\sigma_0$ in the Cartan subalgebra of $su(N)$ is just
the diagonal matrix,
\be
\sigma_0 = {\rm diag}(\sigma_1,\sigma_2,\ldots,\sigma_N)\,,\qquad
\sum_{I=1}^N \sigma_I =0\,.
\ee
Hence, we have the following properties
\be
[\sigma_0,A_a] = \sum_{I\ne J}^N(\sigma_I-\sigma_J)e_{IJ}
A_a^{IJ}\,,\qquad
\tr[\sigma_0,A_a]^2 = -\sum_{I\ne J}^N (\sigma_I-\sigma_J)^2
A_a^{IJ}\bar A_a^{IJ}\,.
\ee
Applying these rules to all fields in the gauge multiplet we
rewrite the expressions (\ref{C4}) and (\ref{C10}) as well as the
Lagrangian for the ghost fields as
\bea
\tr({\cal L}_{\rm bos}+{\cal L}_{\rm gf}) &=& \frac12\sum_{I\ne J}
\left[\bar A^{IJ}_{a}(\Delta+(\sigma_I-\sigma_J)^2) A^{IJ}_a
+\bar\sigma^{IJ}(-\partial^2+(\sigma_I-\sigma_J)^2)\sigma^{IJ}
\right]\,,\\
\tr\, {\cal L}_{\rm ferm}&=&\frac12\sum_{I\ne J}\left[
\lambda^{IJ}(i\gamma^a \hat{\cal D}_a +\frac1{2r}
-i(\sigma_I-\sigma_J))
\bar\lambda^{IJ}
\right]\,,\\
\tr\, {\cal L}_{\rm FP}&=&\sum_{I\ne J}
 b^{IJ}(-\partial^2 +(\sigma_I-\sigma_J)^2) c^{IJ}\,.
\eea
Hence, the one--loop partition function $Z_{\rm SYM}[\sigma_0]$ factorizes
according to the contributions from different fields as
\bea
Z_{\rm SYM}[\sigma_0] &=& Z_A \cdot Z_\sigma \cdot Z_{\rm ferm} \cdot
Z_{b,c}\,,\label{C23}\\
Z_A &=& {\rm Det}^{-\frac12} (\Delta+ (\sigma_I-\sigma_J)^2)\,,\label{C24}\\
Z_\sigma &=& {\rm Det}^{-\frac12}(-\partial^2 +
(\sigma_I-\sigma_J)^2)\,,\\
Z_{\rm ferm} &=& {\rm Det}(i\gamma^a \hat{\cal D}_a -\frac1{2r} +i
(\sigma_I-\sigma_J))\,,\\
Z_{b,c}&=&{\rm Det}(-\partial^2 +(\sigma_I-\sigma_J)^2)\,.
\label{C27}
\eea

The factor $Z_A$ in (\ref{C23}) deserves special attention. The
determinant in (\ref{C24}) is computed in the space of
unconstrained one--forms $A_a$ on $S^3$. This space naturally decomposes into the
divergenceless one--forms $B_a$, $\partial^a B_a=0$, and the
one--forms given by the gradient of a scalar, $\partial_a \phi$,
\be
A_a = B_a+ \partial_a \phi\,,\qquad
\partial^a B_a = 0\,.
\ee
However, the zero mode of the scalar $\phi$ does not contribute to
$A_a$ and, hence, it should be eliminated. Therefore $Z_A$ decomposes
as
\bea
Z_A &=& Z_B \cdot Z_\phi\,,\\
Z_B &=& {\rm Det'}^{-\frac12}(\Delta + (\sigma_I-\sigma_J)^2)\,,\\
Z_\phi &=& {\rm Det'}^{-\frac12}(-\partial^2 +
(\sigma_I-\sigma_J)^2)\,,
\eea
where the determinant in $Z_B$ is computed in the space of
diverdenceless one--forms $B_a$ and $Z_\phi$ is given by the
determinant of the Laplacian in the space of scalar fields $\phi$,
with the zero mode excluded from the spectrum.

It is straightforward to compute the partition function since the
spectra of all the operators in (\ref{C23})--(\ref{C27}) in known (see Appendix B).
The part
\be
Z_B\cdot Z_{\rm ferm} = \prod_{I>J}\left(
\frac{2\sinh(\pi r(\sigma_I-\sigma_J))}{\sigma_I-\sigma_J}
\right)^2
\label{C32}
\ee
of the partition function was computed in \cite{KWY}. Therein, the
determinants of the other fields did not contribute to the
partition function because they do not interact with the
background fiend $\sigma_0$ in the Lorentz gauge (\ref{Lorentz-gauge}).

In our case the modified Lorentz gauge (\ref{f}) effectively gives
the mass term for the scalar $\sigma$ and ghosts and we earn
additional contribution to the partition function depending on
$\sigma_0$,
\be
Z_\sigma\cdot Z_{b,c}\cdot Z_\phi = \frac{
{\rm Det}(-\partial^2 + (\sigma_I-\sigma_J)^2)}{
{\rm Det}^{\frac12}(-\partial^2 + (\sigma_I-\sigma_J)^2)
{\rm Det'}^{\frac12}(-\partial^2 + (\sigma_I-\sigma_J)^2)}\,.
\label{C33}
\ee
All these determinants correspond to the same operator $-\partial^2 +
(\sigma_I-\sigma_J)^2$ acting in the space of scalar fields. Hence,
all the eigenvalues in (\ref{C33}) cancel except for the zero mode because it is
absent in ${\rm Det'}^{\frac12}(-\partial^2 +
(\sigma_I-\sigma_J)^2)$. Thus,
\be
Z_\sigma\cdot Z_{b,c}\cdot Z_\phi =  \prod_{I>J} (\sigma_I-\sigma_J)^2\,.
\ee
This expression cancels the denominator in (\ref{C32}) and we get
exactly the partition function (\ref{ZSYM}) computed
in sect.\ \ref{Zsym} by superfield methods,
\be
Z_{\rm SYM} = \prod_{\alpha>0} 4\sinh^2 (\pi r
(\sigma_I-\sigma_J))\,.
\label{C35}
\ee

Let us now consider the partition function $Z'_{\rm SYM}$
introduced in (\ref{Z'}) and computed in \cite{KWY}. In components, this partition function
is represented by the same functional integral (\ref{C12}), but
with one important difference, namely, the integration over ${\cal
D}\sigma$ runs over the space of scalar fields excluding their
zero modes. When the zero models of $\sigma$ are dropped out,
the gauge fixing function (\ref f) does not
have zero modes as well and, as a consequences, the zero modes are absent
in the Faddeev--Popov ghost fields $b$ and $c$ in (\ref{C15}). As a
result, the zero modes are now absent in all determinants entering
(\ref{C33}),
\be
Z_\sigma\cdot Z_{b,c}\cdot Z_\phi = \frac{
{\rm Det'}(-\partial^2 + (\sigma_I-\sigma_J)^2)}{
{\rm Det'}^{\frac12}(-\partial^2 + (\sigma_I-\sigma_J)^2)
{\rm Det'}^{\frac12}(-\partial^2 + (\sigma_I-\sigma_J)^2)}=1\,.
\ee
So, only (\ref{C32}) contributes to $Z'_{\rm SYM}$,
\be
Z'_{\rm SYM} = \prod_{\alpha>0}\left(
\frac{2\sinh(\pi r(\sigma_I-\sigma_J))}{\sigma_I-\sigma_J}
\right)^2\,.
\label{C37}
\ee
Exactly this partition function was employed in \cite{KWY} in
the localization formula in the $\cN=2$ Chern--Simons theory. The
denominator of (\ref{C37}) gets cancelled in the final stage of
calculations of \cite{KWY} by the Vandermonde determinant in the
integration measure of the scalar $\sigma_0$.

Comparing the formulae (\ref{C35}) and (\ref{C37}) we get the
proof of the identity (\ref{ZZ'}) used in Section
\ref{localization}.


\begin{thebibliography}{99}
\bibitem{BT}
  M.~Blau and G.~Thompson,
  {\it Localization and diagonalization: A review of functional integral techniques for low dimensional gauge theories and topological field theories},
  J.\ Math.\ Phys.\  {\bf 36} (1995) 2192,
  {\tt hep-th/9501075}.
%
\bibitem{Pestun}
  V.~Pestun,
  {\it Localization of gauge theory on a four-sphere and supersymmetric Wilson loops},
  Commun.\ Math.\ Phys.\  {\bf 313} (2012) 71,
  {\tt arXiv:0712.2824 [hep-th]}.
%
\bibitem{FS}
  G.~Festuccia and N.~Seiberg,
  {\it Rigid supersymmetric theories in curved superspace},
  JHEP {\bf 1106} (2011) 114,
  {\tt arXiv:1105.0689 [hep-th]}.
%
\bibitem{Closset1}
  C.~Closset, T.T.~Dumitrescu, G.~Festuccia, Z.~Komargodski and N.~Seiberg,
  {\it Contact terms, unitarity, and F-maximization in three-dimensional superconformal theories},
  JHEP {\bf 1210} (2012) 053,
  {\tt arXiv:1205.4142 [hep-th]}.
%
\bibitem{Closset2}
  C.~Closset, T.T.~Dumitrescu, G.~Festuccia, Z.~Komargodski and N.~Seiberg,
  {\it Comments on Chern-Simons contact terms in three dimensions},
  JHEP {\bf 1209} (2012) 091,
  {\tt arXiv:1206.5218 [hep-th]}.
%
\bibitem{Closset3}
  C.~Closset, T.T.~Dumitrescu, G.~Festuccia and Z.~Komargodski,
  {\it Supersymmetric field theories on three-manifolds},
  JHEP {\bf 1305} (2013) 017,
  {\tt arXiv:1212.3388 [hep-th]}.
%
\bibitem{BKbook} I.L.~Buchbinder, S.M.~Kuzenko, {\it Ideas and Methods of Supersymmetry and Supergravity},
IOP Publishing, Bristol and Philadelphia, 1998, 656 p.
%
\bibitem{GGRS} S.J.~Gates, M.T.~Grisaru, M.~Ro\v cek, W.~Siegel,
{\it Superspace or one thousand and one lessons in supersymmetry},
Benjamin/Cummings, Reading, U.S.A. (1983).
%
\bibitem{Kuzenko1}
  S.M.~Kuzenko, U.~Lindstrom and G.~Tartaglino-Mazzucchelli,
  {\it Off-shell supergravity-matter couplings in three dimensions},
  JHEP {\bf 1103} (2011) 120,
  {\tt arXiv:1101.4013 [hep-th]}.
%
\bibitem{Kuzenko2}
  S.M.~Kuzenko and G.~Tartaglino-Mazzucchelli,
  {\it Three-dimensional N=2 (AdS) supergravity and associated supercurrents},
  JHEP {\bf 1112} (2011) 052, {\tt arXiv:1109.0496 [hep-th]}.
%
\bibitem{Kuzenko3}
  S.M.~Kuzenko, U.~Lindstrom and G.~Tartaglino-Mazzucchelli,
  {\it Three-dimensional (p,q) AdS superspaces and matter couplings},
  JHEP {\bf 1208} (2012) 024,
  {\tt arXiv:1205.4622 [hep-th]}.
%
\bibitem{Kuzenko4}
  D.~Butter, S.~M.~Kuzenko and G.~Tartaglino-Mazzucchelli,
  {\it Nonlinear sigma models with AdS supersymmetry in three dimensions},
  JHEP {\bf 1302} (2013) 121,
  {\tt arXiv:1210.5906 [hep-th]}.
%
\bibitem{KWY}
  A.~Kapustin, B.~Willett and I.~Yaakov,
  {\it Exact results for Wilson loops in superconformal Chern-Simons theories with matter},
  JHEP {\bf 1003} (2010) 089,
  {\tt arXiv:0909.4559 [hep-th]}.
%
\bibitem{KWY1}
  A.~Kapustin, B.~Willett and I.~Yaakov,
  {\it Nonperturbative tests of three-dimensional dualities},
  JHEP {\bf 1010} (2010) 013,
  {\tt arXiv:1003.5694 [hep-th]}.
%
\bibitem{HHL}
  N.~Hama, K.~Hosomichi and S.~Lee,
  {\it Notes on SUSY gauge theories on three-sphere},
  JHEP {\bf 1103} (2011) 127,
  {\tt arXiv:1012.3512 [hep-th]}.
%
\bibitem{Marino11}
  M.~Mari\~no,
  {\it Lectures on localization and matrix models in supersymmetric Chern-Simons-matter theories},
  J.\ Phys.\ A {\bf 44} (2011) 463001,
  {\tt arXiv:1104.0783 [hep-th]}.
%
\bibitem{Ivanov:2013ova}
  E.~Ivanov and S.~Sidorov,
  {\it Deformed supersymmetric mechanics},
  {\tt arXiv:1307.7690 [hep-th]}.
%
\bibitem{Ivanov:2013cea}
  E.~Ivanov and S.~Sidorov,
  {\it Super K\"ahler oscillator from $SU(2|1)$ superspace},
  {\tt arXiv:1312.6821 [hep-th]}.
%
\bibitem{BILS}
  I.A.~Bandos, E.~Ivanov, J.~Lukierski and D.~Sorokin,
  {\it On the superconformal flatness of AdS superspaces},
  JHEP {\bf 0206} (2002) 040, {\tt hep-th/0205104}.
%
\bibitem{K}
  S.M.~Kuzenko, {\it Prepotentials for N=2 conformal supergravity in three dimensions},
  JHEP {\bf 1212} (2012) 021, {\tt arXiv:1209.3894 [hep-th]}.
%
\bibitem{Ivanov92} E.A.~Ivanov, {\it Chern-Simons matter systems with
manifest N=2 supersymmetry}, Phys. Lett. B {\bf 268} (1991) 203.
%
\bibitem{Kuzenko5}
  S.~M.~Kuzenko, U.~Lindstrom, M.~Rocek, I.~Sachs and G.~Tartaglino-Mazzucchelli,
  {\it Three-dimensional N=2 supergravity theories: From superspace to components},
  {\tt arXiv:1312.4267 [hep-th]}.
%
\bibitem{Kuzenko6}
  S.~M.~Kuzenko and J.~Novak,
  {\it Supergravity-matter actions in three dimensions and Chern-Simons terms},
  {\tt arXiv:1401.2307 [hep-th]}.
%
\bibitem{Kuz-14}
  S.~M.~Kuzenko and G.~Tartaglino-Mazzucchelli,
  {\it N=4 supersymmetric Yang-Mills theories in $AdS_3$},
  {\tt arXiv:1402.3961 [hep-th]}.
%
\bibitem{HKLR} N.J.~Hitchin, A.~Karlhede, U.~Lindstr\"om and
M.~Ro\v cek, {\it  Hyperkahler metrics and supersymmetry},
Commun. Math. Phys. {\bf 108} (1987) 535.
%
\bibitem{Cremmer:1978km}
  E.~Cremmer, B.~Julia and J.~Scherk,
  {\it Supergravity theory in eleven dimensions},
  Phys.\ Lett.\ B {\bf 76} (1978) 409.
%
\bibitem{Lu:2002uw}
  H.~Lu, C.~N.~Pope and E.~Sezgin,
  {\it SU(2) reduction of six-dimensional (1,0) supergravity},
  Nucl.\ Phys.\ B {\bf 668} (2003) 237,
  {\tt hep-th/0212323}.
%
\bibitem{Lu:2003yt}
  H.~Lu, C.~N.~Pope and E.~Sezgin,
  {Yang-Mills-Chern-Simons supergravity},
  Class.\ Quant.\ Grav.\  {\bf 21} (2004) 2733,
  {\tt hep-th/0305242}.
%
\bibitem{BF} P.~Breitenlohner and D.Z.~Freedman, {\it Positive energy in Anti-de Sitter
backgrounds and gauged extended supergravity}, Phys. Lett. B {\bf 115} (1982) 197.
%
\bibitem{BPS2}
  I.L.~Buchbinder, N.G.~Pletnev and I.B.~Samsonov,
  {\it Low-energy effective actions in three-dimensional extended SYM theories},
  JHEP {\bf 1101} (2011) 121,
  {\tt arXiv:1010.4967 [hep-th]}.
%
\bibitem{GWth}
  D.~Gaiotto and E.~Witten,
  {\it Janus configurations, Chern-Simons couplings, and the theta-angle
  in N=4 super Yang-Mills theory},
  JHEP {\bf 1006} (2010) 097,
  {\tt arXiv:0804.2907 [hep-th]}.
%
\bibitem{ABJM}
  O.~Aharony, O.~Bergman, D.L.~Jafferis and J.~Maldacena,
  {\it N=6 superconformal Chern-Simons-matter theories, M2-branes and their gravity duals},
  JHEP {\bf 0810} (2008) 091,
  {\tt arXiv:0806.1218 [hep-th]}.
%
\bibitem{Benna}
  M.~Benna, I.~Klebanov, T.~Klose and M.~Smedback,
  {\it Superconformal Chern-Simons theories and AdS$_4$/CFT$_3$ correspondence},
  JHEP {\bf 0809} (2008) 072,
  {\tt arXiv:0806.1519 [hep-th]}.
%
\bibitem{ABJ}
  O.~Aharony, O.~Bergman and D.L.~Jafferis,
  {\it Fractional M2-branes},
  JHEP {\bf 0811} (2008) 043,
  {\tt arXiv:0807.4924 [hep-th]}.
%
\bibitem{Niemi}
  A.~J.~Niemi and G.~W.~Semenoff,
  {\it Axial anomaly induced fermion fractionization and effective gauge theory actions
  in odd dimensional space-times},
  Phys.\ Rev.\ Lett.\  {\bf 51} (1983) 2077.
%
\bibitem{Redlich1}
  A.~N.~Redlich,
  {\it Gauge noninvariance and parity violation of three-dimensional fermions},
  Phys.\ Rev.\ Lett.\  {\bf 52} (1984) 18.
%
\bibitem{Redlich2}
  A.~N.~Redlich,
  {\it Parity violation and gauge noninvariance of the effective gauge field
  action in three-dimensions},
  Phys.\ Rev.\ D {\bf 29} (1984) 2366.
%
\bibitem{Aharony1997}
  O.~Aharony, A.~Hanany, K.~A.~Intriligator, N.~Seiberg and M.~J.~Strassler,
  {\it Aspects of N=2 supersymmetric gauge theories in three-dimensions},
  Nucl.\ Phys.\ B {\bf 499} (1997) 67,
  {\tt hep-th/9703110}.
%
\bibitem{KW}
  B.~Willett and I.~Yaakov,
  {\it N=2 dualities and Z extremization in three dimensions},
  {\tt arXiv:1104.0487 [hep-th]}.
%
\bibitem{Jafferis}
  D.~L.~Jafferis,
  {\it The exact superconformal R-symmetry extremizes Z},
  JHEP {\bf 1205} (2012) 159,
  {\tt arXiv:1012.3210 [hep-th]}.
%
\bibitem{KM1}
  S.~M.~Kuzenko and I.~N.~McArthur,
  {\it On the background field method beyond one loop:
  A Manifestly covariant derivative expansion in superYang-Mills theories},
  JHEP {\bf 0305} (2003) 015,
  {\tt hep-th/0302205}.
%
\bibitem{KM2}
  S.~M.~Kuzenko and I.~N.~McArthur,
  {\it Low-energy dynamics in N=2 super QED: Two loop approximation},
  JHEP {\bf 0310} (2003) 029,
  {\tt hep-th/0308136}.
%
\bibitem{BPS1}
  I.L.~Buchbinder, N.G.~Pletnev and I.B.~Samsonov,
  {\it Effective action of three-dimensional extended supersymmetric matter on gauge superfield background},
  JHEP {\bf 1004} (2010) 124, {\tt arXiv:1003.4806 [hep-th]}.
%
\bibitem{DMP}
  N.~Drukker, M.~Marino and P.~Putrov,
  {\it From weak to strong coupling in ABJM theory},
  Commun.\ Math.\ Phys.\  {\bf 306} (2011) 511,
  {\tt arXiv:1007.3837 [hep-th]}.
%
\bibitem{BPS3}
  I.L.~Buchbinder, N.G.~Pletnev and I.B.~Samsonov,
  {\it Background field formalism and construction of effective action for N=2, d=3 supersymmetric gauge theories},
  Phys.\ Part.\ Nucl.\  {\bf 44} (2013) 234,
  {\tt arXiv:1206.5711 [hep-th]}.
%
\bibitem{Fujitsuka-Higgs}
  M.~Fujitsuka, M.~Honda and Y.~Yoshida,
  {\it Higgs branch localization of 3d N=2 theories},
  {\tt arXiv:1312.3627 [hep-th]}.
%
\bibitem{Benini-Higgs}
  F.~Benini and W.~Peelaers,
  {\it Higgs branch localization in three dimensions},
  {\tt arXiv:1312.6078 [hep-th]}.
%
\bibitem{Gomes12} N.~Doroud, J.~Gomis, B.~Le Floch and S.~Lee,
  {\it Exact results in D=2 supersymmetric gauge theories},
  JHEP {\bf 1305} (2013) 093, {\tt arXiv:1206.2606 [hep-th]}.
%
\bibitem{Benini}
  F.~Benini and S.~Cremonesi,
  {\it Partition functions of N=(2,2) gauge theories on $S^2$ and vortices},
  {\tt arXiv:1206.2356 [hep-th]}.
%
\bibitem{Zabzine1}
  J.~Kallen and M.~Zabzine,
  {\it Twisted supersymmetric 5D Yang-Mills theory and contact geometry},
  JHEP {\bf 1205} (2012) 125,
  {\tt arXiv:1202.1956 [hep-th]}.
%
\bibitem{Zabzine2}
  J.~Kallen, J.~Qiu and M.~Zabzine,
  {\it The perturbative partition function of supersymmetric 5D Yang-Mills theory with matter on the five-sphere},
  JHEP {\bf 1208} (2012) 157,
  {\tt arXiv:1206.6008 [hep-th]}.
%
\bibitem{Hosomichi5}
  K.~Hosomichi, R.-K.~Seong and S.~Terashima,
  {\it Supersymmetric gauge theories on the five-sphere},
  Nucl.\ Phys.\ B {\bf 865} (2012) 376,
  {\tt arXiv:1203.0371 [hep-th]}.
%
\bibitem{Jafferis5}
  D.L.~Jafferis and S.S.~Pufu,
  {\it Exact results for five-dimensional superconformal field theories with gravity duals},
  {\tt arXiv:1207.4359 [hep-th]}.
%
\bibitem{Siegel}
  P.~Dai, R.~-N.~Huang and W.~Siegel,
  {\it Covariant propagator in $AdS_5\times S^5$ superspace},
  JHEP {\bf 1003} (2010) 001,
  {\tt arXiv:0911.2211 [hep-th]}.

\end{thebibliography}
\end{document}